\documentclass[twocolumn,english,superscriptaddress,amsmath,amssymb,aps,prb,floatfix]{revtex4-2}

\usepackage[english]{babel}
\usepackage{booktabs}
\usepackage{graphicx}
\usepackage{dcolumn}
\usepackage{bm}
\usepackage{hyperref}
\hypersetup{colorlinks,
citecolor=blue,
filecolor=blue,
linkcolor=blue,
urlcolor=blue,
breaklinks=true
}

\usepackage[dvipsnames]{xcolor}

\newcommand{\RESP}[1]{{\color{black}#1}}
\begin{document}

\title{Artificial out-of-plane Ising antiferromagnet on the kagome lattice with very small further neighbour couplings}

\newcommand{\PSILMX}{Laboratory for Multiscale Materials Experiments, Paul Scherrer Institute, 5232 Villigen PSI,~Switzerland}
\newcommand{\ETHMesosys}{Laboratory for Mesoscopic Systems, Department of Materials, ETH Zurich, 8093 Zurich,~Switzerland}
\newcommand{\ETHLMI}{Laboratory for Magnetism and Interface Physics, Department of Materials, ETH Zurich, 8093 Zurich,~Switzerland}
\newcommand{\EPFL}{Institute of Physics, Ecole Polytechnique Fédérale de Lausanne (EPFL), CH-1015 Lausanne,~Switzerland}

\author{Jeanne~Colbois}
\email{jeanne.colbois@epfl.ch}
\affiliation{\EPFL}
\author{Kevin~Hofhuis}
\email{kevin.hofhuis@psi.ch}
\affiliation{\ETHMesosys}
\affiliation{\PSILMX}
\author{Zhaochu~Luo}
\affiliation{\ETHMesosys}
\affiliation{\PSILMX}
\author{Xueqiao~Wang}
\affiliation{\ETHMesosys}
\affiliation{\PSILMX}
\author{Ale\v{s}~Hrabec}
\affiliation{\ETHMesosys}
\affiliation{\PSILMX}
\affiliation{\ETHLMI}
\author{Laura~J.~Heyderman}
\affiliation{\ETHMesosys}
\affiliation{\PSILMX}
\author{Fr\'{e}d\'{e}ric~Mila}
\affiliation{\EPFL}

\date{\today}
\begin{abstract}
    Despite their simple formulation, short range classical antiferromagnetic Ising models on frustrated lattices give rise to exotic phases of matter, in particular due to their macroscopic ground state degeneracy. Recent experiments on artificial spin systems comprising arrays of chirally coupled nanomagnets provide a significant strengthening of the nearest neighbour couplings compared to systems with dipolar-coupled nanomagnets. This opens the way to design artificial spin systems emulating Ising models with nearest neighbour couplings. In this paper, we compare the results of an extensive investigation with tensor network and Monte Carlo simulations of the nearest- and further-neighbour ($J_1-J_2-J_{3||}$) kagome Ising antiferromagnet with the experimental spin-spin correlations of a kagome lattice of chirally coupled nanomagnets. Even though the ratios between the further neighbour couplings and the nearest neighbour coupling estimated from micromagnetic simulations are much smaller than for dipolar-coupled nanomagnets, we show that they still play an essential role in the selection of the correlations.
\end{abstract}

\maketitle

\section{Introduction}
Artificial spin systems, or artificial spin ices, are arrays of mesoscopic single-domain nanomagnets whose moments can be probed individually using magnetic microscopy techniques. The moments interact either via their magnetostatic fields or, in the case of connected nanomagnets, via exchange interactions. The flexibility in the design of these arrays makes them extremely useful for the study of frustration and emergent Coulomb phases in two dimensional systems~\cite{Wang2006, Nisoli2013,Heyderman2013}. The field has been continuously growing, showing great promise for the theory of frustrated systems, for experimental work on reprogrammable magnonic crystals~\cite{Rougemaille2019,Skjaervoe2020}, and for applications involving computation to implement computational logic~\cite{Arava2018, Caravelli2020, Gu2015}, as a natural support for artificial neural networks~\cite{Jensen2018,Zhou2020}, or for reprogrammable magnetic robots  \cite{Cui2019}.

When viewed as toy models for frustrated spin systems, the artificial spin systems face the intrinsic difficulty~\cite{Rougemaille2019} that the physics that one would like to study - collective phenomena and exotic emergent models - is exactly that which is hard to explore via the single spin flip dynamics available either through demagnetisation protocols~\cite{Nisoli2013, Ke2008, Nisoli2010} or thermal activity~\cite{Farhan2013, Farhan2013a, Morgan2013, Zhang2013, Arnalds2014, Kapaklis2014}. Additionally, the disorder of the arrays and the large size of the nanomagnets compared to the lattice spacing could in principle affect the simplicity of the modelling with Ising Hamiltonians. Despite these difficulties, great success has been achieved by describing artificial spin systems based on arrays of nanomagnets using Ising Hamiltonians with long range, dipolar couplings. Two prominent examples are the artificial dipolar kagome spin ice~\cite{Farhan2013a}, for which Monte Carlo simulations predict a series of magnetic phase transitions~\cite{Moller2009, Chern2011}, and the artificial dipolar kagome Ising antiferromagnet~\cite{Chioar2014, Chioar2016}, where a weakly first-order phase transition to a long-range ordered ground state has been established~\cite{Chioar2016, Hamp2018, Cugliandolo2020}. The quest for a true ground state degeneracy achieved in artificial spin ices reached an important step when a proposal from M\"oller and Moessner to recover the macroscopic ground state degeneracy of spin ice on the square lattice~\cite{Moller2006} was successfully realised~\cite{Rougemaille2019,Perrin2016,Farhan2019}. On the kagome lattice, however, the quest is still open. To control the couplings between the nanomagnets, new possibilities are being explored, such as the introduction of mixed anisotropy~\cite{Luo2019} or of a soft ferromagnetic underlayer below the moments~\cite{Kempinger2021}. In the present work, motivated by the design of arrays of chirally coupled nanomagnets~\cite{Luo2019} promising the implementation of strong nearest neighbour couplings, we numerically explore nearest- and further- neighbour Ising antiferromagnets on the kagome lattice and compare our theoretical results to those of an experiment on an artificial array of chirally coupled nanomagnets.

The difficulties with further neighbour frustrated spin systems are not restricted to experiments; it is also challenging to study them numerically using Monte Carlo simulations (see for instance the dipolar kagome Ising model~\cite{Chioar2016,Hamp2018}), since ad-hoc updates corresponding to the collective dynamics of the specific model are often needed to combat frustration and critical slowing down (see for instance Ref.~\onlinecite{Mizoguchi2017}). Another technique has shown promising results in the study of classical statistical mechanics: since the pioneering work of Nishino and Okunishi~\cite{Nishino1996}, the transfer matrix techniques applied successfully to classical spin systems in the 1950's and 1980's have found a new formulation in the modern language of tensor networks~\cite{Orus2014,Haegeman2017}. Partition functions, expressed as the trace of a transfer matrix, can be seen as the contraction of a 2D tensor network, which can be computed approximately using techniques initially developed for the study of 1D quantum systems~\cite{Schollwock2011, Fishman2018}, or techniques designed specifically for 2D classical problems~\cite{Nishino1996, Levin2007,Xie2012,Evenbly15}. Of particular interest to us, tensor networks have been used on the anisotropic kagome Ising antiferromagnet in a field~\cite{Li2010}, on 2D and 3D frustrated models to accurately determine their residual entropies~\cite{Vanderstraeten2018}, and more recently on a further neighbour frustrated model on the kagome lattice~\cite{Vanhecke2021}. In this work, we apply tensor network as a useful complement to Monte Carlo simulations for the study of the relevant further neighbour Ising models.

Continuing on the work of Ref.~\onlinecite{Luo2019}, we consider a kagome lattice of chirally coupled nanomagnets. The main idea is to rely on interfacial Dzyaloshinskii-Moriya interactions (DMI) to create regions with out-of-plane anisotropy (OOP regions) connected by chiral domain walls to regions with in-plane anisotropy (IP regions). Arranging the OOP regions on a kagome lattice and separating them with IP regions, one can then use the IP regions to create a very strong nearest neighbour antiferromagnetic coupling between OOP regions. At first sight one could expect the system to behave simply as a nearest neighbour Ising antiferromagnet on the kagome lattice. In Sec. \ref{sec:exp}, we describe this experiment in detail and perform micromagnetic simulations. The effective antiferromagnetic Ising model for the OOP regions that we obtain has very strong nearest neighbour couplings and, relatively, weak next and next-next nearest neighbour couplings - much smaller than in the dipolar case. In the remainder of the paper, we progressively build towards this further neighbour model, providing tensor network and Monte Carlo data for the spin-spin correlations of a series of short range Ising models on the kagome lattice, and comparing these correlations to the experimental results. We start in Sec.~\ref{sec:NN} with the nearest neighbour Ising antiferromagnet on the kagome lattice, which was extensively studied before the present work~\cite{Kano1953, Suto1981, Barry1997, Wills2002, Li2010, Apel2011, Chioar2014, Chioar2016, Muttalib2017}. We discuss its short range spin-spin correlations and we compute its correlation length using tensor networks. We find that even when considering a magnetic field, and despite the strong nearest neighbour couplings in the experiment, this model does not qualitatively explain the experimental observations. In Sec.~\ref{sec:NNN}, we show that even weak second nearest neighbour couplings can partially solve the issue, by changing the order of magnitude of the next nearest neighbour correlations as compared to the next-next nearest neighbour correlations. We consider the effective further neighbour model in a field in Sec.~\ref{sec:J1J2J3ph}, and finally give an overall discussion of the results in Sec. \ref{sec:discussion}.

\section{Experiment: a kagome lattice of chirally coupled nanomagnets}
\label{sec:exp}
We begin with a discussion of the experiments. They rely on the interfacial DMI~\cite{Dzyaloshinskii1958,Moriya1960,Fert1990,Crepieux1998}, 
\begin{equation}
    H_{\text{DMI}} = \mathbf{D}_{i,j} \cdot \left(\mathbf{S}_i \times \mathbf{S}_j\right),
\end{equation}
which is allowed in environments with a lack of inversion symmetry and is induced by spin-orbit coupling. As discussed in Ref.~\onlinecite{Luo2019}, this interaction can be leveraged to induce an effective antiferromagnetic coupling between two OOP regions with well-defined magnetic anisotropy, which take on the role of Ising spins. 

\subsection{Experimental setup}

\subsubsection{Concept}
In order to create these effective antiferromagnetic couplings, we deposit a magnetic trilayer of Pt, Co and Al. These trilayers have a large DMI at the Pt/Co interface and their magnetic anisotropy can be tuned by oxidising the AlOx. Through this oxidisation, one can thus create regions of in-plane (IP) anisotropy and regions of out-of-plane (OOP) anisotropy and control the location of the domain walls between them with left-handed chirality.
\begin{figure}[t]
\centering
\includegraphics[width=0.4\textwidth]{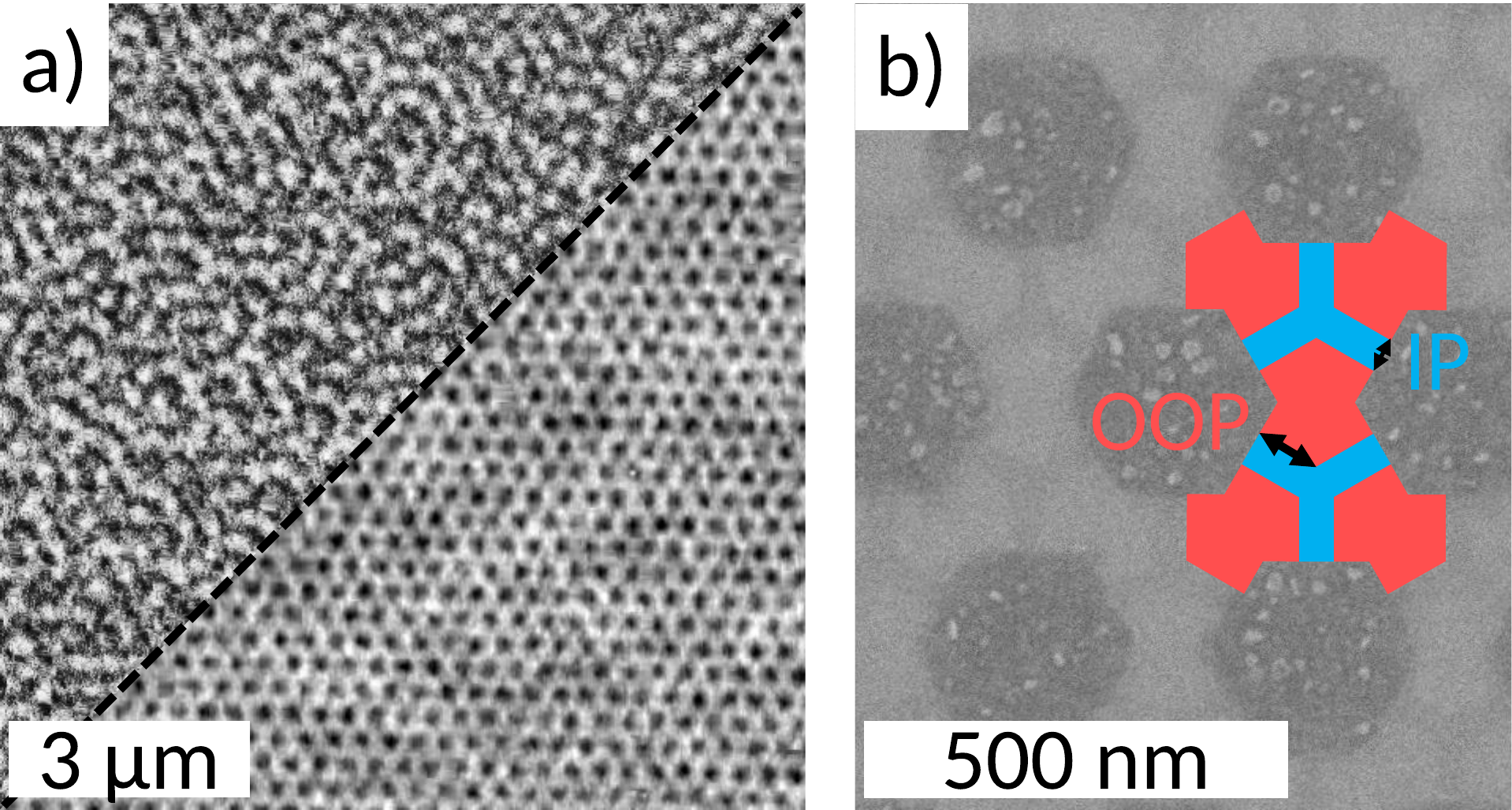}
\caption{\label{fig:expsetup} The samples are extended kagome lattices etched from Pt(6~nm)/Co(1.6~nm)/Al(2~nm) films. (a) Top left: magnetic force microscopy (MFM) image of a sample. The black/white contrast shows the up/down magnetisation of the regions with OOP anisotropy. Bottom right: AFM image of the sample structure. (b) Scanning electron microscopy image. Local Al oxidation can be seen as a grey contrast difference. A schematic of the out-of-plane (OOP) and in-plane (IP) regions is shown. The OOP (red) and IP (blue) anisotropy regions are defined by an OOP edge length of 100~nm an IP region width of 50~nm. The OOP centre to centre distance is then geometrically set and is $\sqrt{3}$OOP~edge~length~$+$~IP~width.}
\end{figure}

Consider a simple system where two regions with OOP anisotropy, which we model as Ising spins, are connected by a region with IP anisotropy. Because of the DMI, it is energetically favourable to have the Ising spins in an antiferromagnetic arrangement. In Ref.~\onlinecite{Luo2019}, it was shown that this antiferromagnetic alignment can be realised in Ising chains and on the square lattice after applying an external magnetic field and provided that the IP regions are small enough. A small kagome sample was also brought to a ground state of the antiferromagnetic nearest neighbour Ising model - namely, an ice-rule obeying state. Here, we apply this procedure to larger kagome systems (Fig.~\ref{fig:expsetup}).

\subsubsection{Details of the sample fabrication and demagnetisation protocol}
The fabrication of the samples is performed using electron beam lithography. First, Pt(6~nm)/Co(1.6~nm)/Al(2~nm) films are deposited on a 200~nm-thick SiNx layer on a silicon substrate by DC magnetron sputtering at a base pressure of $2 \cdot 10^{-8}$~mbar and at a deposition pressure of $3 \cdot 10^{-3}$~mbar. After patterning a polymethyl methacrylate (PMMA) mask with an electron beam writer, the trilayers are milled with Ar ions. The anisotropy of the various regions is determined by whether the Al layer is covered by the mask: the future IP regions are protected by 2~nm~Ta, while the future OOP regions are exposed to oxygen plasma at a power of 30~W and an oxygen pressure of 0.013~mbar. For statistics, 20 samples of size 20~$\times$ ~20~$\mu$m$^2$ are patterned on the same substrate. With an IP length of 50~nm and an OOP edge length of 100~nm, this amounts to approximately 2300 sites per sample.

All 20 samples have been AC demagnetised by a field-based protocol of sinusoidal OOP fields with decreasing amplitudes. The sinusoid amplitude decreased from 900~Oe to 30~Oe linearly in 4000~steps over a period of two hours. At each step, a single sinusoidal period is applied with a frequency of 2~Hz. \RESP{Several demagnetisation protocols were tested, as discussed in Appendix \ref{sec:AppDemag}.}

\subsubsection{Magnetic force microscopy}
Magnetic force microscopy (MFM) measurements are performed after the demagnetisation protocol to extract the OOP spin configurations. A protective layer of PMMA is spin coated on each sample for the MFM measurements. The MFM tips, coated by CoCr, are sensitive to magnetic stray fields. The tip is scanned over the sample at a frequency of 1~Hz and 512~pixels/line. An example of the resulting phase contrast is given in Fig.~\ref{fig:expsetup}. The contrast gives information about the OOP magnetisation, with white (black) for down (up) magnetised OOP regions. The low thickness of magnetic material makes the highly sensitive MFM tip a good candidate for imaging the stray fields of the samples.

\subsection{Micromagnetic simulations and effective model}
\label{sec:micro}
As we will discuss in Sections~\ref{sec:NN} and~\ref{sec:NNN}, the results of tensor networks and Monte Carlo simulations strongly suggest that further neighbour couplings are relevant in this experiment. We see two possible sources of such couplings: the simple dipolar interaction between the regions with out-of-plane (OOP) anisotropy, and a possible interaction (of dipolar nature or some more subtle origin) between the regions with in-plane (IP) anisotropy. Here, we determine the likelihood of these scenarios by performing micromagnetic simulations~\cite{Rougemaille2011,Saccone2019,Farhan2020,Vansteenkiste2014, Leliaert2018}.

\begin{figure}[t]
    \centering
    \includegraphics[width=0.4\textwidth]{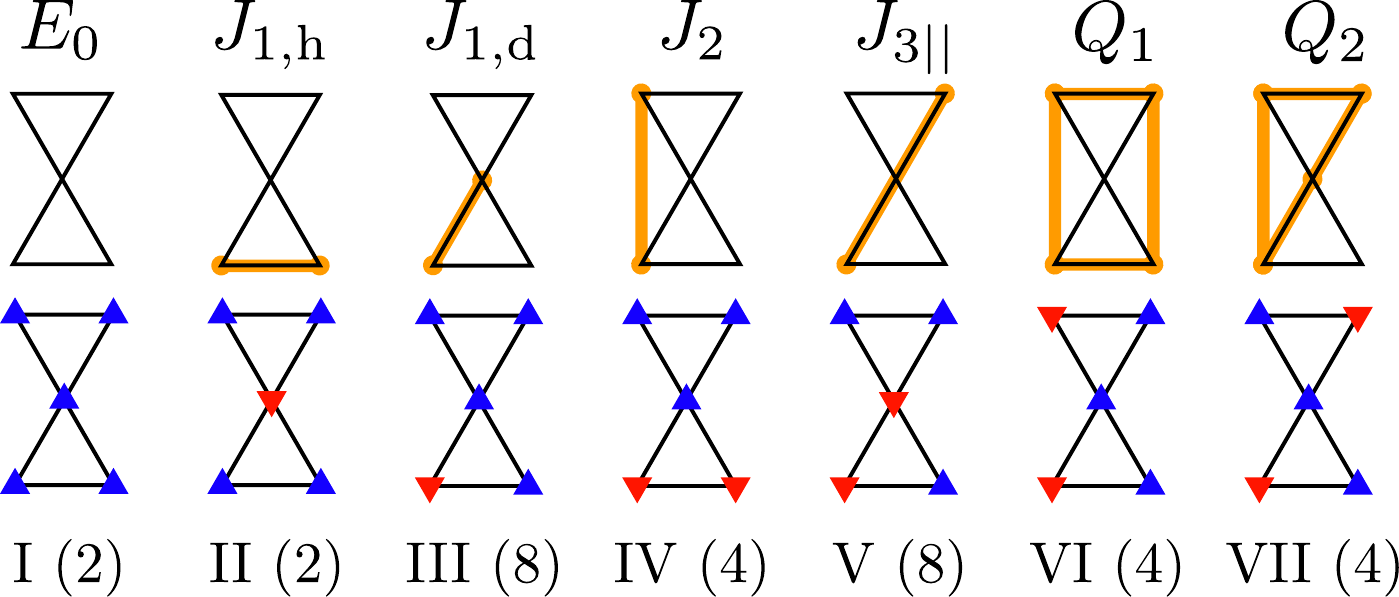}
    \caption{The first row indicates the couplings for the generic model with five sites of Eq. \ref{eq:5sites} (the symmetries of the system and the $\mathbb{Z}_2$ symmetry being imposed). The orange lines indicate the group of spins involved in the interaction corresponding to a given coupling. If the symmetries are imposed, some configurations will have the same energy; we can therefore group configurations by their energy. The second row shows one configuration (blue triangles for up spins, and red triangles for down spins) for each of these seven different groups. Each group is indexed by a Roman number and the Arabic numbers indicate how many configurations are in each group.}
    \label{fig:5sitescouplings}
\end{figure}

Our aim is to find an effective Hamiltonian for the OOP regions, which are modelled with Ising variables. \RESP{Indeed, it is \textit{a priori} unclear how to combine the IP region and the DMI interactions with the dipolar interactions to give an effective model for the Ising spins. We therefore perform micromagnetic simulations for islands (OOP and IP regions) with three and five sites to determine these effective interactions. For this, we consider the most generic spin model for five sites, including three-, four- and five-site interactions, respecting the cluster symmetries. In the micromagnetic computations, we did not find any evidence suggesting $\mathbb{Z}_2$ symmetry breaking. The effective Hamiltonian thus reduces to the following expression, where this symmetry is explicitly imposed by removing terms with an odd number of spins:}
\begin{equation}
\label{eq:5sites}
\begin{split}
	H &= E_0 + J_{1,\text{h}} \sum_{\langle i,j \rangle_{1,\text{h}}} \sigma_i \sigma_j + J_{1,\text{d}} \sum_{\langle i,j \rangle_{1d}} \sigma_i \sigma_j \\
	&+ J_{2} \sum_{\langle i,j \rangle_{2}} \sigma_i \sigma_j + J_{3||} \sum_{\langle i,j \rangle_{3||}} \sigma_i \sigma_j \\
	& + Q_{1} \sum_{\langle i,j,k,l \rangle_{1}} \sigma_i \sigma_j \sigma_k \sigma_l + Q_{2} \sum_{\langle i,j,k,l \rangle_{2}} \sigma_i \sigma_j \sigma_k \sigma_l \\
\end{split},
\end{equation}
\RESP{and} where the groups of spins to which the couplings apply are defined in Fig.~\ref{fig:5sitescouplings}. The $J$ couplings correspond to the usual pair interactions while the $Q$ couplings correspond to four-site interactions.

\subsubsection{Micromagnetic simulation parameters}

We performed micromagnetic simulations using MuMax~$^3$~(v3.10)~\cite{Vansteenkiste2014, Leliaert2018} to determine the (relative) coupling constants of the chirally coupled nanomagnet geometry. The simulations are carried out with a cell size of 0.866~by~0.866~by~1.6~nm$^3$ (1.6~nm being smaller than the exchange length of the system) and with 2048~by~2048~by~1 simulation world size. The cell sizes are chosen such that the angled edges are accurately simulated.

Typical material parameters for cobalt (Co) are used: a saturation magnetisation $M_\text{s} = 600$~kA$/$m, an exchange stiffness $A = 16\times 10^{-12}$~J$/$m, and a magnetocrystalline anisotropy $K_u = 570$~kJ$/$m$^3$ in the OOP region and $K_u = 0$ in the IP region. Interfacial DMI is introduced by setting \RESP{$D_\text{ind} = 0.9 \pm 0.1$~mJ$/$m$^2$~\cite{Luo2019}}. The Gilbert damping $\alpha$ is set to 1.0 to allow the simulation to relax quickly to the state we are interested in, i.e. the state after energy minimisation. The various coupling constants follow from simulating different OOP and IP configurations.

\subsubsection{Results of the micromagnetic simulations}
We first simulate the system in the absence of regions with IP anisotropy, where the interactions are of dipolar nature, similar to Ref.~\onlinecite{Chioar2016}. The detail of the results for each configuration of Fig.~\ref{fig:5sitescouplings} can be found in Appendix~\ref{sec:AppMicromagnetic}.
We find that 
\begin{equation}
\label{eq:OOPres}
\begin{split}
    J_1^{\text{dip}} &= (1.87\pm0.01) \cdot 10^{-20} \text{J}\\
    \frac{J_2^{\text{dip}}}{J_1^{\text{dip}}} &= 0.1188 \pm 0.0006\\ \frac{J_{3||}^{\text{dip}}}{J_1^{\text{dip}}} &= 0.0769 \pm 0.0003
\end{split}
\end{equation}
where the errors are dominated by the error on $J_1$ -- we find two slightly different values for $J_{1,\text{d}}$ and $J_{1,\text{h}}$ \footnote{$J_{1,\text{h}} = 1.868\cdot 10^{-20}$J while $J_{1,\text{d}} = 1.884 \cdot 10^{-20}$J. We use the average of these values for $J_1$, and the difference of these values to the average as an estimate of the error on $J_1$.}. $Q_1$ and $Q_2$ are zero within the error bars. In this case, we indeed recover the dipolar couplings, with a factor of 1.6 correction to the nearest neighbour coupling as compared to the point-dipole approximation, coming from the finite size of the elements and their proximity~\cite{Chioar2014, Chioar2016}).

\begin{figure}[t]
    \centering
    \includegraphics[width =0.45\textwidth]{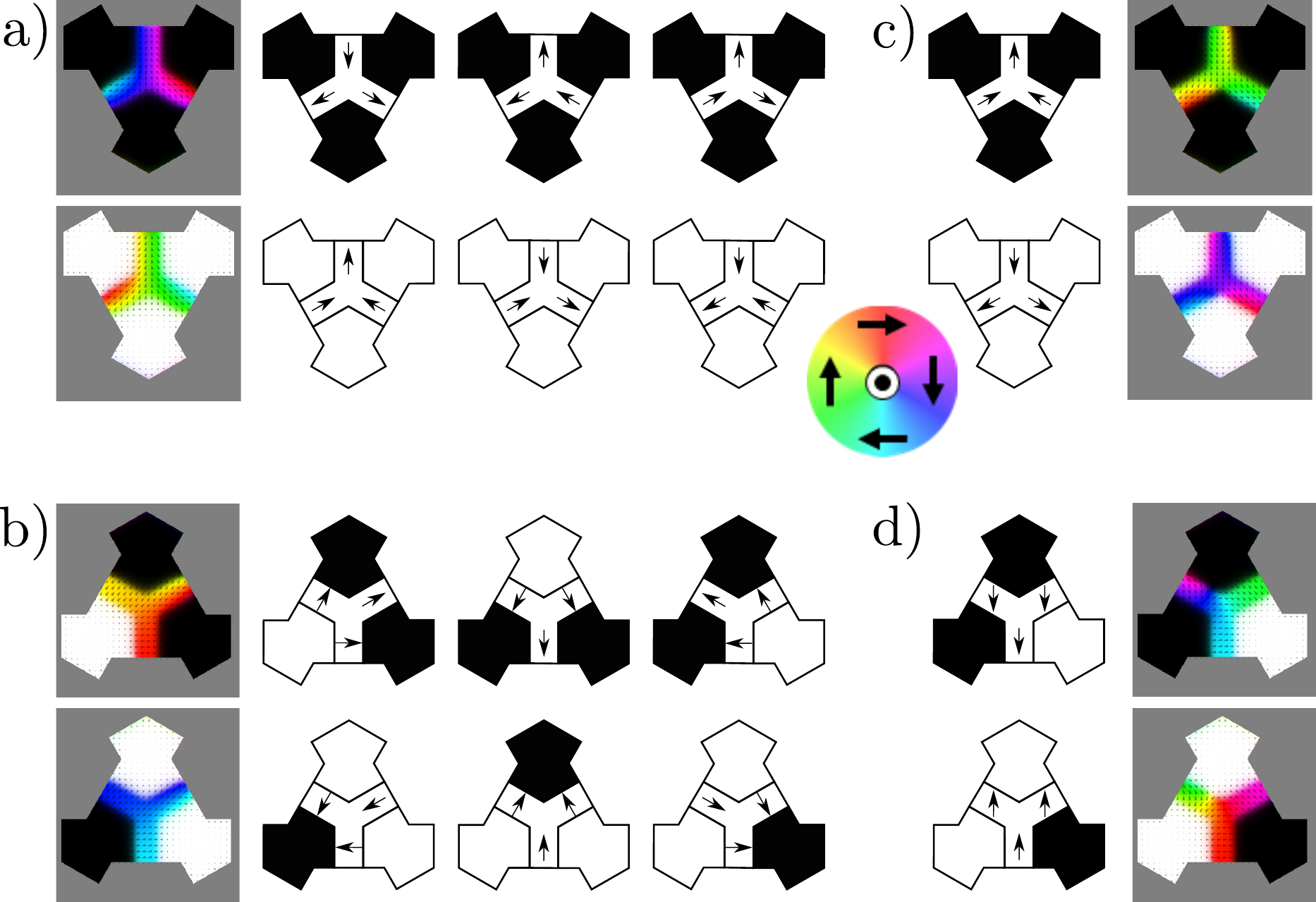}
    \caption{Configurations of the regions with OOP anisotropy and IP anisotropy on a triangle. Black and white OOP regions correspond to up and down Ising spins, respectively. (a) and (b) In the grey squares, the results of the micromagnetic simulations are given; these results are also schematically represented for readability. The colours in the IP region refer to the orientation of the local magnetisation in the plane, as labelled by the coloured disk. (a) Optimal configuration of the IP regions when all the Ising spins are aligned: 1-in 2-out for all up spins, 2-in 1-out for all down spins. These configurations have energy $E = -7.5588 \cdot 10^{-18} \pm 10^{-22}$ J. (b) Optimal configurations of the IP regions when one or two spins are up: the IP region points towards the up spins. These configurations have energy $E = -8.209 \cdot 10^{-18} \pm 5 \cdot 10^{-21}$ J. (c) and (d) Examples of IP configurations which give a higher energy for the OOP configurations corresponding to (a) and (b). Here, the schematic representation shows how the IP region was initialised. (c) A two-in one-out configuration for all up spins, or a one-in two-out configuration for all down spin, is not energetically favourable. (d) If, initially, the IP magnetisation does not point towards the up spin(s), the system tries to relax to the IP configurations shown in (b). }
    \label{fig:3sites}
\end{figure}
\begin{figure}[t]
    \centering
    \includegraphics[width =0.45\textwidth]{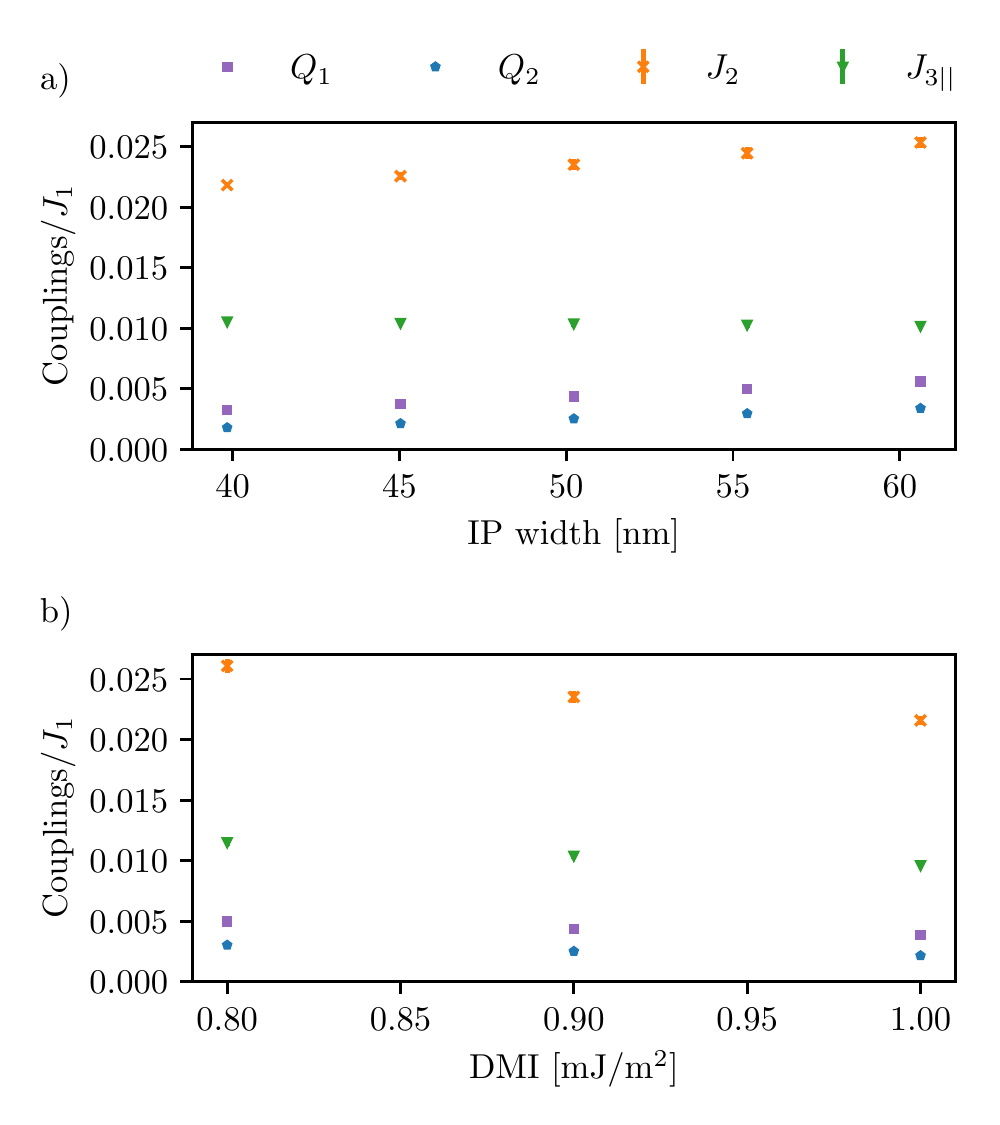}
    \caption{The value of the couplings from Fig.~\ref{fig:5sitescouplings}, relative to $J_1$, as obtained from the micromagnetic configurations. Error bars for $J_2/J_1$ and $J_{3||}/J_1$ are computed from the error bars on $J_1$ and from the change in the value of $J_2$ and $J_{3||}$ when $Q_1$ and $Q_2$ are taken into account versus when they are not.  \RESP{a) Variation of the couplings as a function of the width of the IP domain, with $D = 0.9$~mJ$/$m$^2$. b) Variation of the couplings as a function of the DMI, with an IP width of 50~nm.}}
    \label{fig:micromagcouplings}
\end{figure}

\par We then consider the effect of the IP regions. A small subtlety has to be addressed in this case: \textit{a priori}, for a given OOP configuration (labelled by roman numbers in Fig.~\ref{fig:5sitescouplings}), there can be several IP configurations. We proceed in two steps
\begin{enumerate}
    \item For each configuration of three Ising spins on a triangle, we find the IP configuration that minimises the energy (see Fig.~\ref{fig:3sites}). From this, we can already extract the nearest neighbour coupling
    \begin{equation}
        \label{eq:J1triangle}
        J_1 = (1.63 \pm 0.01) \cdot 10^{-19} \text{J},
    \end{equation}
     (for an IP width of 50~nm) that is almost one order of magnitude larger than in Eq.~\ref{eq:OOPres}. The main source of the error is that the micromagnetic results are not completely rotation-invariant, a difficulty probably related to the use of a square grid for discretization in MuMax$^3$.
    \item For each configuration of five Ising spins on a pair of triangles, we look for the combined configuration of the two IP regions minimising the energy. The optimal configurations and their respective energies can be found in Appendix~\ref{sec:AppMicromagnetic}. The effective model for the OOP region (Ising spins) is based on these energies. 
\end{enumerate}
\RESP{It is important to note that we assume here that the IP region will take the configuration that locally minimises the energy. Because of the limited MFM resolution and the small width of the IP regions, we have not been able to determine from the experimental scans whether this is actually the case.}\\

\RESP{We give the results for various IP widths  and for various values of the DMI in Fig.~\ref{fig:micromagcouplings}. In particular, for an IP width of 50~nm and for $D = 0.9$~mJ$/$m$^2$}, corresponding to our experiment, we find for the full $J_1$ coupling involving dipolar and IP-mediated interactions:
\begin{equation}
    J_1 =(1.6 \pm 0.03) \cdot 10^{-19} \text{J}.
\end{equation}
This result, computed by minimising the energy of pairs of triangles, is in agreement (within the error bars) with the result of the equality in Eq.~\ref{eq:J1triangle} which was obtained by minimising the energy of single triangles.
For the further neighbour couplings, we find:
\begin{equation}
    \frac{J_2}{J_1} \cong 0.0235 \pm 0.0004  
\end{equation}
and  
\begin{equation}
   \frac{J_{3||}}{J_1} \cong 0.0103 \pm 0.0001.
\end{equation}
We see that, as compared to the pure OOP model, when the IP region of width 50~nm is taken into account, the nearest neighbour coupling is increased by almost a factor of 10, but that the further neighbour couplings are affected as well by a factor
\begin{equation}
   \frac{J_2}{J_2^{\text{dip}}} \cong 1.67 \pm 0.01 
\end{equation}
and 
\begin{equation}
   \frac{J_{3||}}{J_{3||}^{\text{dip}}} \cong 1.12 \pm 0.01,
\end{equation}
respectively.
If the IP region had only contributed to the nearest neighbour coupling (and not to further neighbour couplings), the $J_2^{\text{dip}}$ and $J_{3||}^{\text{dip}}$ couplings would have had to be compared not to $J_1^{\text{dip}}$ but to the full $J_1$. In this case, instead of the couplings in Fig.~\ref{fig:micromagcouplings}, for an IP width of 50~nm, we would have found that
\begin{equation}
\label{eq:OOPvsIPJ1}
\begin{split}
    \frac{J_2^{\text{dip}}}{J_1} &= 0.0139 \pm 0.0002\\ \frac{J_{3||}^{\text{dip}}}{J_1} &= 0.0090 \pm 0.0001.
\end{split}
\end{equation}
\RESP{Thus, when the DMI and dipolar interactions are considered for a cluster of five sites, we find that the nearest neighbour couplings are increased by a much larger factor than the second and third neighbour couplings, resulting in an effective model with very small further neighbour interactions as compared to the nearest neighbour couplings. These results vary only slightly with the change in IP width or in DMI, as shown in Fig.~\ref{fig:micromagcouplings}}.

\subsubsection{Sources of a magnetic field}
On top of the effective model discussed above, we have to consider the possible effect of a magnetic field. Indeed, there are two possible sources of a longitudinal field (i.e., a magnetic field parallel to the Ising spin axis) in the experiment: the offset of the demagnetisation field, and the stray field of the MFM measurement tip\RESP{\footnote{A careful look at the MFM results showed some islands changing contrast during the tip scanning process, which suggest a tip-sample interaction}}. The saturation field of the samples (of the order of 4$J_1$) is estimated to be of the order of 2~kOe. The offset in the demagnetisation field is of the order of 10~Oe, and the stray fields from the MFM tip are of order 500~Oe. This means that if there is a field, we expect it to be of the order of $h \sim J_2$ to $h \sim J_1$.

\subsection{Experimental results}
\begin{figure}[t]
\centering
\includegraphics[width=0.48\textwidth]{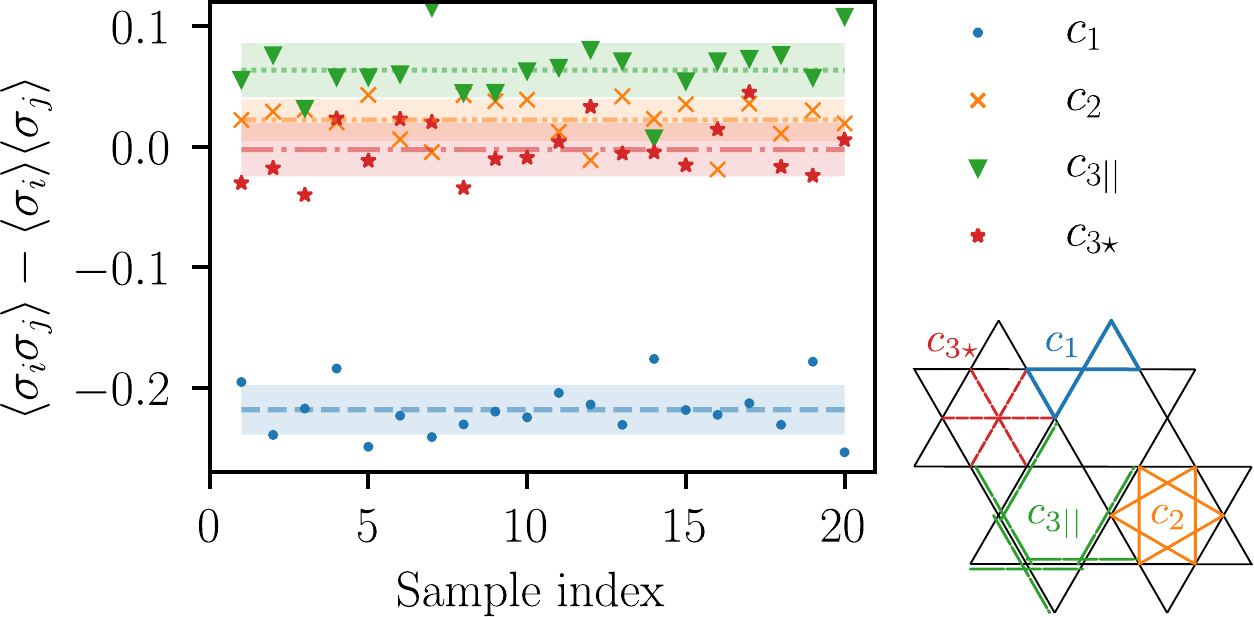}
\caption{\label{fig:expSpinSpin} Short range spin-spin correlations of the 20 kagome samples. The considered spin pairs are defined in the inset. $c_1$ and $c_2$ correspond to nearest and next nearest neighbour correlations, $c_{3||}$ and $c_{3\star}$ to 3rd neighbours. The dashed lines show the mean value and the highlighted areas correspond to one standard deviation around the average.}
\end{figure}
\begin{table}[t]
\caption{\label{tab:expResults}Experimental values for the first spin-spin correlations (See Fig.~\ref{fig:expSpinSpin}).}
\begin{ruledtabular}
\begin{tabular}{cccc}
$c_1$ & $c_2$ & $c_{3||}$& $c_{3\star}$ \\
\colrule
$-0.218 \pm 0.02$ & $0.022 \pm 0.02$ & $0.063 \pm 0.02$ & $-0.003 \pm 0.02$\\
\end{tabular}
\end{ruledtabular}
\end{table}
\begin{table}[t]
\caption{\label{tab:expResultsNumberTriangles}Experimental values for the proportion of triangles with a given net magnetisation.}
\begin{ruledtabular}
\begin{tabular}{cccc}
$r_{m_{\Delta} = +3}$ & $r_{m_{\Delta} = +1}$ & $r_{m_{\Delta} = -1}$ & $r_{m_{\Delta} = -3}$  \\
\colrule
$0.10 \pm 0.024$ & $0.60 \pm 0.03$ & $0.28 \pm 0.04$ & $0.015 \pm 0.008$\\
\end{tabular}
\end{ruledtabular}
\end{table}
To characterise the experimental results, we proceed as in Ref.~\onlinecite{Chioar2014} and we extract the experimental spin-spin correlations in order to compare them against the models under consideration. More precisely, for the four types of spin pairs illustrated in Fig.~\ref{fig:expSpinSpin}, we estimate the connected correlation functions
\begin{equation}
     \langle \sigma_i \sigma_j \rangle - \langle \sigma_i \rangle \langle \sigma_j \rangle
\end{equation}
using, for spin pairs of type $k$, the unbiased estimator
\begin{equation}
    c_k = \tfrac {M}{M-1}\left( \frac{1}{M}\sum_{(i,j)_k}{\sigma_{i}\sigma_{j}}-\frac{1}{M^2}\sum_{i:(i,j)_k}\sigma_i \sum_{j:(i,j)_k} \sigma_j\right)
\end{equation}
where $M$ is the number of spin pairs of type $k$ in the lattice and where $\sum_{(i,j)_k}$ denotes the sum over all spin pairs of type $k$. Note that we use the statistical field theory expressions: ``connected'' correlations means $\langle \sigma_i \sigma_j \rangle - \langle \sigma_i \rangle \langle \sigma_j \rangle$ and ``disconnected'' correlations means $\langle \sigma_i \sigma_j \rangle$. The results for each sample are shown in Fig.~\ref{fig:expSpinSpin}. In this figure, the shaded areas correspond to the values of the respective correlations that are within one standard deviation of the mean over the samples. These means are given in Table~\ref{tab:expResults}.
As shown in Fig.~\ref{fig:expSpinSpin}, the results vary significantly from one sample to the next. It is important to note that the descending order of the correlations is almost systematically $|c_1| > c_{3||} > c_2$, often with $c_2 \gtrsim c_{3\star}$. This qualitative result will drive our analysis.\\

Additionally, we consider two other observables which are characteristic of the experimental results. First, for the net magnetisation of the samples, we get a result significantly different from zero: $m = 0.19 \pm 0.05$; all the samples are magnetised in the same direction. Second, in each sample there are 10\% to 20\% of ferromagnetic triangles (``frustrated'' triangles, which do not respect the two-up one-down, two-down one-up ice rules); we therefore compute the proportion of triangles with a given magnetisation (analogous to the often used charge definition, but without introducing a sign). The results are given in Table~\ref{tab:expResultsNumberTriangles}; overall, the proportion of ice-rule-breaking triangles is $r_{fr} = r_{m_{\Delta} = +3} + r_{m_{\Delta} = -3} = 0.12 \pm 0.03$.

\section{Nearest neighbour model}
Given the very strong nearest neighbour couplings predicted by the micromagnetic simulations, we start by checking whether the experimental results from Sec.~\ref{sec:exp} can be understood using a purely nearest neighbour model,
\begin{equation}
    \label{eq:nnIsingHamiltonian}
    H_{\text{NN}} = J_1 \sum_{\langle i,j \rangle} \sigma_i \sigma_j - h \sum_i \sigma_i.
\end{equation}

We begin by revisiting the well-known case in zero field using tensor networks. Because of the significant magnetisation of the experimental results, we then consider how the picture is modified when the external magnetic field is turned on, a case which was has been extensively studied~\cite{Moessner2000,Moessner2001,Udagawa2002,Moessner2003,Li2010} but for which we did not find data regarding the first spin-spin correlations as a function of temperature.

\label{sec:NN}
\subsection{Nearest neighbour Ising antiferromagnet in zero field}

Introduced in Ref.~\onlinecite{Syozi1951}, the kagome lattice is a natural playground for frustrated models. In particular, the nearest neighbour Ising antiferromagnet on this lattice is known to have no order at any temperature, exhibiting a macroscopic ground state degeneracy with an extremely large residual entropy $S = 0.50183...$~\cite{Kano1953}. Later, it was shown by A. S\"ut\"o that the spin-spin correlations in the nearest-neighbour antiferromagnetic Ising model on the kagome lattice decay exponentially~\cite{Suto1981}, and that
\begin{equation}
    \left| \left< S_i S_j \right> \right| \leq 4 \times 0.74^{|i-j|}
\end{equation}
which implies an upper bound for the correlation length \footnote{Note that the magnetisation is zero in the ground state of the model, hence the connected or disconnected correlations have the same value}:
\begin{equation}
    \xi \leq \xi_{\text{S\"ut\"o}} -1/\ln{0.74} = 3.32109.
\end{equation}
An exact result was obtained more recently for the correlation length between ``middle spins'' from transfer matrix computations using Toeplitz determinants~\cite{Apel2011}:
\begin{equation}
    \label{eq:corrlengthexact}
    \xi = - \frac{2}{\ln\left( 10 - \sqrt{96}\right)} \cong 1.250559...
\end{equation}

\begin{figure}[t]
    \centering
    \includegraphics[width = 0.48\textwidth]{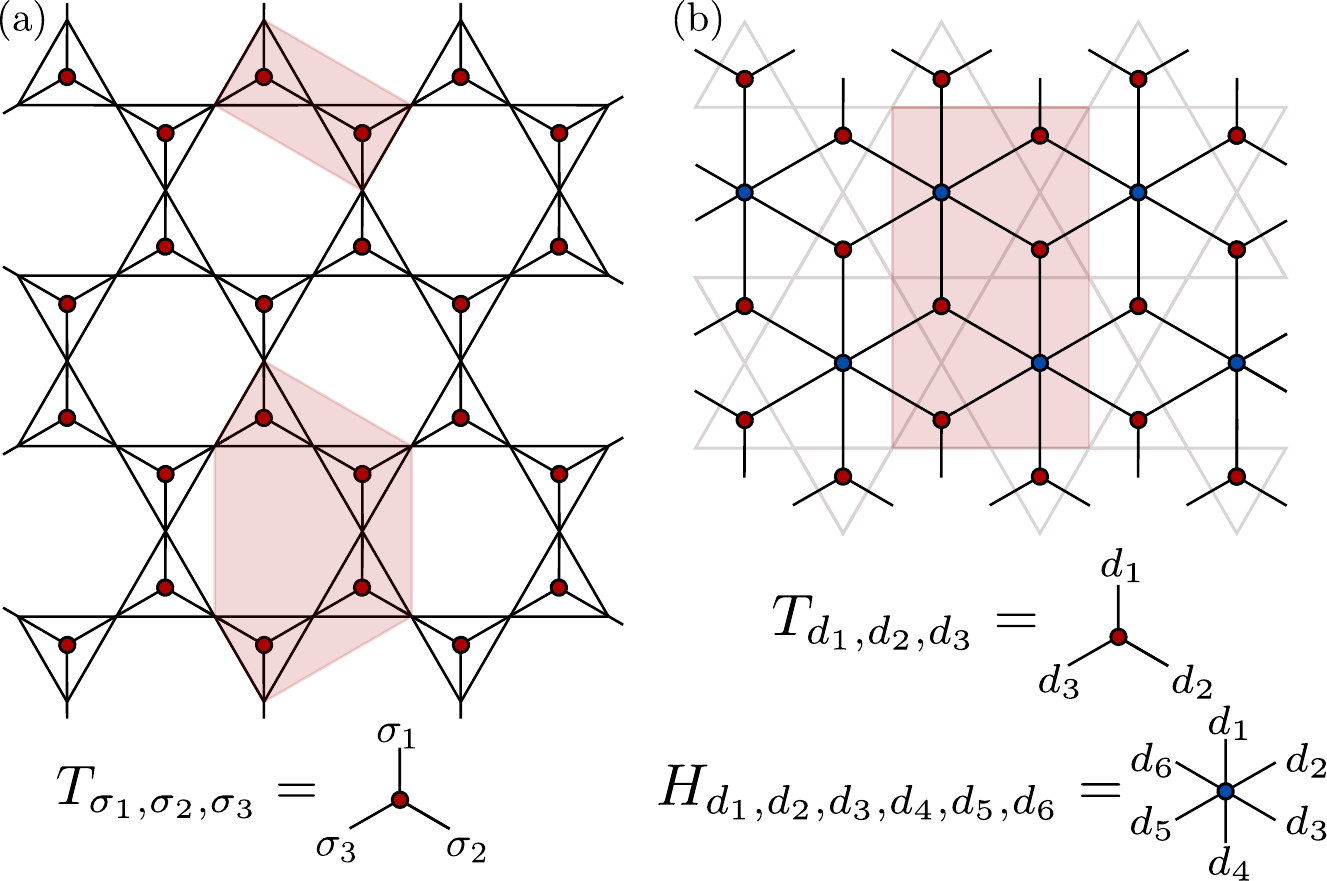}
    \caption{Tensor network formulations for the partition function of the nearest neighbour model on the kagome lattice. In highlight, we indicate the regions that can be grouped in a single tensor such that the tensor network on the honeycomb or dice lattice is reduced to a square lattice tensor network which can then be contracted using vumps. (a) ``Direct'' construction on the honeycomb lattice, with the tensor on the triangle given by Eq.~\ref{eq:weightJ1}, and which is easily extended to the nearest neighbour model in a field. (b) ``Dual'' construction on the dice lattice, with the tensors given in Eqs.~\ref{eq:triangledimers}, \ref{eq:hexdimers}. This construction is easily extended to the next nearest neighbour model in zero field.}
    \label{fig:partitionfunction}
\end{figure}

It is well known that the correlation length can be computed from the two leading eigenvalues of the transfer matrix $\lambda_1, \lambda_2$ (see for instance~\cite{Baxter1982}) as
\begin{equation}
    \frac{1}{\xi} = \ln{\frac{\lambda_2}{\lambda_1}}.
\end{equation}
We write the partition function of the nearest neighbour model on the kagome lattice as a tensor network as depicted in Fig.~\ref{fig:partitionfunction}. We contract these networks using the vumps algorithm (for ``variational uniform matrix product states'')~\cite{ZaunerStauber2018,Fishman2018}, which finds the leading eigenvector of the (infinite) 1D transfer matrix using as a variational Ansatz a translationally invariant matrix product state (MPS). 

In the ``direct'' construction, the tensor
\begin{equation}
    \label{eq:weightJ1}
    T_{\sigma_i, \sigma_j, \sigma_k} = e^{-\beta J_1 (\sigma_i \sigma_j + \sigma_j \sigma_k + \sigma_k \sigma_i + 1)}
\end{equation}
describes all the Boltzmann weights on a triangle, with a shift corresponding to the ground state energy on a triangle. In the ``dual'' construction, we first define a classical dimer variable on each nearest neighbour bond $\langle i, j \rangle$ 
\begin{equation}
    d_{i,j} := \sigma_i \sigma_j.
\end{equation}
When $d_{i,j}=1$, there is a dimer, and when $d_{i,j}=-1$, there is no dimer.
The tensor on each triangle is thus 
\begin{equation}
    \label{eq:triangledimers}
    T_{d_{i,j}, d_{j,k}, d_{k,i}} = \begin{cases}
    e^{-\beta J_1 (d_{i,j}+ d_{j,k} + d_{k,i} + 1)} & \, d_{i,j}d_{j,k}d_{k,i} =1\\
    0 & \text{ otherwise}
    \end{cases},
\end{equation}
The tensor on each hexagon $H_{d_1, d_2, d_3, d_4, d_5, d_6}$ only imposes that the number of dimers is even, such that the dimer configuration maps to a spin configuration:
\begin{equation}
    \label{eq:hexdimers}
    H_{d_1, d_2, d_3, d_4, d_5, d_6} = \begin{cases}
    1 & \, \text{ if } d_1 d_2 d_3 d_4 d_5 d_6 = 1\\
    0 & \text{ otherwise}
    \end{cases}
\end{equation}
\\
\begin{table}[t]
\caption{\label{tab:corr}First spin-spin correlations in the ground state of the nearest neighbour model (See Fig.~\ref{fig:ExactResults}).}
\begin{ruledtabular}
\begin{tabular}{l|cccc}
 & $c_1$ & $c_2$ & $c_{3||}$& $c_{3\star}$ \\
\colrule
Exact~\cite{Barry1997} & $-1/3$ & 0.1234... & 0.1014 ... & -0.0743 \\
Direct TN & $-1/3$ & 0.12343725 & 0.10144577& -0.07480837\\
Dual TN & $-1/3$ & 0.12343725 & 0.10144577 & -0.07480837\\
\end{tabular}
\end{ruledtabular}
\end{table}

In both constructions, we recover the exact ground state entropy of the kagome Ising antiferromagnet~\cite{Kano1953} to the 14th decimal place. For the first few correlations in the ground state, the two tensor constructions agree to the 3rd decimal place with the exact results~\cite{Barry1997, Muttalib2017}(Table \ref{tab:corr}).

\begin{figure}[t]
    \centering
    \includegraphics[width = 0.48\textwidth]{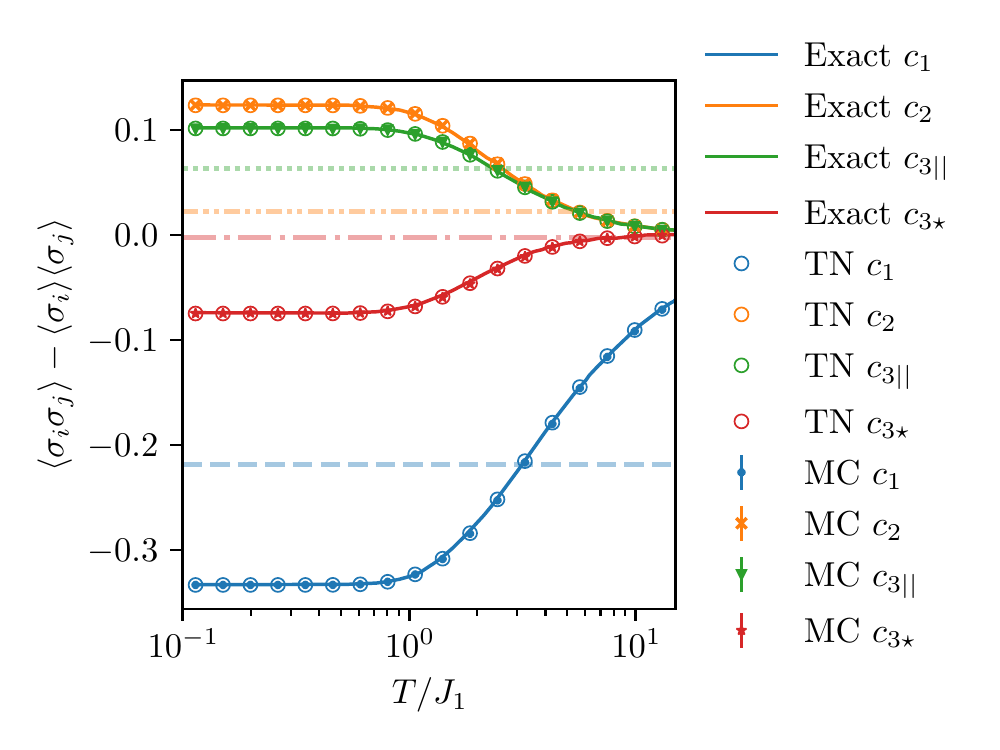}
    \caption{First few correlations in the nearest neighbour Ising antiferromagnet: comparison between the exact results (Ref.~\onlinecite{Barry1997}, extracted from Ref.~\onlinecite{Muttalib2017}), and the experimental results. As a benchmark, we also show the results of the two numerical methods (TN: tensor networks, with (final) MPS bond dimension $\chi = 13$, MC: Monte Carlo with number of sites $N = 576$). The dashed lines correspond to the four experimental correlations results from Fig.~\ref{fig:expSpinSpin}. The results for the nearest neighbour correlations are compatible with the experiment at relatively large effective temperatures ($T \sim 3 J_1$), but the order of $c_2$ and $c_{3||}$ in the experiment is inverted as compared to the theoretical values in the nearest neighbour model. }
    \label{fig:ExactResults}
\end{figure}

As a function of the temperature, the comparison between the tensor networks and the exact solution for the first few correlations is given in Fig.~\ref{fig:ExactResults}. We also use it as a benchmark for the Monte Carlo simulations, performed with a dual worm algorithm~\cite{Rakala2017} (Appendix~\ref{sec:AppMC}).

\begin{figure}[t]
    \centering
    \includegraphics[width = 0.45\textwidth]{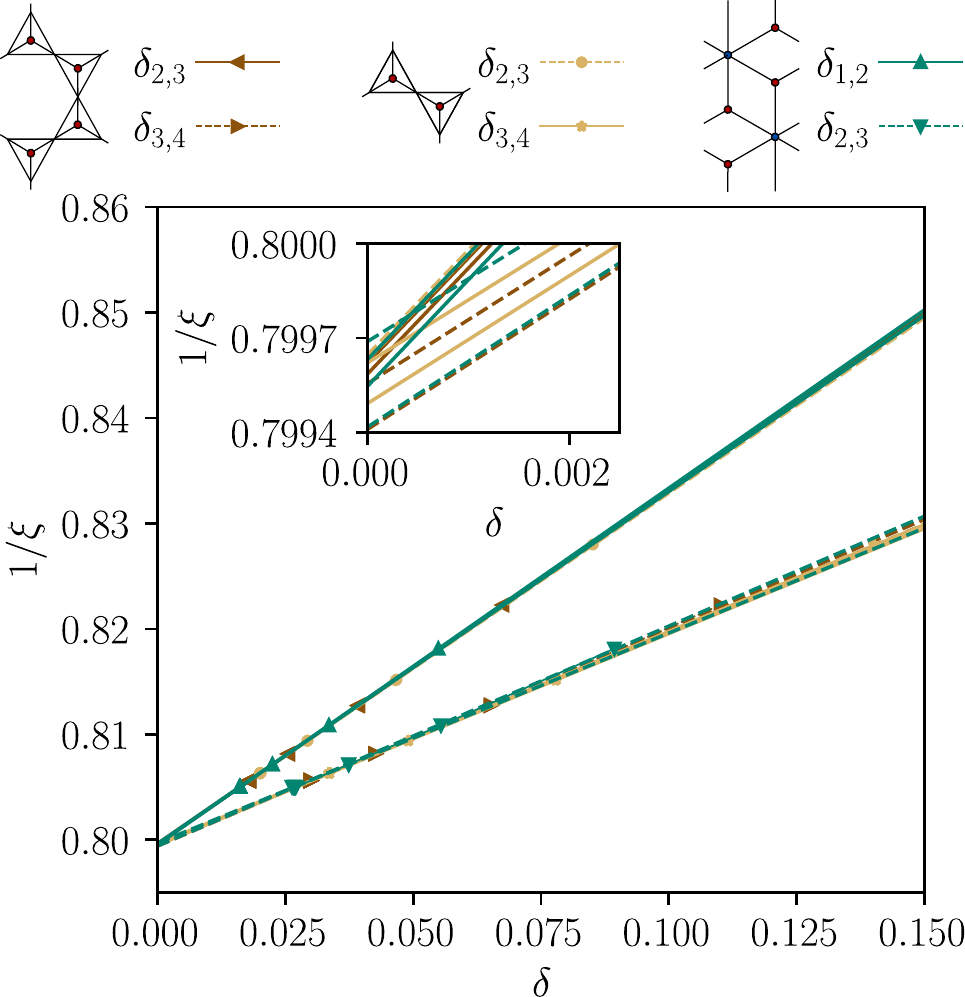}
    \caption{For two slightly different tensor network constructions, we plot one over the correlation length (Eq. \ref{eq:correlationlength}) as a function of the logarithm of another ``gap'' in the transfer matrix spectrum (Eq. \ref{eq:delta}). The legend indicates, for each set of points, to which unit cell it is associated and which gap $\delta$ in the transfer matrix spectrum is selected. The correlation length corresponds to the correlations along a line of kagome lattice nearest neighbours. For each set of points we do two fits, one with all the points and one without the largest $\delta$. The inset shows the crossing of each of these fit with the axis $\delta = 0$, giving an estimate for the actual correlation length in the limit of an infinite bond dimension. }
    \label{fig:CorrelationLength}
\end{figure}
To extract the correlation length from the direct tensor network, we follow Ref.~\onlinecite{Rams2018} and compute the first eigenvalues of the transfer matrix
\begin{equation}
    \lambda_j = e^{-(\epsilon_j + i \phi_j)P}
\end{equation}
with $|\lambda_1| > | \lambda_2| > |\lambda_3| \geq |\lambda_4| \geq ... $, where $\epsilon_j$ and $\phi_j$ correspond respectively to the log of the absolute value and to the phase, and where $P$ stands for the periodicity of the MPS in units of the number of lattice sites. The correlation length is thus given by
\begin{equation}
    \label{eq:correlationlength}
    \frac{1}{\xi} = \epsilon_2 - \epsilon_1.
\end{equation}
To extract the correlation length in the dual tensor network formulation, one can use the fact that the product of two spins is given by the product along a path of all the dimer variables separating them. We thus define a correlation tensor $C$ as the contraction along $d_c$ of the two tensors
\begin{align}
    T^{c}_{d_{\text{top},2}, d_{r}, d_{c}} &= d_{\text{top},2}T_{d_{\text{top},2}, d_{r}, d_{c}},\\
    H^{c}_{d_{\text{top}, 1}, d_{c}, d_3, d_4, d_5, d_6} &= d_{\text{top},1}H_{d_{\text{top},1}, d_c, d_3, d_4, d_5, d_6}.
\end{align}
The correlation length is then given by
\begin{equation}
    \frac{1}{\xi} = \epsilon - \epsilon_1
\end{equation}
with $\lambda = e^{-(\epsilon + i \phi)P}$ corresponding to the leading eigenvalue of the transfer matrix based on $C$ (note that here, the periodicity of the MPS compared to the lattice is $P = 2$). In both the direct and the dual cases, we define
\begin{equation}
    \label{eq:delta}
    \delta_{i,j} = \epsilon_j - \epsilon_i
\end{equation}
the other gaps in the transfer matrix spectrum, which must go to zero in the limit of infinite bond dimensions to produce corrections to the form of the correlations~\cite{Rams2018}. In Fig.~\ref{fig:CorrelationLength}, the value of the correlation length as a function of such gaps for different bond dimensions is illustrated; in the direct construction this is shown for two different orientations for the contractions displayed in Fig.~\ref{fig:partitionfunction}. We are limited to a maximal bond dimension of $\chi = 14$, after which the Schmidt values fall below numerical precision. From using the various constructions, selecting various gaps, and making the fits with all the points and all the points but one, we can finally extract the correlation length along the nearest neighbour chains and the errors on its estimation as
\begin{equation}
    \xi = 1.2507 \pm 0.0003
\end{equation}
in units of the lattice spacing. This is extremely short, consistent with the upper bond from A. S\"ut\"o's computation, and matches within the error bars the exact solution given in~Eq.~\ref{eq:corrlengthexact}.

In Fig.~\ref{fig:ExactResults}, we compare the experimental results for the correlations to the nearest neighbour Ising antiferromagnet at all temperatures. At this stage, we should recall that the samples are not at all expected to be thermally active. Here, the temperature is introduced as a Lagrange parameter for the energy, in an attempt to account for the non-zero percentage of ice-rule-breaking triangles in a non-biased way \footnote{As discussed in Ref.~\onlinecite{Rougemaille2019}, the notion of effective temperature mostly describes whether a given snapshot of a configuration of the nanomagnets is characteristic of some state at equilibrium, or conversely, whether an out-of-equilibrium description is required. In the same spirit, it is now quite standard in artificial spin ices to consider an energy-based effective temperature corresponding to the canonical distribution describing effectively the vertex population~\cite{Nisoli2010}. This is the approach that we followed here.}. Accordingly, corresponding to the non-zero proportion of frustrated triangles, one can see that the experimental value for the nearest neighbour correlations correspond to a finite effective temperature.

The description with the nearest neighbour model fails in two ways. First, because there is no $\mathbb{Z}_2$ symmetry breaking, the proportions of triangles of given magnetisation (Table~\ref{tab:expResultsNumberTriangles}) and the finite magnetisation cannot be recovered. Second, the relative order of $c_2$ and $c_{3||}$ in the experiment is inverted as compared to the nearest neighbour model: despite the strong value of the nearest neighbour couplings as compared to the further neighbour couplings from Fig.~\ref{fig:micromagcouplings}, considering only nearest neighbour couplings does not allow for a valid qualitative description of the experiment from the point of view of the descending order of the spin-spin correlations.

\subsection{Nearest neighbour Ising antiferromagnet in a field}

The experiments exhibit a finite magnetisation and an imbalance of the number of triangles with a given magnetisation. A simple way to account for this in the model is to introduce the corresponding Lagrange parameter, that is, the magnetic field. It is thus natural, as a next step, to test whether a longitudinal magnetic field lifting (partially) the ground state degeneracy of the nearest neighbour ground state could be enough to explain not only the magnetisation and the proportion of triangles with a given magnetisation but also the result that $c_{3||} \gtrsim 2 c_2$.

We use a tensor network contraction to compute the correlations systematically as a function of field and temperature. Before doing so, we take advantage of this construction to study the ground state phase diagram of this kagome Ising antiferromagnet in a field.

\begin{figure}[t]
    \centering
    \includegraphics[width = 0.45\textwidth]{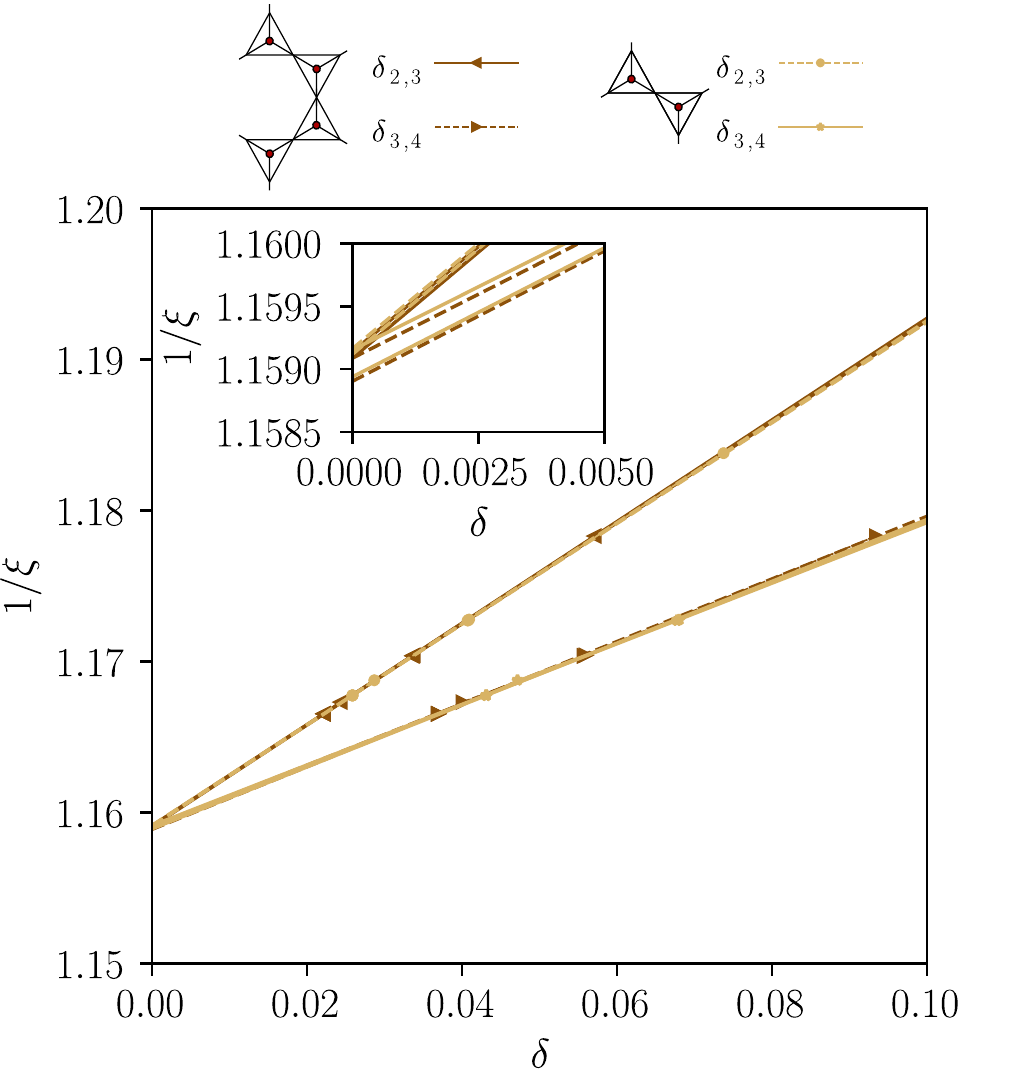}
    \caption{Determining the correlation length Eq.~\ref{eq:corrlengthh=4J} (for the connected correlations) at $h = 4J_1$. See the caption of Fig.~\ref{fig:CorrelationLength} for the detail.}
    \label{fig:CorrelationLength=4j}
\end{figure}

\begin{table}[t]
\caption{\label{tab:corrJ1h}First spin-spin correlations in the critical ground state of the nearest neighbour model in a field.}
\begin{ruledtabular}
\begin{tabular}{l|cccc}
 & $c_1$ & $c_2$ & $c_{3||}$& $c_{3\star}$ \\
\colrule
Ref.~\onlinecite{Moessner2003} & $-4/9$ & 0.36756  &  0.07688 & -0.30396  \\
Direct TN ($\chi = 80$) & $-4/9$ & 0.36755  & 0.07689& -0.30396\\
\end{tabular}
\end{ruledtabular}
\end{table}

\begin{table}[t]
\caption{\label{tab:corrJ1hcrit}First spin-spin correlations in the ground state in the nearest neighbour model in a longitudinal field $h = 4J_1$. For the tensor network, the Schmidt values decay extremely fast and fall below numerical precision for $\chi > 10$. For the Monte Carlo, we show results with a number of sites $N = 2304$, at temperature $T/J = 7\cdot 10^{-3}$.}
\begin{ruledtabular}
\begin{tabular}{l|cccc}
 & $c_1$ & $c_2$ & $c_{3||}$& $c_{3\star}$ \\
\colrule
Direct TN  & -0.1602714 & 0.04322263 & 0.0382691& -0.019949696\\
MC & -0.16025(6) & 0.04322(8) & 0.03819(8) & -0.0198(3) \\
\end{tabular}
\end{ruledtabular}
\end{table}
The ground state of the nearest neighbour model in a field (Eq.~\ref{eq:nnIsingHamiltonian}) is known to exhibit a magnetisation plateau $m = \frac{1}{3}$ for fields $0 < \frac{h}{J_1} < 4$, where each triangle bears two spins up and one spin down. The configurations in this plateau can be exactly mapped to a hardcore dimer model on the honeycomb lattice (placing a dimer on each down spin), leading to a macroscopic ground state degeneracy with a residual entropy corresponding to one third of the nearest neighbour antiferromagnet on the triangular lattice~\cite{Moessner2001,Moessner2003, Wannier1950, Wannier1973, Udagawa2002,Loh2008}. The \textit{connected} correlations are critical, decaying as $1/r^2$ (\cite{Moessner2003,Loh2008} and references therein). In the ground state, the correlations have been tabulated (Fig. 3 of Ref.~\onlinecite{Moessner2003}).

The model can be studied using the tensor network expression for the partition function from Fig.~\ref{fig:partitionfunction}(a), with the slight modification that
\begin{equation}
    \label{eq:weightJ1h}
    T_{\sigma_1, \sigma_2, \sigma_3} = \begin{cases}
    e^{-\beta J_1 (\sum_{\langle i,j\rangle}\sigma_i \sigma_j + 1) + \beta h (\sum_i \sigma_i -1)/2} & \, h \leq 4J_1 \\
     e^{-\beta J_1 (\sum_{\langle i,j\rangle}\sigma_i \sigma_j  -3) + \beta h (\sum_i \sigma_i -3)/2} & \, h \geq 4J_1
     \end{cases}
\end{equation}
(see as well Ref.~\onlinecite{Li2010} for a similar construction and a contraction with TRG, which was not applied to the case we present here). 
In the magnetisation plateau, with bond dimension $\chi = 80$, we find indeed
\begin{equation}
S_{h < 4J_1} = 0.1076886 \pm 10^{-7}
\end{equation}
which corresponds to one third of the triangular Ising antiferromagnet entropy. We compare the correlations to the result of Ref.~\onlinecite{Moessner2003} in Table \ref{tab:corrJ1h}.

Similar to what happens on the square and triangular lattices~\cite{Metcalf1978,Hwang_2007}, at the critical field $h/J_1 = 4$, a number of additional configurations contribute to the ground state. It has been noted before that at low temperature, this leads to a special value of the magnetisation $m = 3/5$~\cite{Semjan2020}. We find a slightly different value which is consistent between our tensor network computations in the ground state and our MC simulations at $T/J_1 = 7 \cdot 10^{-3}$:
\begin{align}
    m^{\text{TN}}_{h = 4 J_1} &= 0.599660907836 \pm 10^{-12}\\
    m^{\text{MC}}_{h = 4 J_1} &= 0.59968 \pm 8 \cdot 10^{-5}
\end{align}
For the residual entropy, we find
\begin{equation}
    S_{h = 4 J_1} = 0.387800244253 \pm 10^{-12}.
\end{equation}
The value of the residual entropy is consistent with Ref.~\onlinecite{Semjan2020}, although with a significant improvement in the precision owing to the small bond dimension required to compute the entropy~\footnote{We also confirmed our results by using a slightly different tensor network formulation based on ground state local rules in the spirit of Ref.~\onlinecite{Vanhecke2021}.}. The values of the first few correlations at small but non-zero temperature for $h/J_1 = 4$ are given in Table \ref{tab:corrJ1hcrit}, where it can be seen that the Monte Carlo and tensor network computations agree. A similar analysis as the one performed in zero field (Fig.~\ref{fig:CorrelationLength=4j}) yields a finite correlation length 
\begin{equation}
    \label{eq:corrlengthh=4J}
    \xi_{h = 4J_1} = 0.8627 \pm 0.0001
\end{equation}
in units of the lattice spacing. This small correlation length is the reason why the tensor network results are obtained with such high precision even with extremely small bond dimension (here, the largest bond dimension is $\chi = 10$ as for larger bond dimensions, the Schmidt values decay below numerical precision).
For $\frac{h}{J_1} > 4$, the ground state is the fully ferromagnetic state. 

\begin{figure}[t]
    \centering
    \includegraphics[width = 0.48\textwidth]{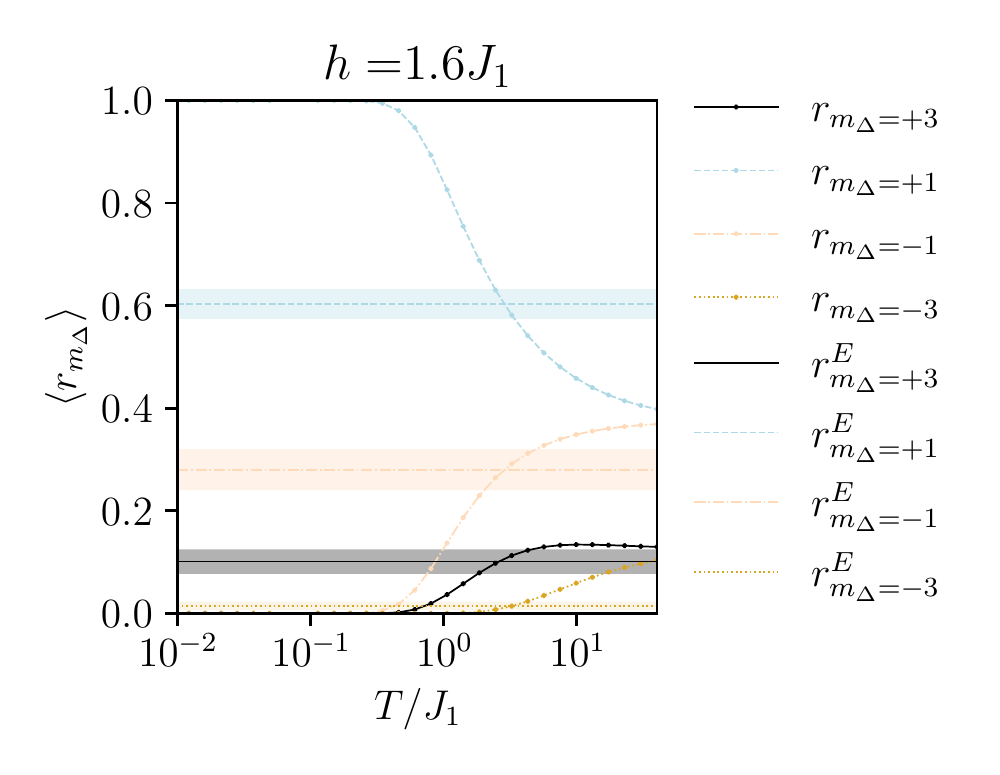}
   \caption{Tensor networks (maximal bond dimension $\chi = 80$) results for the proportion of triangles of given magnetisation at a field of $h = 1.6J_1$. For different fields spaced regularly every $0.2J_1$, this is the one at which the prediction from the tensor networks simulations are closest to the experimental results. The optimal temperature is $T = (2.8 \pm 0.2)J_1$. The data for the proportion of triangles of each type at regularly spaced fields can be found at~\cite{Zenodo}.}
    \label{fig:trianglesratiosJ1h}
\end{figure}

\begin{figure}[t]
    \centering
    \includegraphics[width = 0.48\textwidth]{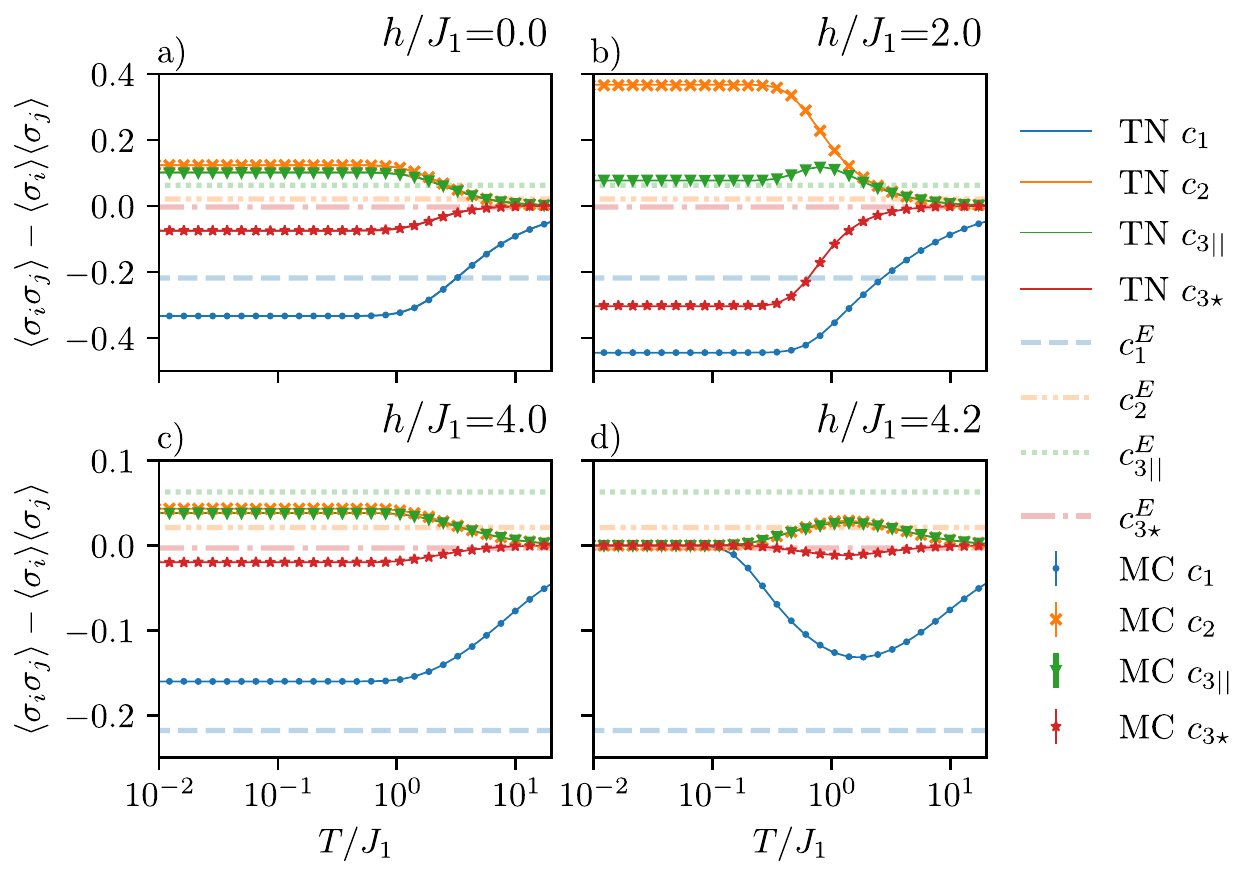}
   \caption{Tensor networks (maximal bond dimension $\chi = 80$) and Monte Carlo ($N = 2304$ sites) results for the first few correlations at specific magnetic fields. Note the difference in the vertical axis for subplots (a) $h = 0$ and (b) $h = 2J_1$ versus (c) critical field $h = 4J_1$ and (d) $h = 4.2 J_1$. See as well Tables \ref{tab:corrJ1h} and \ref{tab:corrJ1hcrit} for the values in the ground state. Notice that in the second row, we have zoomed in on the values of the correlations. The data for the correlations at regularly spaced fields can be found at~\cite{Zenodo}.}
    \label{fig:overviewcorrJ1h}
\end{figure}

 We first check that by introducing a field, we can account for the number of triangles with a given magnetisation. For this, we compute the proportions $r_{m_{\Delta} = +3}$, $r_{m_{\Delta} = +1}$, $r_{m_{\Delta} = -1}$ and $r_{m_{\Delta} = -3}$ for regularly spaced fields (every $0.2 J_1$). As shown in Fig.~\ref{fig:trianglesratiosJ1h}, we find that we obtain the best fit at $h = (1.6 \pm 0.2)J_1$ for a temperature $T = (2.8 \pm 0.2)J_1$. As a direct consequence, in this region, the magnetisation and the nearest neighbour correlations match the experimental results.

 We compute the correlations as a function of field and temperature using the tensor network construction. For a few fields we compute the spin-spin correlations with Monte Carlo simulations using replicas in field and temperature (similar to Refs.~\onlinecite{Semjan2020, Soldatov2019}) to confirm the tensor network results; this is plotted in Fig.~\ref{fig:overviewcorrJ1h} (a plot for the magnetisation can be found in Appendix~\ref{sec:AppJ1h}). From the systematic computation of the spin-spin correlations, we find that at all fields the $c_2$ correlations remain larger or equal to the $c_{3||}$ correlations. In particular, at large temperatures, $c_{3||}$ is essentially equal to $c_2$. Thus, even though combining the magnetic field and the temperature allows us to account for the proportion of triangles of given magnetisation, and therefore account for both the magnetisation and the percentages of frustrated triangles, the nearest neighbour model is not sufficient to fully account for the experimental results. This is possible because the proportion of each type of triangles is mostly related to the nearest neighbour correlations as well as the magnetisation, which are controlled respectively by $J_1/T$ and $h/T$, but the further neighbour spin-spin correlations are related to correlations between the triangles. This is somewhat surprising, since the next-nearest neighbour coupling is only about 2\% of $J_1$.

\section{Next nearest neighbour Ising antiferromagnet}
\label{sec:NNN}

Since the nearest neighbour model does not fully describe the experimental correlations, and since the micromagnetic simulations predict a small but non-zero value for the next nearest neighbour couplings, we now consider whether such small couplings are sufficient to explain the inversion of the relative order of the $c_{2}$ and $c_{3||}$ correlations. Further neighbour couplings are known to lift the degeneracy either partially~\cite{Takagi1993, Chern2012, Mizoguchi2017} or completely~\cite{Wolf1988, Takagi1993} (depending on whether they are ferro- or antiferromagnetic). For now, we ignore the problem of the magnetisation and consider the Hamiltonian~\cite{Takagi1993}\footnote{Notice the opposite sign convention and factor two.}
\begin{equation}
    H_{NNN} = J_1 \sum_{\langle i,j \rangle} \sigma_i \sigma_j + J_2 \sum_{\langle i,j \rangle_2} \sigma_i \sigma_j
\end{equation}
where $\langle i,j \rangle_2$ stands for next nearest neighbour spin pairs as illustrated by $c_2$ in the inset of Fig.~\ref{fig:expSpinSpin}. The key question is then how large the next nearest neighbour coupling has to be to change the descending order of the further neighbour correlations and explain the experimental observation that $c_{3||}\gtrsim 2 c_2$.

Here, we focus on the case with antiferromagnetic next nearest neighbour couplings for the Ising model. It should be noted that, upon changing the sign of the couplings and multiplying them by 2, the model maps onto the spin-ice model on the kagome lattice with \textit{ferromagnetic} next nearest neighbour couplings~\cite{Wills2002,Chern2012}. We study the model with an ad-hoc Monte Carlo algorithm~\cite{Rakala2017} and our dual tensor network construction from Fig.~\ref{fig:partitionfunction}(b). The expression of the tensor on the triangle Eq.~\ref{eq:triangledimers} remains unchanged, while Eq.~\ref{eq:hexdimers} becomes
\begin{equation}
    \label{eq:hexdimersJ1J2}
    H_{d_1, d_2, d_3, d_4, d_5, d_6} = \begin{cases}
    e^{-\beta J_2 (\sum_{i=1}^6 d_i d_{i+1} + 2)} & \, \prod_i d_i =1\\
    0 & \text{ otherwise}
    \end{cases}
\end{equation}
where $d_7 = d_1$.

The $J_2$ couplings form a set of three interpenetrating kagome sublattices. The ``two-up one-down, two-down one-up'' ice rule can be satisfied simultaneously on each triangle on the initial kagome lattice as well as on each triangle on these three kagome lattices. The ground state energy per site is thus~\cite{Takagi1993, Hamp2018}
\begin{equation}
    E_{\text{G.S.}} = -\frac{2}{3}J_1 - \frac{2}{3}J_2.
\end{equation}
\begin{table}[t]
\caption{\label{tab:corrJ1J2}First spin-spin correlations in the ground state of the next nearest neighbour model. For the Monte Carlo simulations, we show results with the number of sites $N = 1296$ and at a temperature $T/J = 7\cdot 10^{-3}$.}
\begin{ruledtabular}
\begin{tabular}{l|cccc}
 & $c_1$ & $c_2$ & $c_{3||}$& $c_{3\star}$ \\
\colrule
Dual TN  ($\chi = 144$) & -1/3 & -1/3 & 0.5726&  0.5933\\
MC & -1/3 & -1/3 & $0.567 \pm 0.005$ & $0.586 \pm 0.005$ \\
\end{tabular}
\end{ruledtabular}
\end{table}\\
Imposing these rules only leads to a partial lifting of the ground state degeneracy, and from Pauling estimates one gets a residual entropy per site~\cite{Hamp2018} $S_{J_1, J_2} \cong \ln\left(2 \left(\frac{3}{4}\right)^{(4N/3)}\right) = 0.3096$. 
From the contraction of the tensor network, we get
\begin{equation}
    S_{J_1, J_2} = 0.285299 \pm 1.4\cdot10^{-6},
\end{equation}
where the error is estimated from the difference between the value at maximal MPS bond dimension ($\chi = 144$) and the result of the fit in the infinite bond dimension limit. The result matches what we obtain with the method of Ref.~\onlinecite{Vanhecke2021}, and our Monte Carlo thermodynamic integration result $S_{J_1, J_2} \cong 0.285 \pm 0.001$ (Appendices~\ref{sec:AppMC},\ref{sec:J1J2app}). In Table~\ref{tab:corrJ1J2}, we also give the first spin-spin correlations in the ground state, as obtained with both methods.

\begin{figure}[t]
    \centering
    \includegraphics[width = 0.48\textwidth]{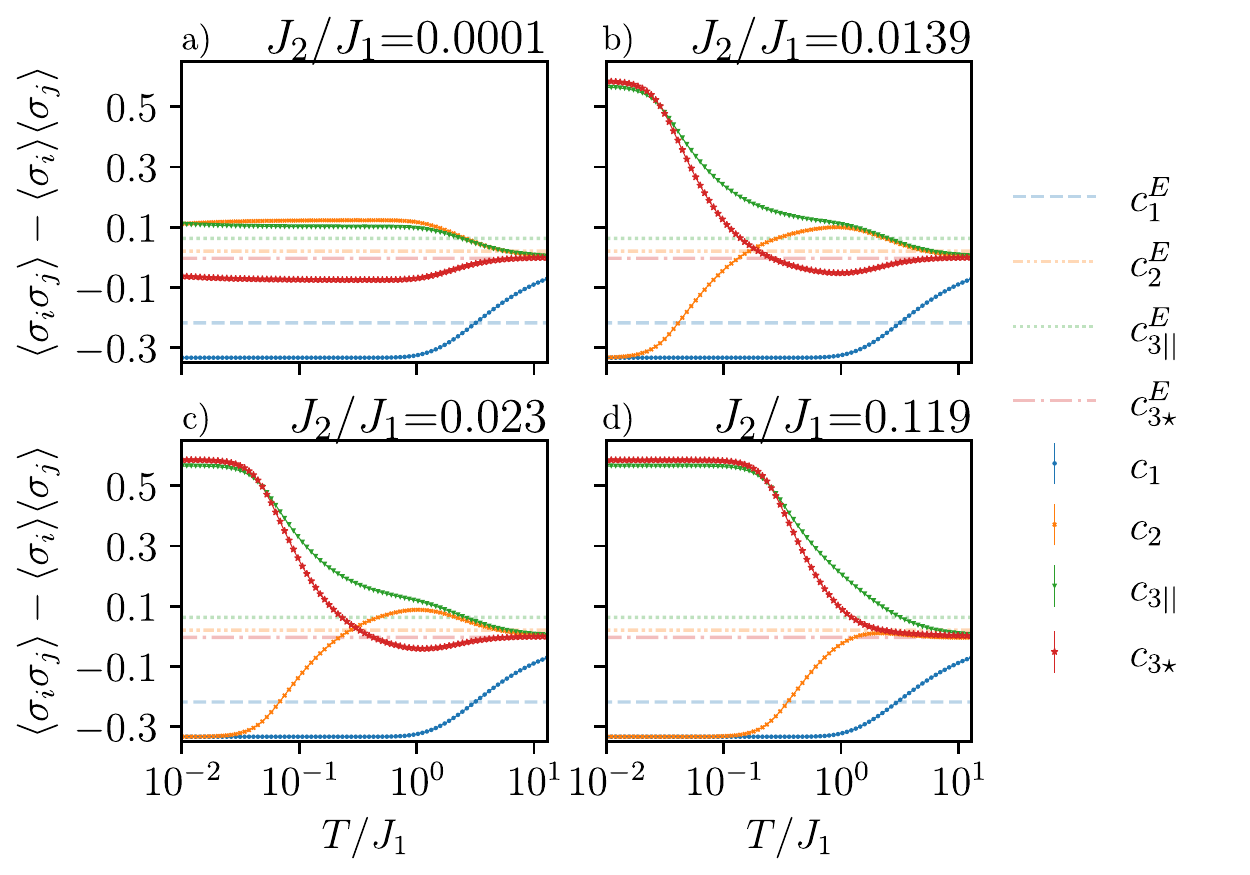}
    \caption{Overview of the behaviour of the first spin-spin correlations as a function of the temperature for various values of $J_2$, with Monte Carlo simulations for two system sizes ($N = 576$ and $1296$); the smaller system size is shown with a line while the larger one is shown with symbols. The dashed lines show the experimental values for comparison. The temperature at which the ground state correlations are reached depends on the value of $J_2/J_1$. (a) For small $J_2$, we recover the nearest-neighbour model correlations at intermediate temperatures. (b) and (c) Upon increasing $J_2$, the region of temperature where $c_2 > c_{3||}$ is pushed towards high temperatures, and (d) eventually disappears. When $J_2$ is large, the $c_2$ correlations remain negative at any temperature. An overview is shown in Fig.~\ref{fig:J2Crossings}. }
    \label{fig:J2Correlations}
\end{figure}

Takagi and Mekata~\cite{Takagi1993} predicted a KT transition to this critical ground state phase at temperatures of order of $J_2$. Correspondingly, for small values of $J_2$, the specific heat exhibits two broad peaks corresponding to the two stages of the loss of entropy (first for imposing the ice rule on nearest neighbour triangles and then for imposing it on further neighbour triangles), while for larger values of $J_2$, the two features merge into one~\cite{Takagi1993,Wills2002} (Appendix~\ref{sec:J1J2app}). The behaviour of the spin-spin correlations corresponds to this picture (Fig.~\ref{fig:J2Correlations}), since we find that for small values of $J_2$ the correlations take their nearest neighbour model value for a range of temperatures before going to their ground state values, while for larger values of $J_2$ the competition between $J_2$ and $J_1$ significantly affects the correlations even at large temperatures. Note that Fig.~\ref{fig:J2Correlations} shows the values of the correlations with the Monte Carlo simulations for those ratios of $J_2$ to $J_1$ corresponding to the three scenarios discussed in Sec.~\ref{sec:micro} ($J_2^{\text{dip}}/J_1$, $J_2/J_1$, $J_2^{\text{dip}}/J_1^{\text{dip}}$).
\\
\begin{figure}[t]
    \centering
    \includegraphics[width = 0.48\textwidth]{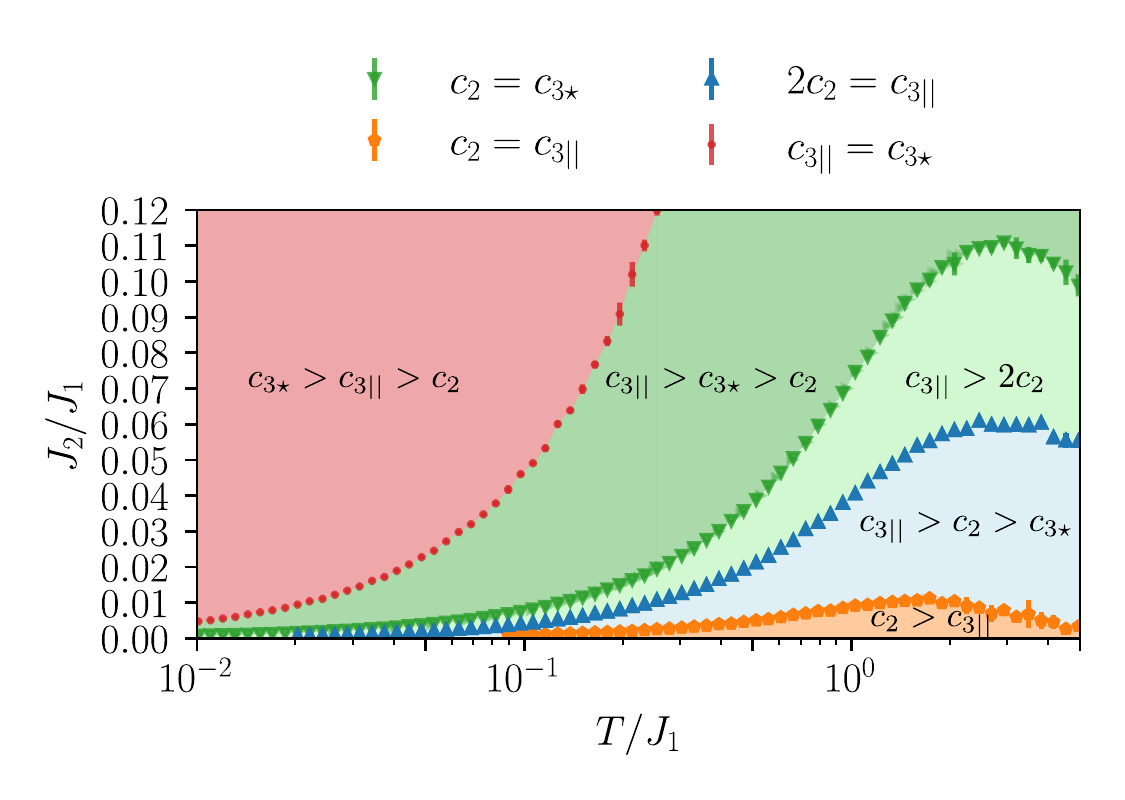}
    \caption{Map of the descending order of the correlations as a function of the values of the temperature $T/J_1$ and the next nearest neighbour coupling $J_2/J_1$ (obtained with MC simulations with $N = 576$ sites). In the experiment, the correlations satisfy $c_{3||} > c_2 > c_{3\star}$, with $c_{3||} \gtrsim 2c_2$ (light green region). For small enough $J_2$, there is a temperature region with $c_2 > c_{3||}$ (orange region), compatible with the nearest neighbour physics in terms of the descending order of the correlations. Importantly, the smallest value of $J_2/J_1$ in this graph is $J_2/J_1 = 0.001$: for arbitrarily small $J_2$ the crossing $c_2 = c_{3||}$ happens at arbitrarily small temperatures (Fig.~\ref{fig:J2Correlations}). Note as well that at large temperatures, all the correlations are close to zero.}
    \label{fig:J2Crossings}
\end{figure}

Simulating the model systematically for a range of values of $J_2$, we can map out the values of the next nearest neighbour coupling and of the temperature where the spin-spin correlations are in a certain descending order. This is shown in Fig.~\ref{fig:J2Crossings}, where one can see that -- in agreement with the above discussion -- for very small $J_2/J_1 \lesssim 0.01$, there is a range of temperatures where the descending order of the correlations is compatible with the nearest neighbour model. For $J_2/J_1 \gtrsim 0.01$, one can see a broad region of temperatures and couplings where the spin-spin correlations are in the same relative order as in the experiment ($|c_1| > c_{3||} > c_2 \gtrsim c_{3\star}$), and within this region, a non-negligible region where $c_{3||} > 2c_2$. In particular, for the micromagnetic value $J_2 = 0.023 J_1$, the temperature range where $c_{3||} > 2c_2$ is $ 0.3 \lesssim T/J_1\lesssim 0.6$. If $J_2 \gtrsim 0.06 J_1$, the region where $c_{3||} > 2c_2$ extends all the way to $T = 5J_1$.

Although this model does not involve a magnetic field, it provides evidence that, in order to recover qualitatively the descending order of the spin-spin correlations, further neighbour couplings have to play a role. Additionally, it provides an idea of how sensitive the correlations are to these further neighbour couplings.

\section{Results in the $J_1-J_2-J_{3||}$ model}
\label{sec:J1J2J3ph}
The micromagnetic simulations predict third neighbour couplings that are almost half of the second neighbour couplings. Since there is a competition between these couplings, we have to check that, in their presence, the predicted descending order of the correlations is still $c_{3||}> c_2 > c_{3\star}$. Additionally, we have seen in Sec.~\ref{sec:NN} that the proportion of triangles with a given magnetisation can be accounted for by introducing a longitudinal magnetic field.
We thus want to consider the following Hamiltonian:
\begin{equation}
    H = J_1 \sum_{\langle i,j\rangle_1} \sigma_i \sigma_j+J_2 \sum_{\langle i,j\rangle_2} \sigma_i \sigma_j+J_{3||} \sum_{\langle i,j\rangle_{3||}} \sigma_i \sigma_j -h \sum_i \sigma_i.
\end{equation}
Because of the combination of further neighbour frustration and the presence of the field, this model is challenging to study using Monte Carlo simulations. At the same time, writing a converging tensor network formulation for such frustrated systems is far from trivial. Here, we present a study based on an understanding of the ground state using both methods, and results for the correlations as a function of the temperature using Monte Carlo simulations for small system sizes.

\subsection{Ground state of the $J_1-J_2-J_{3||}$ model with and without a field}
In zero field, the location of the phase boundaries and the value of the ground state energy for the ground state phase diagram has already been established~\cite{Wolf1988}, using exact ground state lower bounds computed with Kanamori's method~\cite{Kaburagi75}. For antiferromagnetic couplings $J_2,\,J_{3||}>0$, there are four different ground state phases. Our micromagnetic values for the couplings lie well within one of these phases, where
\begin{equation}
    \label{eq:GSJ1J2J3p_H0}
    E_{\text{G.S.}} = -\frac{2}{3}J_1 - \frac{2}{3} J_2 + \frac{2}{3}J_{3||}.
\end{equation}
As we will show, this phase exhibits a macroscopic ground state degeneracy.

In the ground state of frustrated systems, the standard formulation of tensor network partition functions based on Boltzmann weights typically fails~\cite{zhu2019,Liu2020,Vanhecke2021} - this is also the case for spin glasses. There are two possible approaches to this issue: one can either work around the problem of the multiplication of big and small numbers by adapting the tensor network algorithms to working with the logarithms of the Boltzmann weights~\cite{Liu2020}, or one can use the insight that this failure is related to the presence of frustrated couplings and try to find the ground state local rule to implement it at the level of the tensor, thereby ``relieving'' the frustration~\cite{Vanhecke2021}. Here, we follow this second approach to construct a contractible tensor network that we can use to study the ground state.

In this further neighbour model, there is a special line for $J:=J_2=J_{3||}$ where the problem can be elegantly studied using a charge representation~\cite{Mizoguchi2017}. For positive $J$, this line coincides with two successive phase boundaries in the ground state of the $J_1-J_2-J_{3||}$ model. For $0 <J< J_1/3$, a classical spin liquid with an unusual residual entropy $S \cong 0.32$ (the ``hexamer'' classical spin liquid) was found in Ref.~\onlinecite{Mizoguchi2017}. The effect of a longitudinal magnetic field on this phase has been recently studied in a follow-up work and was shown to give rise, in the ground state, to a number of magnetisation plateaus with finite residual entropy~\cite{Tokushuku2020}. We use these phases to check our tensor network construction in the presence of a magnetic field, and find that the entropies we obtain for the various ground state phases are in agreement with the existing results (Table~\ref{tab:J1J2J3ph_comp}).
\begin{table}[t]
\caption{\label{tab:J1J2J3ph_comp} Residual entropy in the ground state for $J_2=J_{3||}\lesssim0.2$ for the various ground state phases.
When indicated by the value of the field, we are looking at a phase boundary, and when indicated by the value of the magnetisation plateau, we are looking at the phase between these boundaries. }
\begin{ruledtabular}
\begin{tabular}{l|cc}
&  TN ($\chi_{\text{max}} = 120$)& Refs.~\onlinecite{Mizoguchi2017,Tokushuku2020}\\
\colrule
$h = 0$ & $0.322273784 \pm 10^{-9}$& 0.32\\
$m = 1/9$& $0.125616 \pm10^{-6} $ & 0.12(6)\\
$h = 6 J_{3||}$ &  $0.313746908 \pm 10^{-9}$& -\\
$m = 1/3$ & $0.107688 \pm 10^{-6}$& $S_{\text{TLIAF}}/3$\\
$h = 4(J_1 + J_2) - 6J_{3||}$ & $0.26718206 \pm 10^{-8}$& -\\
\end{tabular}
\end{ruledtabular}
\end{table}

As a first step in the tensor network construction, linear programming is used to build ground state energy lower bounds based on splitting the Hamiltonian into terms defined on clusters~\cite{Huang2016,Vanhecke2021}. We can thus determine the ground state energy in the various phases by comparing the lower bound from this method and the upper bound from the Monte Carlo simulations; if they match we have a proof for the value of the ground state energy, and we know that the tensor network will describe the complete ground state manifold.

\begin{figure*}[t]
    \centering
    \includegraphics[width =0.7\textwidth]{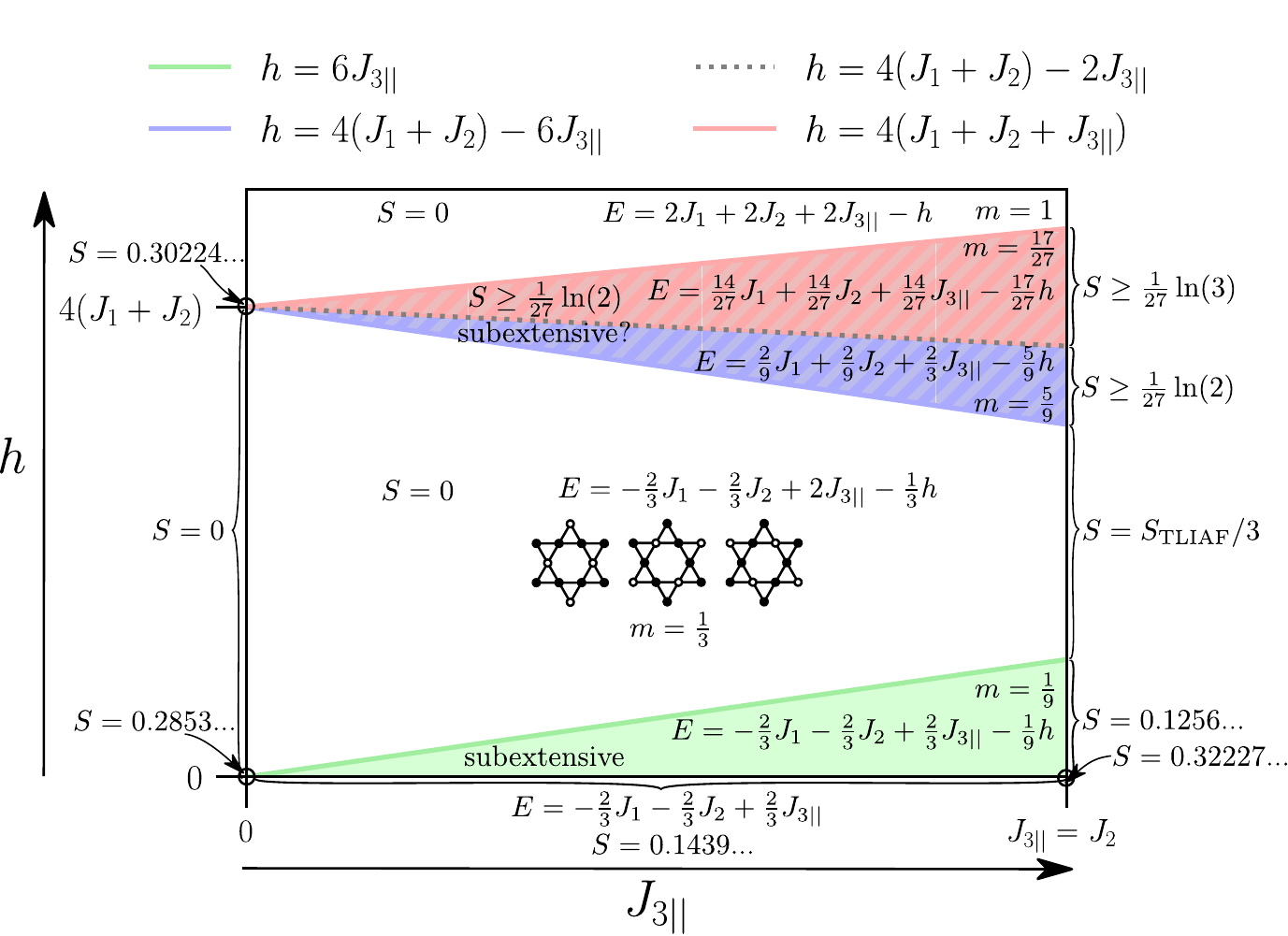}
    \caption{Candidate ground state phase diagram for the further neighbour model in a magnetic field, at a fixed value of $J_2 < \frac{1}{3}J_1$. The two hatched regions could only be studied with Monte Carlo computations, whereas the rest of the phase diagram was determined both with Monte Carlo simulations and with tensor networks. When the values are indicated next to arrows, they correspond to a specific point (for instance, we give the result for $J_{3||} = J_2$ and $h = 0$) and when they are indicated next to brackets they correspond to a specific line (for instance, we give the results for $J_{3||} = J_2$ and various ranges of the field, which are different from the results for $J_{3||} < J_2$). The ground state entropies are computed from tensor networks contractions, except for the two inequalities on the $J_2 = J_{3||}$ line, studied in Ref.~\onlinecite{Tokushuku2020}, and the lower bound in the $m = 17/27$ plateau, estimated from the Monte Carlo results. The bottom left corner corresponds to the $J_1-J_2$ model. A magnetic field in this model immediately lifts the degeneracy to the long range ordered stripe phase. For finite $J_{3||}$, there is a macroscopic ground state degeneracy in zero field. With increasing field, there is first a magnetisation plateau at $m = 1/9$. Then, the long range ordered stripe phase is selected, followed by a plateau at $m = 5/9$ and another one at $m = 17/27$ before saturation. For each phase we give the ground state energy, which is rigorously proved everywhere except in the two hatched regions.}
    \label{fig:GSJ1J2J3ph}
\end{figure*}

Our results for the ground state phase diagram for $J_{3||} < J_2$ are summarised in Fig.~\ref{fig:GSJ1J2J3ph}. In zero field, we recover the ground state energy of Eq.~\ref{eq:GSJ1J2J3p_H0}, and from the tensor network construction, we find a residual entropy
\begin{equation}
    \label{eq:SJ1J2J3p}
    S_{J_1-J_2-J_{3||}} = 0.143949 \pm 6 \cdot 10^{-6}
\end{equation}
which shows that some sort of classical spin liquid is available even when $J_{3||} \neq J_2$.

When introducing a finite magnetic field, the system enters an $m = 1/9$ plateau which survives as long as $h < 6J_{3||}$. In this plateau, the ground state energy is given by
\begin{equation}
    E_{1/9} = -\frac{2}{3}J_1 -\frac{2}{3}J_2 + \frac{2}{3}J_{3||} - \frac{1}{9} h.
\end{equation}
We find a zero residual entropy in the thermodynamic limit by contracting the tensor network, but from the Monte Carlo simulations, we find that the ground states differ by non-local updates. In simulations with periodic boundary conditions, these updates correspond to strings of spins that cross the sample and close in on themselves through the periodic boundary conditions (Appendix~\ref{sec:AppJ1J2J3h}). If one looks at periodic boundary conditions by placing the lattice on a torus, then these updates are winding around the torus. Together with the tensor network result, this suggests that the ground state degeneracy is macroscopic, but with a sub-extensive residual entropy, growing with the linear system size.

The $m = 1/3$ plateau corresponds to long-ranged ordered strings of nearest neighbour up spins separated by down spins, with a ground state energy
\begin{equation}
    E_{1/3} = -\frac{2}{3}J_1 -\frac{2}{3}J_2 + 2J_{3||} - \frac{1}{3} h.
\end{equation}
The phase boundary between the $m=1/9$ and $m = 1/3$ plateau is thus found at $h = 6J_{3||}$. 
At this boundary, the ground states can have various magnetisations, corresponding to mixtures of states of both plateaus. Different sizes in the Monte Carlo simulations select different average magnetisations (Appendix~\ref{sec:AppJ1J2J3h}); we therefore refrain from stating a value for the magnetisation here.

The $m = 1/3$ plateau extends until $h = 4 (J_1 + J_2) - 6 J_{3||}$ where we find a phase boundary. Above this field, the largest clusters that we used to split the Hamiltonian give an energy lower bound which does not match the exact ground state energy \footnote{Meaning the corresponding ground state tiles fail to tile the lattice, and the tensor network contraction fail. Larger clusters would be required, probably made of 6 to 12 stars from what we see of the Monte Carlo simulations; this requires some adaptation of the code, which we did not yet tackle.}. However, we have evidence from Monte Carlo simulations (Fig.~\ref{fig:J1J2J3m}) that suggests the presence of $m = 5/9$ and $m = 17/27$ plateaus corresponding to the ones found at $J_2 = J_{3||}$ in Ref.~\onlinecite{Tokushuku2020}. The corresponding phase boundaries are at $h = 4 (J_1 + J_2) - 2J_{3||}$ and $h = 4 (J_1 + J_2 +J_{3||})$, which means that this region is extremely reduced for our micromagnetic values of $J_1, J_2$ and $J_{3||}$. In the $m=5/9$ plateau, with periodic boundary conditions in the Monte Carlo, we find again that ground states differ by strings of spins winding the torus, suggesting a sub-extensive residual entropy, whereas some local moves can be seen in the $m = 17/27$ plateau, providing a lower bound for the residual entropy $S \geq \frac{1}{27} \ln(2)$ (Appendix~\ref{sec:AppJ1J2J3h}). 

\subsection{Effect of the temperature and spin-spin correlations}
\begin{figure}[t]
    \centering
    \includegraphics[width =0.45\textwidth]{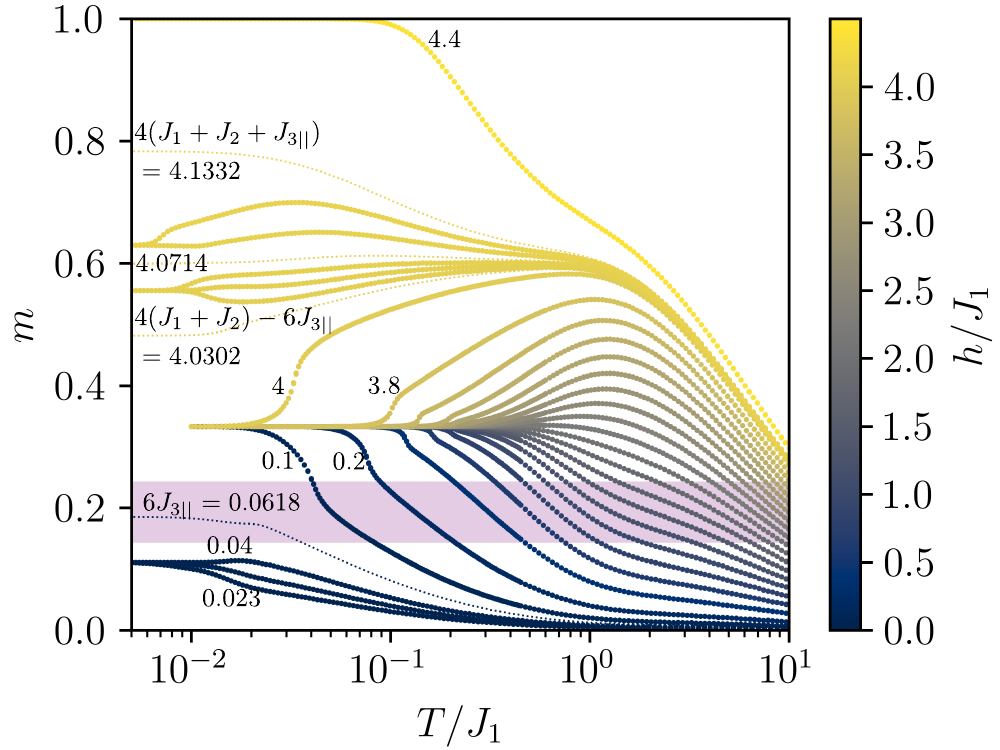}
    \caption{Qualitative behaviour of the magnetisation as a function of the field and temperature in the $J_1-J_2-J_{3||}-h$ model. The points come from Monte Carlo simulations with $N = 900$ in the $m = 1/3$ plateau, with $N = 144$ in the $m = 5/9$ plateau and with $N = 376$ in the other regions. The dotted lines are also from Monte Carlo simulations but correspond to the magnetisation at the transition between plateaus; they are only given as a guide to the eye as the results strongly depend on the system size (Appendix~\ref{sec:AppJ1J2J3h}). The highlighted region corresponds to the region within one standard deviation of the experimental magnetisation. The annotations indicate the value of the magnetic field for various curves.}
    \label{fig:J1J2J3m}
\end{figure}

We finally give a qualitative discussion of the effects of temperature on the spin-spin correlation in this further neighbour model in a field, and compare our predictions to the experimental results. In the rest of this section we consider $J_2 = 0.023 J_1$ and $J_{3||} = 0.0103 J_1$, corresponding to the micromagnetic simulations results. Fig.~\ref{fig:J1J2J3m} gives a qualitative picture of the magnetisation as a function of field and temperature. Because the problem is a challenge for our Monte Carlo simulations (where we use the single-spin-flip algorithm combined with replicas in magnetic field and temperature), we only focus on small system sizes where needed. In Fig.~\ref{fig:J1J2J3m}, the selection of the various magnetisation plateaus of the ground state with increasing fields is shown. With increasing temperatures, we find around the $m = 1/3$ and $m = 1/9$ plateau a behaviour similar to the one in the $J_1-h$ model around the $m = 1/3$ plateau (Appendix~\ref{sec:AppJ1h}): if the field is large, intermediate temperatures will give a larger average magnetisation than the plateau value, whereas if the field is small, the magnetisation immediately decreases with increasing temperatures.

\begin{figure}[t]
    \centering
    \includegraphics[width = 0.48\textwidth]{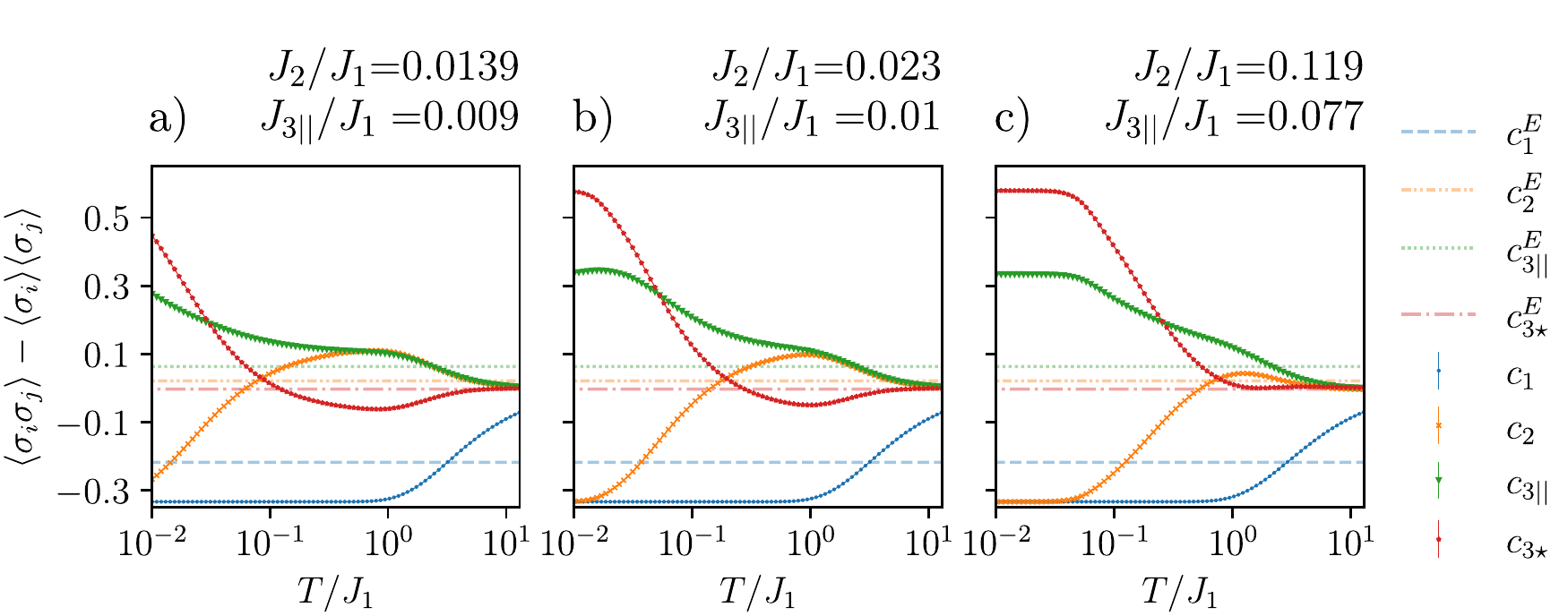}
    \caption{Overview of the behaviour of the first few correlations as a function of the temperature for the values of $J_2$ and $J_{3||}$ from micromagnetic simulations ((a): dipolar corrections to the nearest neighbour model, (b) actual couplings with IP part, (c) purely dipolar model (no IP regions)), with Monte Carlo simulations for 2 system sizes ($N = 576, 1296$). The smaller system size is shown with a line while the larger one is shown with symbols.}
    \label{fig:J2J3pCorrelations}
\end{figure}
\begin{figure}[t]
    \centering
    \includegraphics[width =0.48\textwidth]{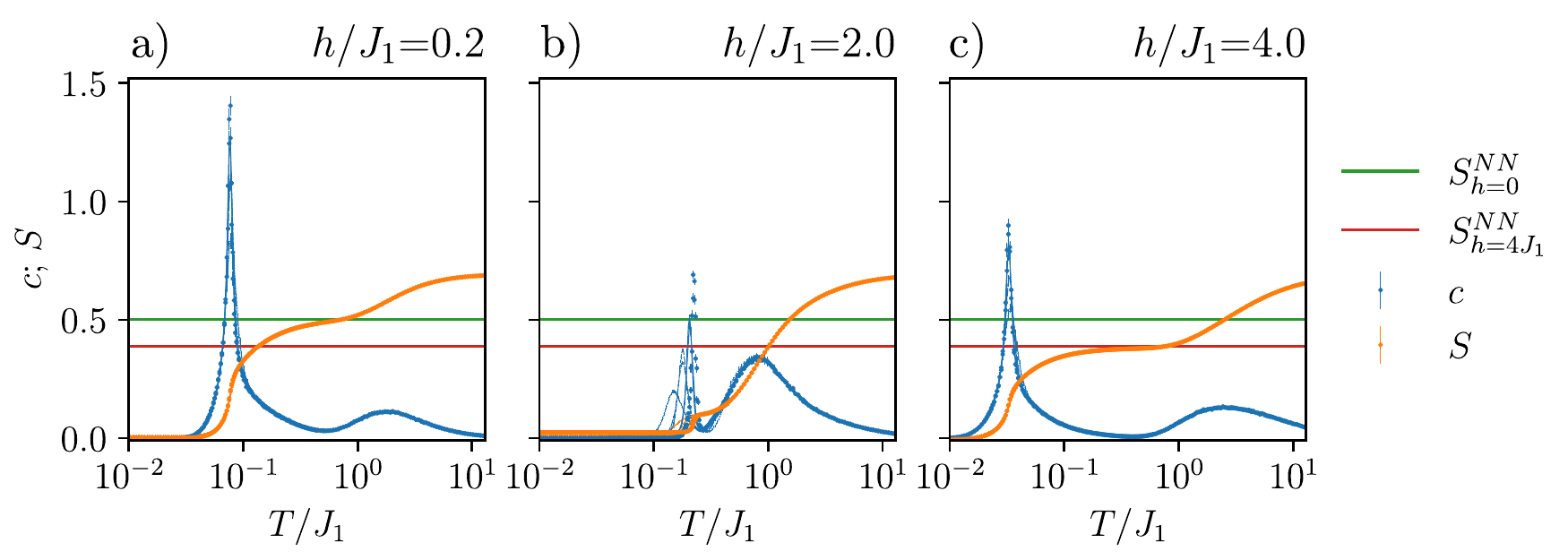}
    \caption{Specific heat and residual entropy in the $m = 1/3$ plateau for $N = 36$ to $N = 1296$ (markers). (a) At small fields there is an intermediate temperature region where the entropy corresponds to that of the nearest neighbour model in zero field, whereas (c) at large fields there is an intermediate temperature region where the entropy corresponds to the nearest neighbour model in a field at $h = 4J_1$. (b) We note that at intermediate fields there are strong finite size effects and the loss of residual entropy at the transition is not captured correctly by our simulations - the residual entropy in the ground state should be zero.}
    \label{fig:J1J2J3pspecificheat}
\end{figure}
\begin{figure}[t]
    \centering
    \includegraphics[width =0.48\textwidth]{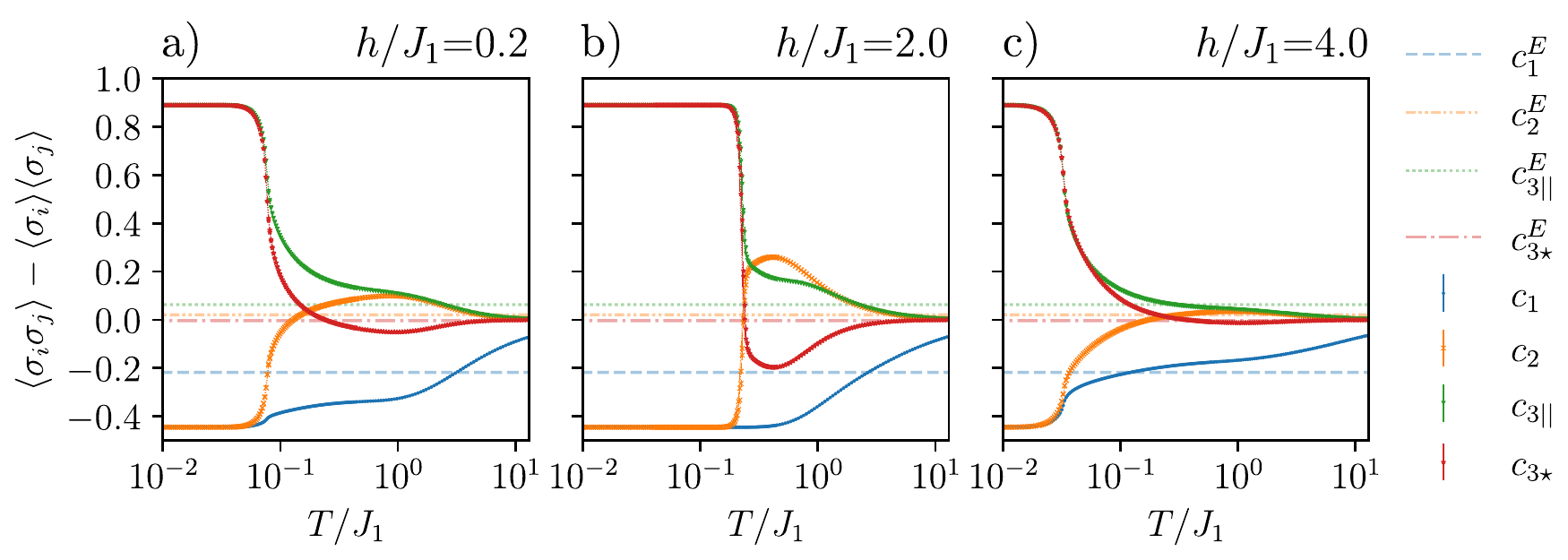}
    \caption{Spin-spin correlations in the $m = 1/3$ plateau for $N = 900$ (dotted lines) and $N = 1296$ (markers). The dashed lines indicate the experimental correlations, for comparison. (a) For small magnetic fields (Region~\textbf{A} in Fig.~\ref{fig:J1J2J3phcrossings}) there is an intermediate range of temperatures where the correlations correspond to the model in zero field (Fig.~\ref{fig:J2J3pCorrelations}). (b) At intermediate field, the $c_2$ correlations again become larger than the $c_{3||}$ correlations. (c) At large fields, there is an intermediate temperature region where the correlations are similar to those in the $J_1-h$ model at $h = 4 J_1$, but with an effect of the further neighbour couplings which inverts the $c_2$ and $c_{3||}$ correlations.}
    \label{fig:J1J2J3pOneThirdCorrels}
\end{figure}
\begin{figure}[t]
    \centering
    \includegraphics[width =0.45\textwidth]{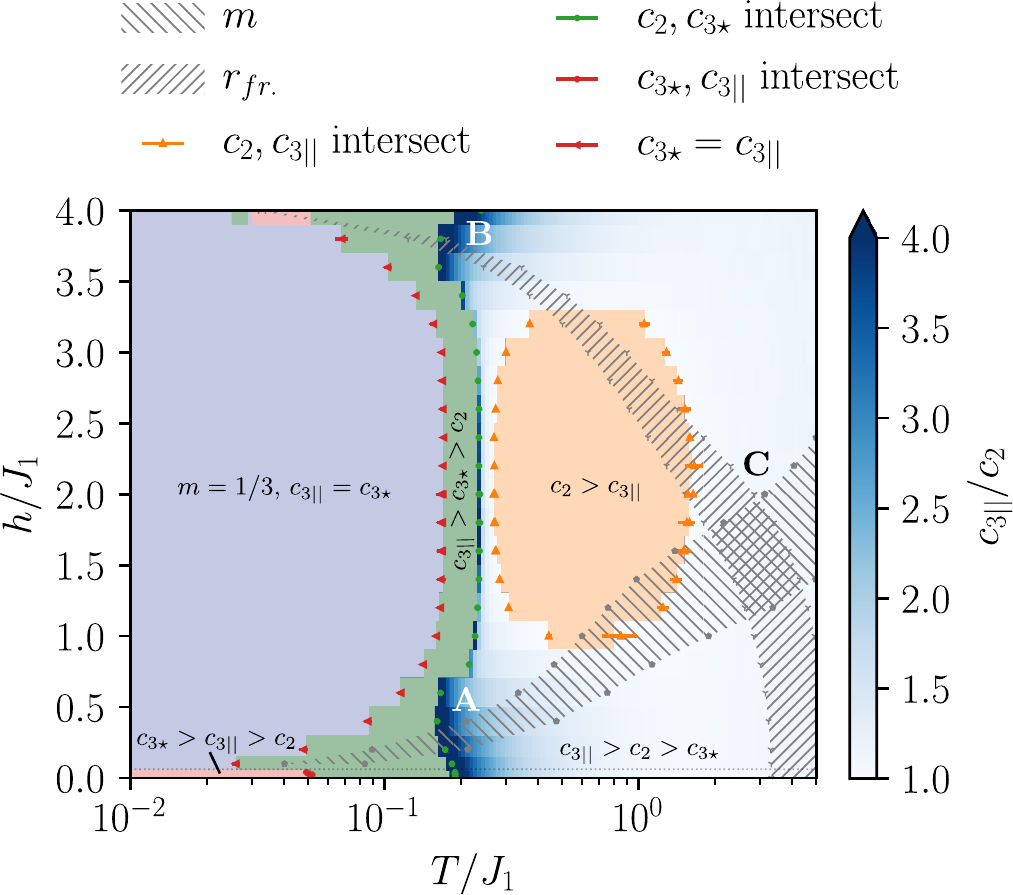}
    \caption{Monte Carlo simulation results for the $J_1-J_2-J_{3||}-h$ model as a function of the temperature. (A), (B) and (C) : the letters are placed just above three regions of interest, discussed in the main text. The various coloured areas correspond to the values of temperature and fields where the correlations are in a certain descending order according to the Monte Carlo simulations. The ``pixels'' correspond to the 22 field points and 214 temperatures. With increasing temperature: in light blue, the region where the system is in the $m = 1/3$ plateau ground state, with $c_{3||} = c_{3\star} = 8/9$; in light red, the region outside this plateau where $c_{3||} < c_{3\star}$; in green, the region where the $c_2$ correlations are larger than $c_1$ but still smaller than $c_{3\star}$; with the colour map, the region where the descending order of the correlations is the same as in the experiment: $c_{3||} \geq c_2 > c_{3\star}$; in orange, the region where $c_2 > c_{3||} > c_{3_\star}$. The hatched regions indicate the values of the temperature and field where the Monte Carlo simulations results for the magnetisation $m$ and the proportion of ferromagnetic (``frustrated'') triangles $r_{fr.}$ agree within the errors with the experimental values. The colour map indicates the ratio of $c_{3||}$ to $c_2$ in the region where the descending order of the correlations is the same as in the experiment; experimentally, this ratio is of the order of $2$ to $3$ in most samples (but $c_2$ can be negative in some samples, Fig.~\ref{fig:expSpinSpin}).}
    \label{fig:J1J2J3phcrossings}
\end{figure}

In analysing the spin-spin correlations, we first consider Fig.~\ref{fig:J2J3pCorrelations}, presenting the results in zero field. Although, in the ground state, the results are quite different from the $J_1-J_2$ model (Fig.~\ref{fig:J2Correlations}), the qualitative result that there is a temperature range where $c_{3||} \gtrsim 2c_2$ remains correct. This has to be contrasted to the nearest neighbour model results, and shows that the experimental results can only be understood by taking into account further neighbour couplings, and that $J_2$ plays the important role, while $J_{3||}$ is simply not large enough to suppress the effect of $J_2$. It can also be seen in this figure that these results would be valid for a wide range of values of the couplings, including the dipolar case truncated to $3||$ neighbours.

Besides the descending order of the correlations, a qualitative characteristic of the experimental results is the presence of a non-zero magnetisation, which we try to account for by introducing a magnetic field. We have seen that if there is a field in the experiment, we expect it to be very roughly of the order of $J_2 \lesssim h \lesssim J_1$; in addition, when considering the percentages of triangles with a given magnetisation, the nearest neighbour model predicts $h \sim 1.6 J_1$. It thus seems sufficient to focus mainly on the two first magnetisation plateaus. This is confirmed by the comparison between the experimental magnetisation and the Monte Carlo simulations prediction for the magnetisation in Fig.~\ref{fig:J1J2J3m}. This result shows that it is sufficient to study the behaviour in temperature for fields $6J_{3||} < h < 4(J_1+J_2)-6J_{3||}$ corresponding to the $ m = 1/3$ plateau. Figs.~\ref{fig:J1J2J3m}, \ref{fig:J1J2J3pspecificheat}, and~\ref{fig:J1J2J3pOneThirdCorrels} give a consistent picture of the behaviour of the model as a finite temperature is applied on the $m = 1/3$ plateau ground state. At small fields, there is an intermediate temperature region where the system essentially behaves as the further neighbour model in zero field: the residual entropy of the nearest neighbour model is recovered (Fig.~\ref{fig:J1J2J3pspecificheat}) but the $c_2$ and $c_{3||}$ correlations are already inverted as compared to the nearest neighbour model (Fig.~\ref{fig:J1J2J3pOneThirdCorrels}). With increasing fields, the transition to the long range ordered ground state (stripe phase) happens at higher temperatures, with stronger and stronger finite size effects, reaches a maximum, then decreases again (Figs.~\ref{fig:J1J2J3m} and~\ref{fig:J1J2J3phcrossings}). For $h = 4J_1$, at intermediate temperatures, we find approximately the residual entropy of the nearest neighbour model (Fig.~\ref{fig:J1J2J3pspecificheat}), with again an effect of the further neighbour couplings on the $c_2$ and $c_{3||}$ correlations (Fig.~\ref{fig:J1J2J3m}).

We want to verify that the zero field $c_{3||}~>~c_2$ region is preserved in the presence of a magnetic field. In order to compare to the experimental results, we show a map of the descending order of the first four spin-spin correlations in Fig.~\ref{fig:J1J2J3phcrossings}, for $J_2 = 0.023 J_1$ and $J_{3||} = 0.0103 J_1$, as a function of the field and temperature. The region where the descending order of the correlations in the experiment ($c_{3||} > c_2 > c_{3\star}$) is reproduced by the Monte Carlo simulations is shown by a colour map, which gives the $c_{3||}/c_2$ ratio. For small fields $0 \leq h/J_1 \lesssim 0.5$ and temperatures $0.1 \lesssim T/J_1 \lesssim 1$, the $c_{3||}>c_2$ region indeed survives (with a reasonably large ratio of $c_{3||}/c_2$). At fields ranging from $0.8 J_1$ to $3.2 J_1$, $c_2$ becomes larger than $c_{3||}$ for $0.15 \leq T/J_1 \leq 1.5$, which is consistent with the results of the nearest neighbour model in a field, for which in this region the $c_2$ correlations are much larger than the $c_{3||}$ correlations. At large fields, $c_2$ decreases again such that $c_{3||}>c_2$ is valid for fields $3.5 \lesssim h/J_1 \lesssim 4 $. Thus, even in the presence of an external magnetic field in the further neighbour model, there are still regions that account for the order of spin-spin correlations in the experiment, with a reasonable ratio of $c_{3||}$ to $c_2$.

\begin{figure}[t]
    \centering
    \includegraphics[width =0.48\textwidth]{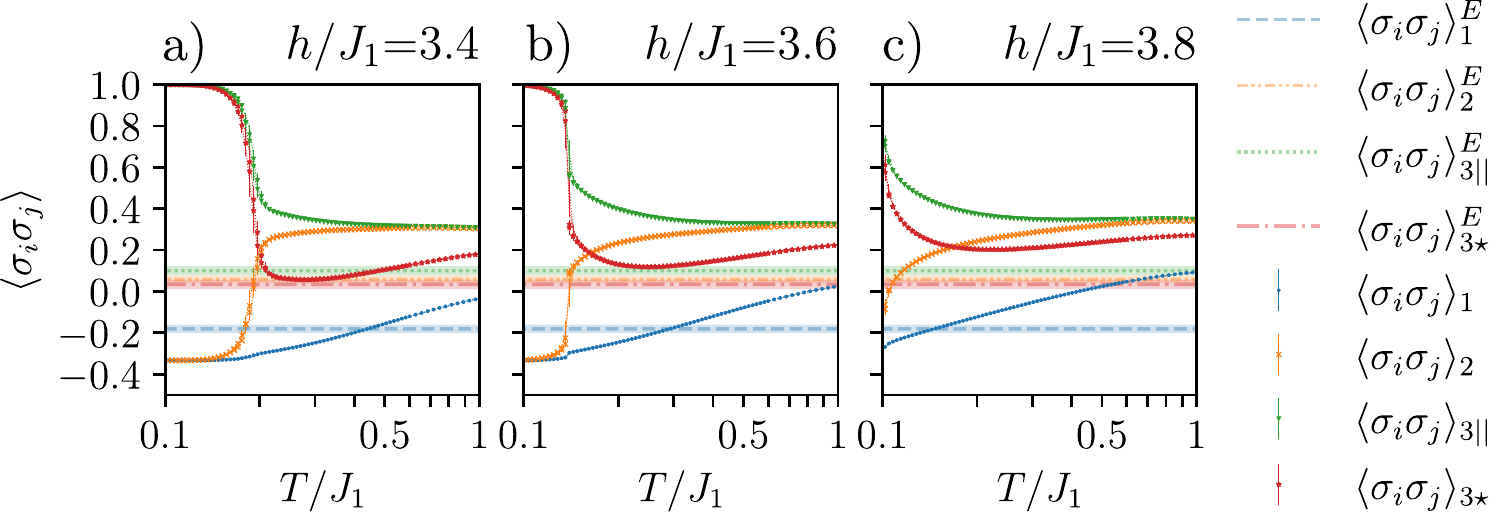}
    \caption{(Region~\textbf{B} in Fig.~\ref{fig:J1J2J3phcrossings}) Behaviour of the first few (disconnected) correlations at large magnetic field, in the window where $c_{3||}/c_2$ is of the same order of magnitude as in the experiment. The proportion of frustrated triangles is directly related to $\langle \sigma_i \sigma_j \rangle_1$; in this region, both are in agreement with the experimental value. Since the magnetisation is very different from the experimental value, we plot the disconnected correlations (i.e. the correlations without removal of the squared magnetisation) for a valid comparison.} 
    \label{fig:J1J2J3phmean}
\end{figure}
\begin{figure}[t]
    \centering
    \includegraphics[width =0.48\textwidth]{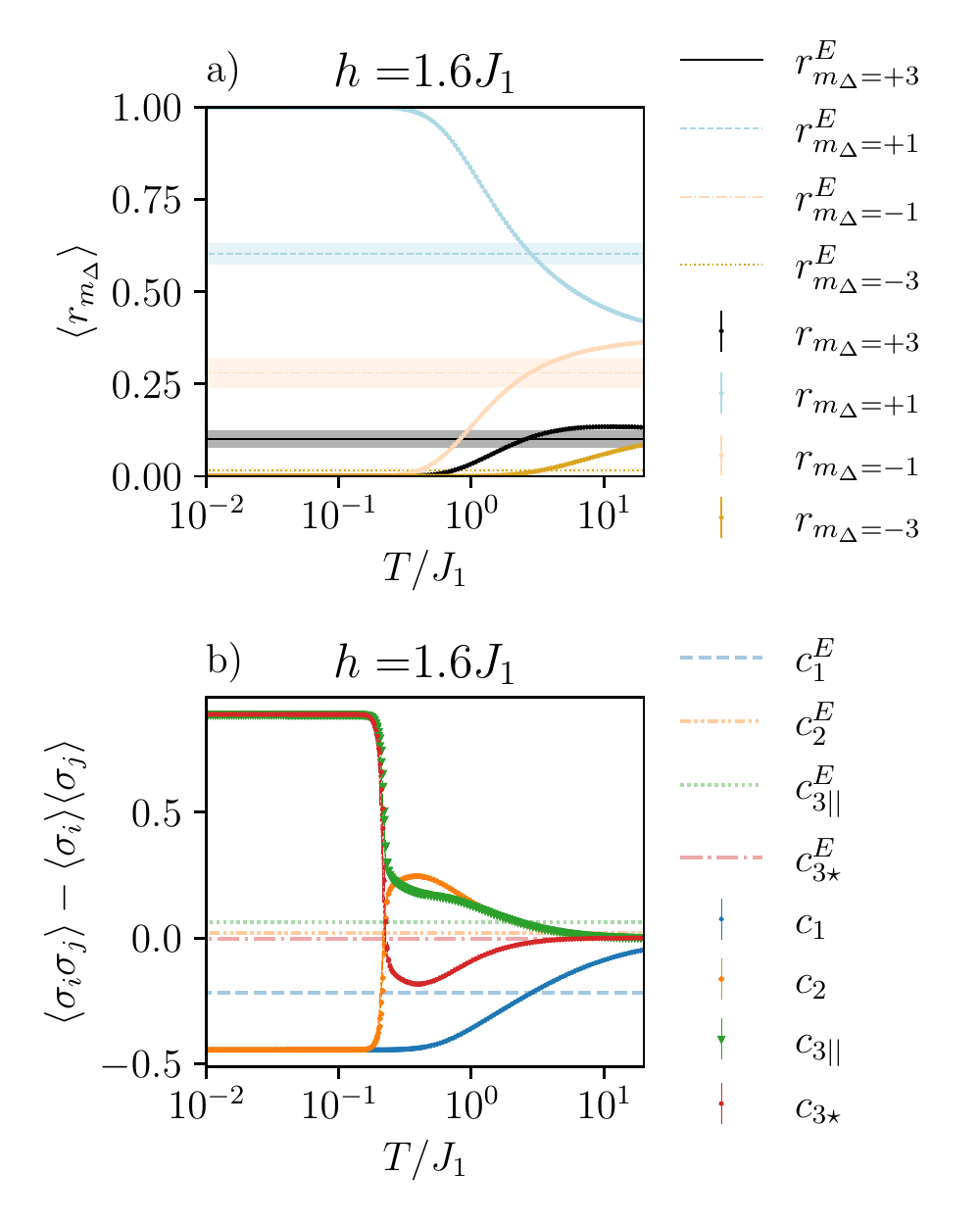}
    \caption{(Region~\textbf{C} in Fig.~\ref{fig:J1J2J3phcrossings}) Monte Carlo results for two sizes ($N = 900, 1296$). (a) Expectation value for the proportion of triangles with magnetisation $m = -3, -1, +1, +3$ as a function of the temperature for an intermediate field. We recover the result of the nearest neighbour model in a field (Fig.~\ref{fig:trianglesratiosJ1h}. (b) Behaviour of the first few correlations at the same field. At intermediate temperatures, $c_2 > c_{3||}$, and at large temperatures, $c_2 \lesssim c_{3||}$ (the two are essentially equal). } 
    \label{fig:J1J2J3phC}
\end{figure}

In Fig.~\ref{fig:J1J2J3phcrossings}, the hatched regions correspond to the values of field and temperature where the theoretical magnetisation and the proportion of ferromagnetic triangles agree within the error bars with the experimental values. Although it demonstrates that it is not possible to account for all the experimental results at once, this graph highlights three regions of interest:
\begin{itemize}
    \item region~\textbf{A} at low field $h/J_1\sim 0.2$~to~$0.5$ and temperatures $0.2 \lesssim T/J_1 \lesssim 0.4$ where the Monte Carlo results for the magnetisation match the experimental results and where the ratio of $c_{3||}$ to $c_2$ is similar to the experimental value, but where the Monte Carlo simulations predict zero ice-rule-breaking triangles. The corresponding correlation results are illustrated in the panel (a) ($h = 0.2J_1$) of Fig.~\ref{fig:J1J2J3pOneThirdCorrels}; it can be seen that they are quite different from the experimental results.
    \item region~\textbf{B} at large field $3.4 \lesssim h/J_1\lesssim 3.8$ and intermediate temperatures $0.15 \lesssim T/J_1 \lesssim 0.5$, where the combined effect of the temperature and magnetic field reproduces the experimental proportion of ice-rule-breaking triangles, and where the ratio of $c_{3||}$ to $c_2$ is again similar to the experimental value. However, in this region, the Monte Carlo simulations predict $\langle r_{m_{\Delta}=-1}\rangle = \langle r_{m_{\Delta} \rangle =-3} =0 $ (all triangles are aligned with the field). Correspondingly, the magnetisation predicted by the Monte Carlo simulations is around twice the experimental results ($0.35 \lesssim m \lesssim 0.5$ instead of $m \sim 0.2$). The disconnected correlations for this region are shown in Fig.~\ref{fig:J1J2J3phmean}; the further neighbour correlations are larger than their experimental counterpart. 
    \item region~\textbf{C} at intermediate field $1 \lesssim h/J_1 \lesssim 2$ and high temperatures $2 \lesssim T/J_1 \lesssim 4$ where the number of triangles and the magnetisation are both compatible with the experiment. Just as in the nearest neighbour model in a field, requiring that all the percentages of triangles with given magnetisation match with the experiment gives $h \cong 1.6 J_1$ (Fig.~\ref{fig:J1J2J3phC}a). However, Figs.~\ref{fig:J1J2J3phcrossings} and~\ref{fig:J1J2J3phC}b both show that, despite taking into account the further neighbour couplings, the correlations $c_{3||}$ and $c_2$ remain essentially equal in this region.
\end{itemize}

\section{Discussion}
\label{sec:discussion}
From the analysis of the experimental data, we extract the main observation that $|c_1| > c_{3||} > c_2 \gtrsim c_{3\star}$. There are two additional qualitative observations: the systematic presence of a non-negligible proportion of ice-rule-breaking triangles and of a finite magnetisation for all the samples, the magnetisation of all the samples having the same sign. These two observations correspond to having precise proportions of triangles with magnetisation $-3, -1, +1$ and~$+3$, given in Table~\ref{tab:expResultsNumberTriangles}.

From the micromagnetic simulations, we obtain that the effect of the IP regions in the samples is first and foremost to increase the nearest neighbour couplings significantly (approximately by a factor of 10), and second to slightly increase the second and third neighbour couplings, resulting in a system where the next nearest neighbour coupling $J_2$ is of the order of 2\% of the nearest neighbour coupling $J_1$, and the third nearest neighbour coupling $J_{3||}$ is of the order of 1\% of $J_1$.

Basing our analysis on the Monte Carlo and tensor networks simulations, we have asked the following questions:
\begin{enumerate}
    \item Considering an equilibrium distribution characterised by two Lagrange parameters - temperature and field - and corresponding to a model involving only nearest neighbour couplings, can we reproduce the experimental proportions of triangles having a given magnetisation?
    \item Considering this equilibrium distribution, is it possible to describe the descending order of the first spin-spin correlations in the experiment, or does one have to consider a model involving further neighbour couplings? If so, is there a limit on how small these couplings have to be to become negligible?
    \item Finally, does the resulting equilibrium fully describe the experimental results, including the percentages of triangles with $m = +3, +1, -1, -3$ and the actual values of the correlations?
\end{enumerate}

The data in Fig.~\ref{fig:trianglesratiosJ1h} allows us to easily determine that a simple model, the nearest neighbour Ising antiferromagnet in a field, admits a region of field and temperature matching the experimental results for the number of triangles with a given magnetisation. However, the corresponding temperature is quite large ($T~\sim~2.8J_1$).

When taking into consideration the experimental spin-spin correlations and comparing them to the nearest-neighbour Ising antiferromagnet (Figs.~\ref{fig:ExactResults}~and~\ref{fig:overviewcorrJ1h}), it is quite clear that the system is not accurately modelled by an equilibrium distribution if one simply neglects the further neighbour couplings altogether. Instead, the simulations of the $J_1-J_2$ model confirm that even a small next nearest neighbour coupling plays an essential role, affecting the correlations in a significant way (Figs.~\ref{fig:J2Correlations},~\ref{fig:J2Crossings} and~\ref{fig:J2J3pCorrelations}). In the ground state, this is obviously the case because further neighbour couplings lift partially the macroscopic ground state degeneracy of the nearest neighbour model. What our results show is that, in addition, the correlations are also affected up to large temperatures, even for small values of $J_2$. For intermediate temperatures ($0.1 J_1 \lesssim T \lesssim 0.5J_1$), this modification can be sufficient to explain the descending order of the experimental correlations with respect to one another for $J_2 \gtrsim 0.02$ (Fig.~\ref{fig:J2Correlations}); for large temperatures ($T \sim 3J_1$), the next nearest neighbour couplings need to be larger ($J_2 \gtrsim 0.06$) to obtain the same result. The micromagnetic simulations predict a third neighbour ($J_{3||}$) coupling of the order of $J_2/2$ (Fig.~\ref{fig:micromagcouplings}). This coupling naturally competes with the $J_2$ coupling; in the ground state, it further lifts the degeneracy, leading to a reduced residual entropy (Eq.~\ref{eq:SJ1J2J3p}), and it affects the correlations (Fig.~\ref{fig:GSJ1J2J3ph}). At finite temperature, this third neighbour coupling modifies the correlations as compared to the $J_1-J_2$ model, and therefore has to be taken into account. Despite this competition, a region with $c_{3||} \gtrsim 2 c_2$ is still present (Fig.~\ref{fig:J2J3pCorrelations}).

In the further neighbour model, the magnetic field plays a similar role as in the nearest neighbour model: combined with the temperature, it selects the proportion of triangles with a certain magnetisation, setting the overall magnetisation (Fig.~\ref{fig:J1J2J3m}) and the number of frustrated triangles (Fig.~\ref{fig:J1J2J3phcrossings}). Additionally, it affects the correlations both in the ground state and at finite temperature. In the ground state, depending on the value of the field, the degeneracy gets either partially or completely lifted (Fig.~\ref{fig:GSJ1J2J3ph}) and, at experimentally relevant fields, the third neighbour correlations can vary from $c_{3||} = 26/81$ ($m = 1/9$ plateau) to $c_{3||} = 8/9$ ($m = 1/3$ plateau). At finite temperature, the region where $c_{3||}$ is larger than $2 c_2$ is preserved for two ranges of magnetic fields, illustrated by regions \textbf{A} and \textbf{B} in Fig.~\ref{fig:J1J2J3phcrossings}.

While the data in Fig.~\ref{fig:J1J2J3phcrossings} demonstrates that the descending order of the correlations in the experiment can be recovered provided that further neighbour couplings are considered, this data also illustrates the impossibility of explaining simultaneously the entirety of the results based on these equilibrium distributions: the ratio of $c_{3||}$ to $c_2$ in most of the samples cannot be explained at the same time as the value of the magnetisation and the proportion of ice-rule-breaking triangles. Based on the available information about the experiment, it is difficult to know in which further direction to push the models. Nevertheless, three interesting regions can be observed, at intermediate temperatures for small field (region~\textbf{A}) and large field (region~\textbf{B}), and at large temperatures for intermediate fields (region~\textbf{C}). 

In region~\textbf{A}, the magnetisation matches the experimental results, but there is no ice-rule-breaking triangle, and correspondingly the nearest neighbour correlations take their ground state value, which is qualitatively very different from the experiment. To fully account for the experimental results, an additional mechanism would thus have to be invoked. Starting from a region with a smaller magnetic field than region~\textbf{A}, a possibility could be a source of disorder that would not be described by the temperature, such as a disorder in the coupling strengths generated by the IP regions \RESP{ or the presence of sites whose magnetisation is not well defined}, pinning ice-rule-breaking triangles at ``domain walls'' between low-energy states grown on the lattice, thus reducing the spin-spin correlations and fixing the magnetisation.

Region~\textbf{B} is different, in that the percentage of ferromagnetic triangles, and correspondingly the expectation value $\langle \sigma_i \sigma_j \rangle_1$, correspond to the experimental results, but that the magnetisation and the expectation values $\langle \sigma_i \sigma_j \rangle_k$ for the further neighbours are larger than the experimental result. Again, in absence of additional insight regarding the experiment, we can only suggest hypotheses involving another effect. In this case, it could perhaps be that, by switching domains of a certain size, the demagnetisation protocol would preserve the number of ice-rule-breaking triangles, while reducing the further neighbour correlations and the magnetisation.

Finally, region~\textbf{C} is interesting because the experimental proportion of triangles with a given magnetisation is well reproduced by the magnetic field and the temperature: no additional mechanism would be needed to describe the magnetisation or the number of ice-rule-breaking triangles. However, the second and third neighbour correlations $c_2$ and $c_{3||}$ are essentially equal (when $J_2$ and $J_{3||}$ take the values predicted by the micromagnetic simulations), which does not seem to be in full agreement with the experiments. Looking at Fig.~\ref{fig:J2Crossings}, one can see that in the $J_1$-$J_2$ model and for temperatures $T/J_1 \sim 3$, the next nearest neighbour couplings should be of order $J_2 \sim 0.06J_1$ to $J_2 \sim 0.11 J_1$ to reproduce the experimental correlations. This would correspond to at least a factor of $3$ for $J_2/J_1$ as compared to the micromagnetic simulations. This seems unlikely, but we have to note that the micromagnetic simulations are performed for an idealised nanomagnet, and that the use of a square grid in the simulation did create a small difference between the nearest neighbour couplings in two different directions ($J_{1,\text{h}} = 1.868\cdot 10^{-20}$J while $J_{1,\text{d}} = 1.884 \cdot 10^{-20}$J). However, the insight on the value of $J_2$ based on Fig.~\ref{fig:J2Crossings} needs to be taken into account very carefully as it would be affected by the presence of third neighbour couplings and a magnetic field. We have not performed a systematic scan for a range of values of $J_{3||}$ vs $J_2$; we can therefore not conclude what range of couplings would provide a complete match with the experimental results in this region. Another difficulty with this region is conceptual: the corresponding effective temperature seems large and could be interpreted as an attempt to model disorder \RESP{in the coupling strengths}; in this context, it is difficult to decide how far a precise comparison between the predicted and experimental correlations in this region can be pushed to draw meaningful conclusions.

\section{Summary and outlook}
In this paper, we have presented Monte Carlo and tensor network results for a series of short range antiferromagnetic Ising models on the kagome lattice, in the presence of a longitudinal field, computing the first few spin-spin correlations systematically. These models are combined to give the $J_1-J_2-J_{3||}-h$ model, for which we established a candidate ground state phase diagram as well as the temperature dependence of the magnetisation and first correlations. In light of these simulations together with micromagnetic computations, we have studied the experimental results obtained from an array of chirally coupled nanomagnets on the kagome lattice.

The micromagnetic simulations show that arrays of chirally coupled nanomagnet using Dzyaloshinskii-Moriya interactions are a good basis for investigating models with extremely strong nearest neighbour antiferromagnetic couplings. Indeed, even though short range further neighbour couplings are strengthened as well, the micromagnetic simulations suggest that they increase by a much smaller factor, resulting in an effective model which is significantly different from the dipolar Ising case.

Our Monte Carlo and tensor network simulations show that even very small further neighbour couplings significantly affect the first few spin-spin correlations, and that the next nearest neighbour coupling $J_2$ cannot be neglected even if it is as small as 2\% of the nearest neighbour coupling $J_1$. Additionally, our results show how the correlations are affected by the introduction of a third neighbour coupling $J_{3||}$ and a magnetic field $h$. Their effect on the spin-spin correlations is summarised in Figs.~\ref{fig:GSJ1J2J3ph} (for the ground state) and~\ref{fig:J1J2J3phcrossings} (for the descending order of the correlations as a function of the temperature).

Although Fig.~\ref{fig:J1J2J3phcrossings} shows that it is not possible to obtain a complete quantitative description of all the results of the experiment, it allows to spot two regions of interest (\textbf{A} and~\textbf{B}) at small and at large fields which seem to contain the essence of the experimental results (namely the descending order of the correlations) and might serve as a support to explain the observations modulo the introduction of an additional ingredient (a source of defects at weak field and a mechanism of reduction of the magnetisation at large field ). A third region (\textbf{C}) shows that a combination of an intermediate field and a surprisingly large temperature (suggesting the presence of disorder \RESP{in the coupling strength due to local changes in the DMI or in the IP width, or disorder corresponding to the presence of nanomagnets whose local magnetisation is not well determined from our measurements}) can reproduce the experimental proportions of triangles with a given magnetisation, but not the difference between the second and third neighbour correlations.

We hope that our results can motivate and support further research both on the experimental and theoretical side. Experimentally, our contribution is to make a step in the direction of tuning couplings for frustrated Ising models on the kagome lattice. Realising a similar system but where the in-plane part separating three nearest neighbour sites on the kagome lattice would have a triangular hole might change the effective model. \textit{A priori}, one could expect that the nearest neighbour couplings would remain large but that the effective second and third neighbour couplings would be smaller~\footnote{This would perhaps give results similar to Eqs.~\ref{eq:OOPvsIPJ1}.}. This is, however, a technical challenge because of the small size of the triangle that needs to be created. It is also not clear whether this could help to recover the nearest neighbour correlations, since $J_2$ should still be of the order of 1\% of $J_1$; as Figs.~\ref{fig:J2Correlations} and ~\ref{fig:J2J3pCorrelations} show, the results will also depend on the ratio of $J_{3||}$ to $J_2$. \RESP{Another possible direction of improvement would be to have measurements allowing the determination of the configuration of the IP regions as well as the OOP regions \footnote{For instance, by combining X-PEEM with MFM.}, to confirm the micromagnetic simulations predictions and check that the IP regions do not get stuck in configurations which do not minimise the energy.}

On the theoretical side, we have uncovered the phase diagram of the $J_1>J_2>J_{3||}$ model in a field, exhibiting an interesting range of macroscopically degenerate phases, some with sub-extensive residual entropy, hopefully motivating further investigation. Besides the Hamiltonian considered here, \RESP{ we see two main directions for more involved models. The first is to determine whether the observed effects could be due to disorder in the coupling strengths or the presence of vacant sites, by modelling the effect that such disorder would have on the spin-spin correlations. The second, assuming absence of disorder in coupling strength, would be to perform micromagnetic simulations on larger structures to determine the impact that the IP regions have on effective long range couplings in this system, and, in particular, to see if these long range couplings are different from the well-studied dipolar model. Comparing the values of these longer range couplings to the predicted short range four site couplings (Fig.~\ref{fig:5sitescouplings}) seems to be an essential step to build a more accurate model including the next order of interactions}. \\

{\it Acknowledgements.} Jeanne Colbois and Fr\'ed\'eric Mila would like to thank the QUTE group at Ghent University for giving them access to their vumps implementation, and in particular Frank Verstraete, Laurens Vanderstraeten and Bram Vanhecke for a number of fruitful discussions on tensor networks and frustration. They are also grateful to Andrew Smerald for relevant discussions on the $J_1-J_2$ model and Monte Carlo simulations. Jeanne Colbois also thanks Natalia Chepiga, Jonathan D'Emidio, Mithilesh Nayak and Patrick Emonts for interesting blackboard discussions. Jeanne Colbois and Frédéric Mila are supported by the Swiss National Science Foundation (Project Number : 200020\_182179). Laura J. Heyderman and Kevin Hofhuis acknowledge financial support from the Swiss National Science Foundation (Project Number : 200020\_172774). Xueqiao Wang acknowledges the ThinkSwiss Research Scholarship. Ale\v{s} Hrabec was funded by the European Union's Horizon 2020 research and innovation program under Marie Skłodowska-Curie grant agreement number 794207 (ASIQS). Part of the computations have been performed using the facilities of the Scientific IT and Application Support Center of EPFL (SCITAS).
\appendix
\begin{figure*}
    \centering
    \includegraphics[width =0.65\textwidth]{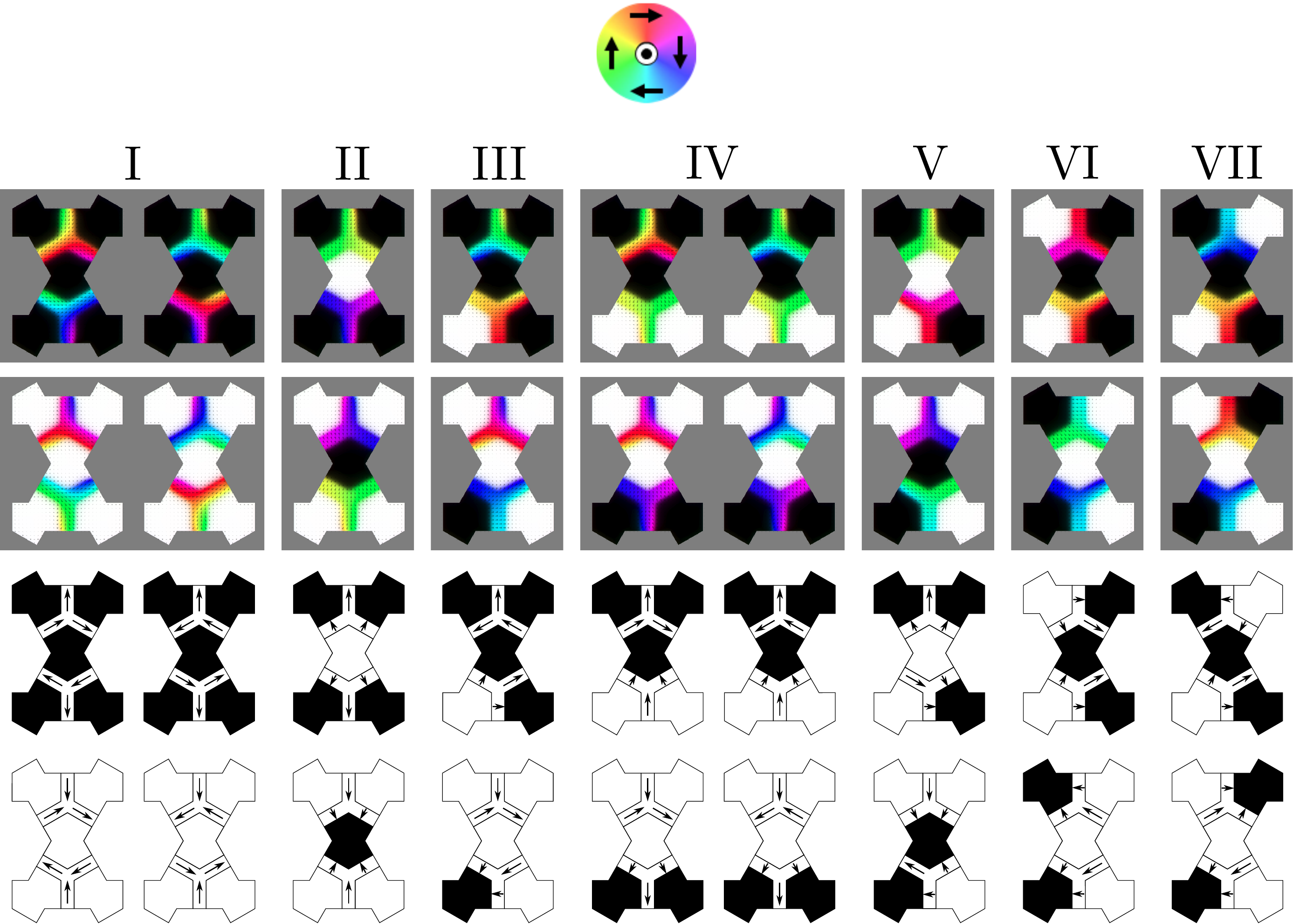}
    \caption{First row: micromagnetic simulations results (after relaxation) for each OOP configuration from Fig.~\ref{fig:5sitescouplings}. For each OOP configuration, we show the IP configuration that minimises the energy. The coloured disk at the top gives the legend for the IP magnetisation. For the OOP regions, black means up Ising spin, white means down Ising spin. Second row: IP configurations that minimise the energy when all the spins are flipped. Here, we do not show the micromagnetic results for applying rotations and/or mirror symmetries, but they have been used in determining the average energy and the error bars for the results in Table~\ref{tab:MicromagneticResults}. Third and fourth rows: sketch illustrating the OOP and IP configurations from the first and second rows, for readability.} 
    \label{fig:5sitesconfigurations}
\end{figure*}
\begin{table*}[t]
\caption{\label{tab:MicromagneticResults} Results of the micromagnetic simulations for the energies of the various configurations from Fig.~\ref{fig:5sitescouplings}, for an IP width of 50~nm.}
\begin{ruledtabular}
\begin{tabular}{@{}rlll@{}}
\multicolumn{1}{l}{} & Expression of the Energy                                                             & Energy (pure OOP)~[$10^{-17}$J]        & Energy (OOP + IP)~[$10^{-17}$J]    \\
\colrule
I                    & $E_0 + 4J_{1,\text{d}} + 2J_{1,\text{h}} + 2J_2 + 2J_{3||} + Q_1 + 4 Q_2$ & $-1.645897 \pm 6\cdot 10^{-7}$  & $-1.2408601 \pm 5 \cdot 10^{-7}$ \\
II                   & $E_0 - 4J_{1,\text{d}} + 2J_{1,\text{h}} + 2J_2 + 2J_{3||} + Q_1 - 4 Q_2$ & $-1.6609718 \pm 2\cdot 10^{-7}$ & $-1.367404 \pm 1 \cdot 10^{-6}$  \\
III                  & $E_0 +2 J_{1,\text{d}} - Q_1 - 2 Q_2$                                     & $-1.65413964\pm 8\cdot 10^{-7}$ & $-1.3064207 \pm 4 \cdot 10^{-7}$ \\
IV                   & $E_0 + 2J_{1,\text{h}} - 2J_2 - 2J_{3||} + Q_1 $                          & $-1.6549028\pm 2\cdot 10^{-7}$  & $-1.3062638 \pm 4 \cdot 10^{-7}$ \\
V                    & $E_0 - 2J_{1,\text{d}} - Q_1 + 2 Q_2$                                     & $-1.6616731 \pm 7\cdot 10^{-7}$ & $-1.3693708 \pm 5 \cdot 10^{-7}$ \\
VI                   & $E_0 - 2J_{1,\text{h}} + 2J_2 - 2J_{3||} + Q_1$                           & $-1.661483 \pm 2\cdot 10^{-7}$  & $-1.3699101 \pm 6 \cdot 10^{-7}$ \\
VII                  & $E_0 - 2J_{1,\text{h}} - 2J_2 + 2J_{3||} + Q_1$                           & $-1.661797 \pm 9\cdot 10^{-7}$  & $-1.3707522 \pm 5 \cdot 10^{-7}$ \\ 
\end{tabular}
\end{ruledtabular}
\end{table*}

\section{Demagnetisation protocol}
\label{sec:AppDemag}
\begin{figure}
\centering
\includegraphics[width=0.5\textwidth]{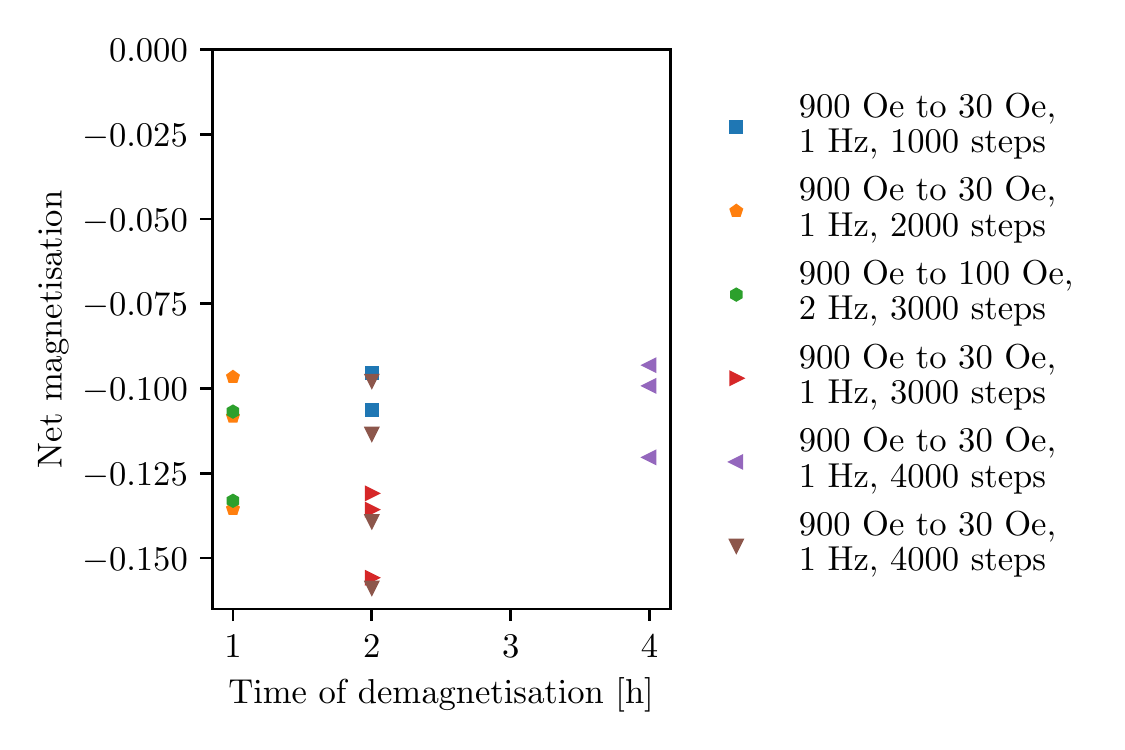}\\
\caption{
\label{fig:demag}\RESP{MFM optimisation of the demagnetisation protocol, displaying various 1, 2, and 4 hours protocols. Different colours at each protocol time indicate different protocols, and each point represents an individual array whose net magnetisation is computed. The net magnetisation cannot be driven closer to zero than -0.09. The more extended protocols do not show a clear trend towards reducing the net magnetisation. The larger spread at 2 hours is due to more samples being tested. In order to be efficient in measuring our 20 arrays, the 2 hours protocol was chosen.}}
\end{figure}

\RESP{Several demagnetisation protocols with different parameters were tested using the net magnetisation of a specific array on a sample as a measure of success. However, the spread in the net magnetisation for different samples was larger than any improvement in the magnetisation achieved with the protocols when going from a 1 hour protocol to 4 hours (Fig. \ref{fig:demag}), and a zero net magnetisation was not reached. This is different from results for dipolar coupled nanomagnets, where it has been found advantageous to use longer protocols \cite{Wang2007}, as is demonstrated for instance by Ref.~\onlinecite{Parakkat2019}. While we did not observe such a clear trend in our demagnetisation protocols, we should note that our samples and protocols are significantly different from the usual dipolar coupled nanomagnets as our samples have chirally interacting IP and OOP regions. In addition, we do not rotate the sample in the demagnetising field. In our chirally coupled systems, rotating the sample would induce strong interactions between external fields and the IP regions.}

\section{Micromagnetic simulations results}
\label{sec:AppMicromagnetic}
The aim is to estimate the values of the various coupling constants relative to the full nearest neighbour coupling, to determine if they are large enough to explain the inversion of the $c_2$ and $c_{3||}$ correlations observed in Fig.~\ref{fig:expSpinSpin} as compared to the nearest neighbour models in Figs.~\ref{fig:ExactResults},~\ref{fig:overviewcorrJ1h}. To do this, we simulate the various configurations of Fig.~\ref{fig:5sitescouplings} for the Ising spins, and let the system relax. We then compare their respective energies to extract the effective couplings.
For the configurations of Fig.~\ref{fig:5sitescouplings}, indexed by roman numbers, we obtain the micromagnetic results shown in Fig.~\ref{fig:5sitesconfigurations}. In Table~\ref{tab:MicromagneticResults}, we summarise the expression of the energy (based on Fig.~\ref{fig:5sitescouplings}), and the value of the energy in the micromagnetic simulations, first in the pure out-of-plane case (dipolar couplings, no IP region), and second in the complete case shown in Fig.~\ref{fig:5sitesconfigurations}. The errors in the energies are estimated from the differences in the energies of configurations that belong to the same group.

\section{Monte Carlo simulations}
\label{sec:AppMC}

The Monte Carlo simulations algorithms depend on the problems we are studying. In zero field, we use the algorithm of Rakala and Damle~\cite{Rakala2017}, which consists in mapping spin configurations to dimer configurations on the dual (dice) lattice, and building loop updates in this dual space which respect local detailed balance and can thus be accepted once they close. In the presence of a field, the dual updates are typically rejected, and we resort to a simpler single-spin-flip update. This is, however, not sufficient, as those updates are typically rejected at low temperature. To alleviate the problem, we use replicas in field and in temperature (known as parallel tempering in the case of the replicas in temperature)~\cite{Loh2008,Soldatov2019,Hukushima1996}. When we cannot compare to the tensor networks results, the simulations are always completed for a number of sizes until convergence, and for at least two independent runs for each size. The correlations are always computed as an average over the sample, and then an ensemble average over the set of measurements. The errors are estimated from a binning analysis~\cite{Ambegaokar2009}. Each Monte Carlo step (MCS) consists of $(N_{\text{spin}},N_{\text{worm}}, N_{\text{rep}})$ single spin flip, worm and replica updates. The parameters for the various Figures are as follows:

\begin{enumerate}
    \item Fig. \ref{fig:ExactResults} : $ (8N_{\text{sites}},8N_{\text{sites}},1)$; thermalisation - 1024 MCS, measurements - 16384 spaced by 3 MCS;
    \item Fig. \ref{fig:overviewcorrJ1h} : $ (8N_{\text{sites}},8N_{\text{sites}},1)$; thermalisation - 1024 MCS, measurements - 16384 spaced by 3 MCS;
    \item Figs. \ref{fig:J2Correlations} and \ref{fig:J2entropy} $ :  (4N_{\text{sites}},4N_{\text{sites}},1)$; thermalisation - 4096 MCS, measurements - 32768 spaced by 8 MCS;
    \item Fig.~\ref{fig:J2Crossings} $ : (4N_{\text{sites}},4N_{\text{sites}},1)$; thermalisation - 4096 MCS, measurements - 8192 spaced by 8 MCS;
    \item Fig.~\ref{fig:J1J2J3m} $ (4N_{\text{sites}},0,1)$ : thermalisation - 262144 MCS, measurements - 16384 spaced by 16 MCS (except for the 1/3 plateau, see parameters for Figs.~\ref{fig:J1J2J3pOneThirdCorrels} and \ref{fig:J1J2J3pspecificheat}).
    \item Figs.~\ref{fig:J2J3pCorrelations} and \ref{fig:J2J3pentropy} : $(4N_{\text{sites}},4N_{\text{sites}},1)$; thermalisation - 16384 MCS, measurements - 32768 spaced by 8 MCS;
    \item Figs.~\ref{fig:J1J2J3pspecificheat},~\ref{fig:J1J2J3pOneThirdCorrels} and~\ref{fig:J1J2J3phmean}: $ (4N_{\text{sites}},0,1)$: thermalisation - 65536 MCS, measurements - 8192 spaced by 8 MCS;
    \item Fig.~\ref{fig:J1J2J3phC} : $ (4N_{\text{sites}},0,1)$: thermalisation - 65536 MCS, measurements - 8192 spaced by 8 MCS
\end{enumerate}

The residual entropy is computed from a thermodynamic integration of the energy~\cite{Roma2004}. At inverse temperature $\beta$, one has:
\begin{equation}
    S(\beta) = \ln(2) + \beta E(\beta) - \int_0^\beta E(\beta') d\beta',
\end{equation}
where here $E$ stands for the energy per site.
Because in practice the integration is only done up to $\beta_0 = 1/T_{\text{max}}$, the residual entropy at high temperature is estimated analytically based on the high temperature expansion of the specific heat:
\begin{equation}
    c_{\text{Hight T}}(\beta) \cong \beta^2 (2J_1^2 + 2J_2^2+ 2J_{3||}^2 + h^2).
\end{equation}

\section{More about the $J_1-h$ model}
\label{sec:AppJ1h}
\begin{figure}[t]
    \centering
    \includegraphics[width = 0.48\textwidth]{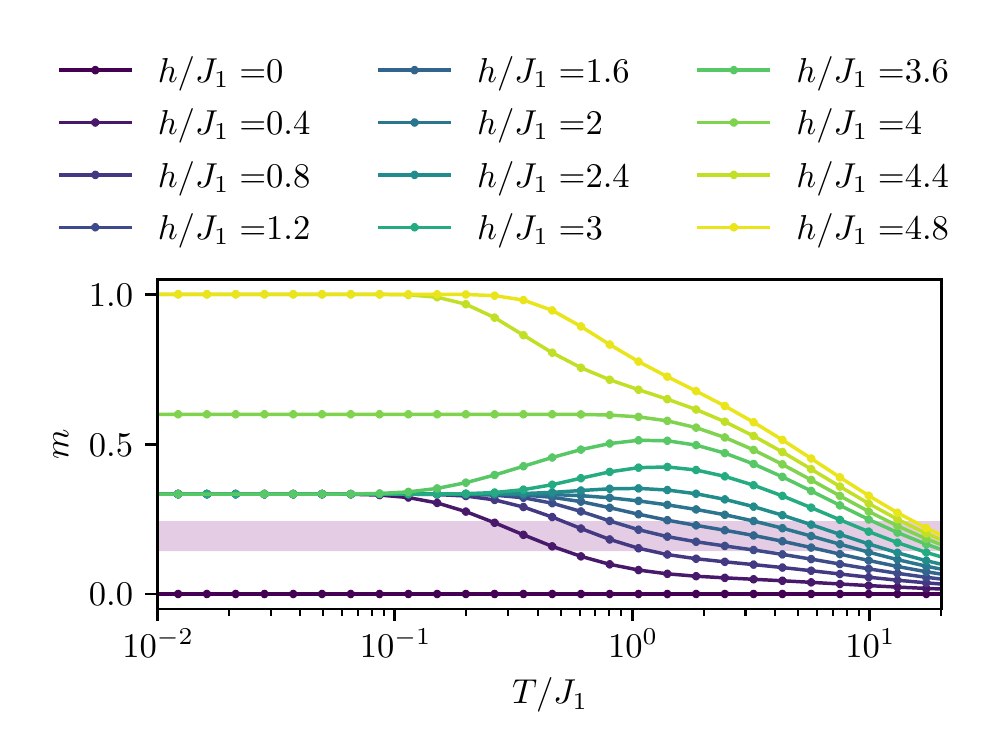}
    \caption{Magnetisation as a function of the field and the temperature for the nearest neighbour Ising antiferromagnet in a magnetic field.}
    \label{fig:J1hm}
\end{figure}

In Fig.~\ref{fig:J1hm}, we give an overview of the magnetisation as a function of the field and the temperature for the nearest neighbour Ising antiferromagnet on the kagome lattice. The highlighted region corresponds to the magnetisation within one standard deviation of the experimental value. The results are obtained from systematic tensor network contractions as described in Sec.~\ref{sec:NN}. For each field and temperature we computed as well the first spin-spin correlations and found that $c_2 \geq c_{3||}$ everywhere. 

\begin{figure}[t]
    \centering
    \includegraphics[width = 0.48\textwidth]{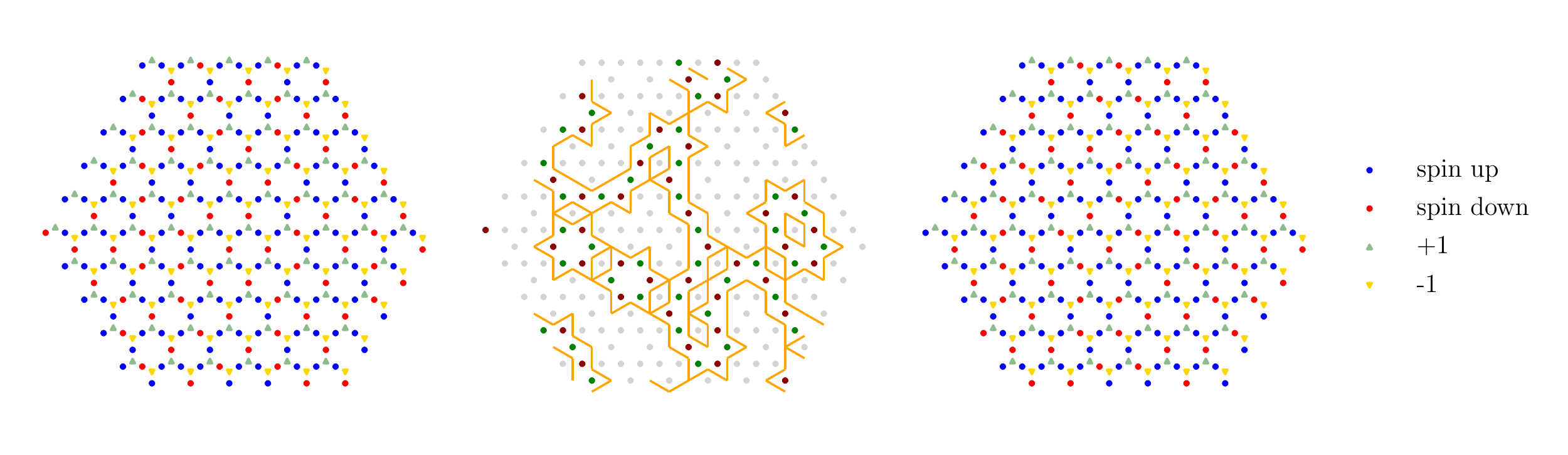}
    \caption{Example of two ground state configurations in the nearest neighbour model in a magnetic field $0 < h < 4 J_1$ (leftmost and rightmost panels), and the difference between the two configurations (central panel). The orange lines delimit the regions where spins are flipped. They correspond to updates in the classical dimer configuration on the (dual) dice lattice. The green dots correspond to spins that are flipped from up to down, and the dark red dots to spins that a flipped from down to up. Comparing the two ground states, it can be seen that the underlying charge configuration is long-range ordered, while the spin configuration is not. }
    \label{fig:J1hGS}
\end{figure}

In Fig.~\ref{fig:J1hGS}, Monte Carlo ground state configurations of the nearest neighbour in a field for the $m = 1/3$ plateau are illustrated. For these configurations, it can be easily verified that the underlying charge configuration of the nearest neighbour model in the 1/3 plateau is long range ordered (corresponding to the 2-up 1-down rule).
\section{More about the $J_1-J_2$ model}
\label{sec:J1J2app}
\begin{figure}[t]
    \centering
    \includegraphics[width = 0.48\textwidth]{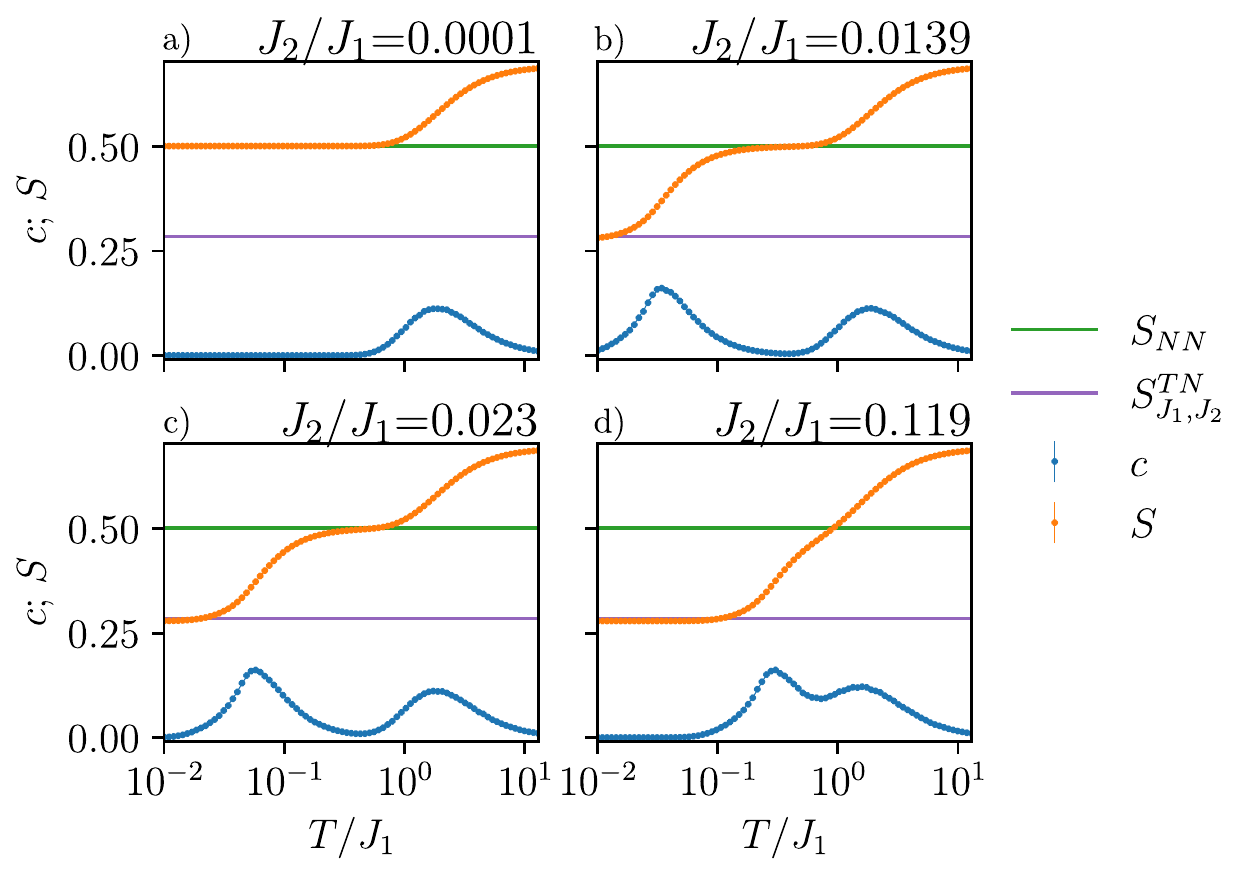}
    \caption{Overview of the behaviour of the specific heat and entropy as a function of the temperature for various values of $J_2$, with Monte Carlo simulations for two system sizes ($N = 576, 1296$). The smaller system size is shown with a line while the larger system size is shown with symbols. }
    \label{fig:J2entropy}
\end{figure}
In Fig. \ref{fig:J2entropy}, the specific heat and the entropy per site as a function of the temperature for two different system sizes in the $J_1-J_2$ model are shown. The selected values of $J_2$ correspond to the panels of Fig. \ref{fig:J2Correlations}. As expected, we find two features in the specific heat - two regimes where a loss of entropy takes place. The feature at larger temperatures, which does not change from one panel to the next, corresponds to the nearest neighbour model, as can be seen from the fact that it leads to an entropy of $S \cong 0.502$. The second feature corresponds to the selection, within the ground state of the nearest neighbour model, of those configurations that respect an ice rule for the $J_2$ kagome sublattices.

\section{More about the $J_1-J_2-J_{3||}$ model in zero field}
\label{sec:AppJ1J2J3}

\begin{figure}[t]
    \centering
    \includegraphics[width = 0.48\textwidth]{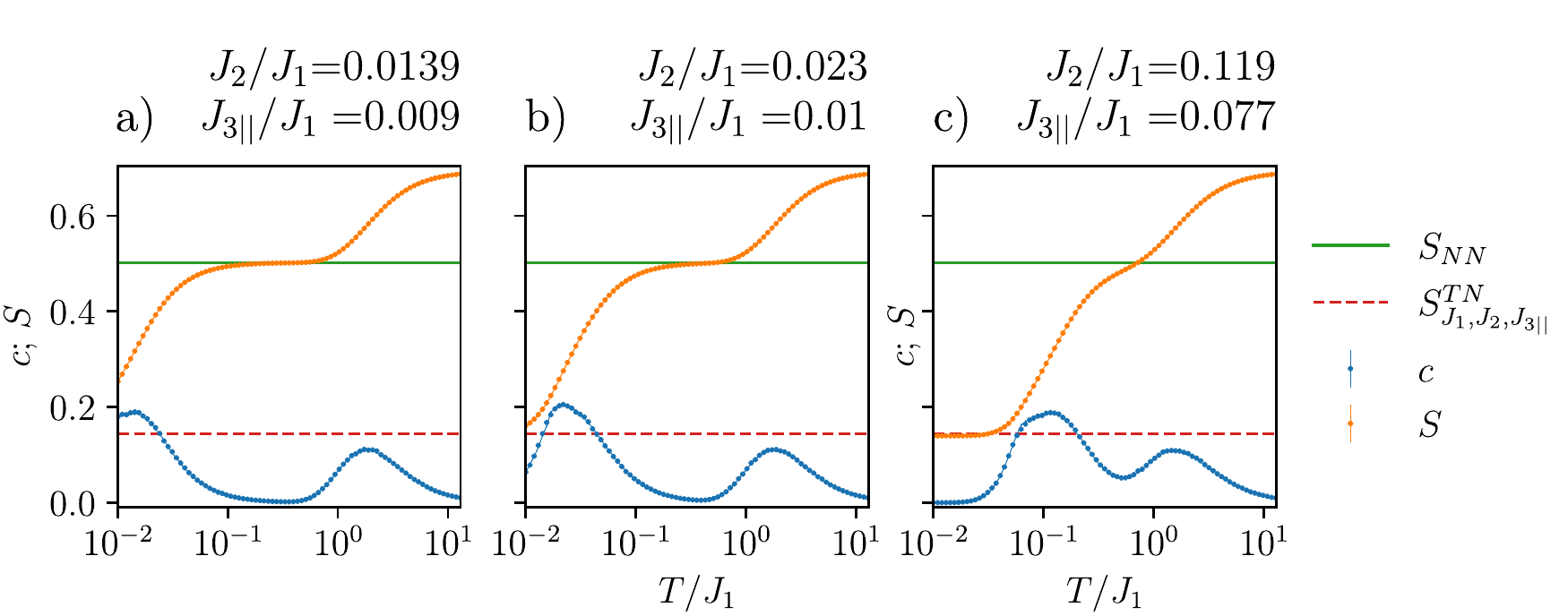}
    \caption{Overview of the behaviour of the specific heat and entropy as a function of the temperature for the values of $J_2$ and $J_{3||}$ from Sec.~\ref{sec:micro}, with Monte Carlo simulations for 3 system sizes ($N = 576, 1296$). The smaller system size is shown with a line while the larger is shown with symbols. }
    \label{fig:J2J3pentropy}
\end{figure}
\begin{figure}[t]
    \centering
    \includegraphics[width = 0.4\textwidth]{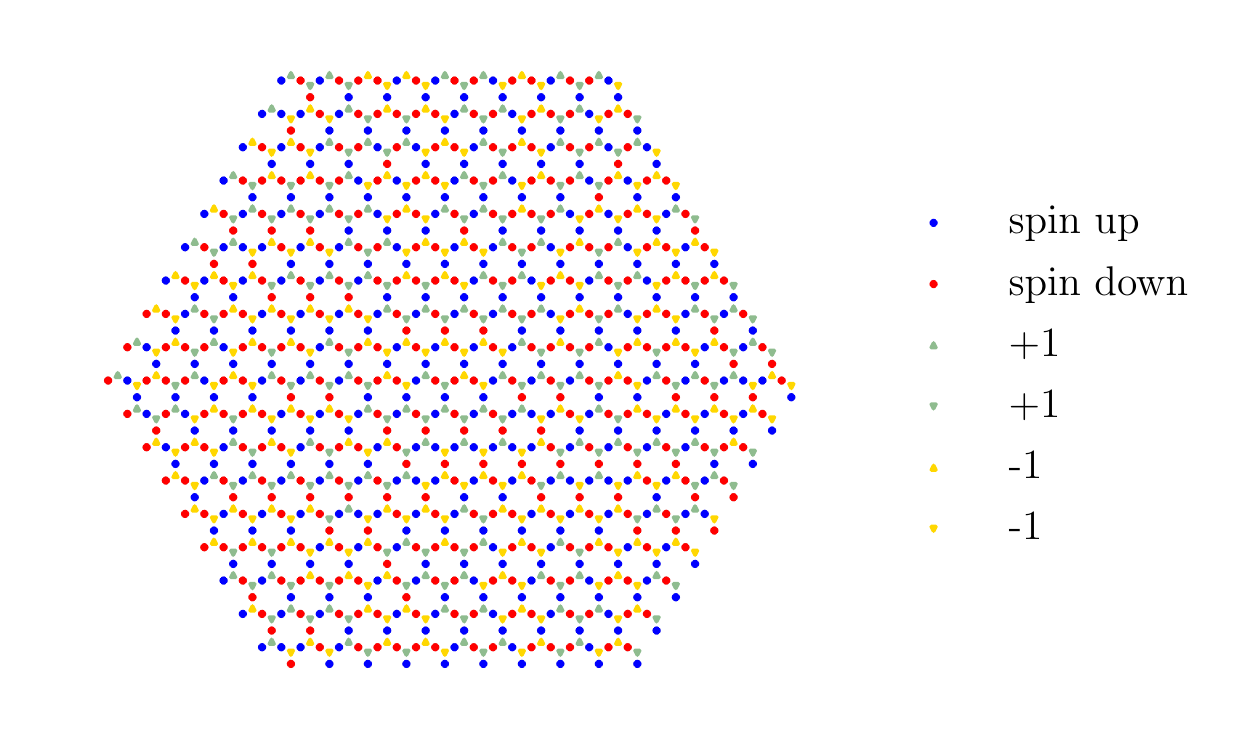}
    \caption{An example of a ground state configuration for the $J_1-J_2-J_{3||}$ model, including the corresponding charge configuration.}
    \label{fig:J1J2J3pGS}
\end{figure}
In Fig. \ref{fig:J2J3pentropy}, we show the specific heat and the entropy per site as a function of the temperature for two different system sizes in the $J_1-J_2-J_{3||}$ model in zero field. The selected values of $J_2$ and $J_{3||}$ correspond to the panels of Fig. \ref{fig:J2J3pCorrelations}, i.e. to the values that have been discussed in the micromagnetic simulations section. Again, we find two peaks in the specific heat. Exactly as in the $J_1-J_2$ case, the feature at larger temperatures corresponds to the nearest neighbour model. The second feature is actually a combined effect of the $J_2$ and $J_{3||}$ couplings. The $J_2$ ice rule is imposed, but additionally, the $c_{3||}$ correlations are restricted to the minimum that they can reach in the $J_1-J_2$ set of ground states. An example ground state is shown in Fig.~\ref{fig:J1J2J3pGS}.

\section{Detail of the $J_1-J_2$-$J_{3||}-h$ ground states}
\label{sec:AppJ1J2J3h}
\begin{figure}[t]
    \centering
    \includegraphics[width = 0.48\textwidth]{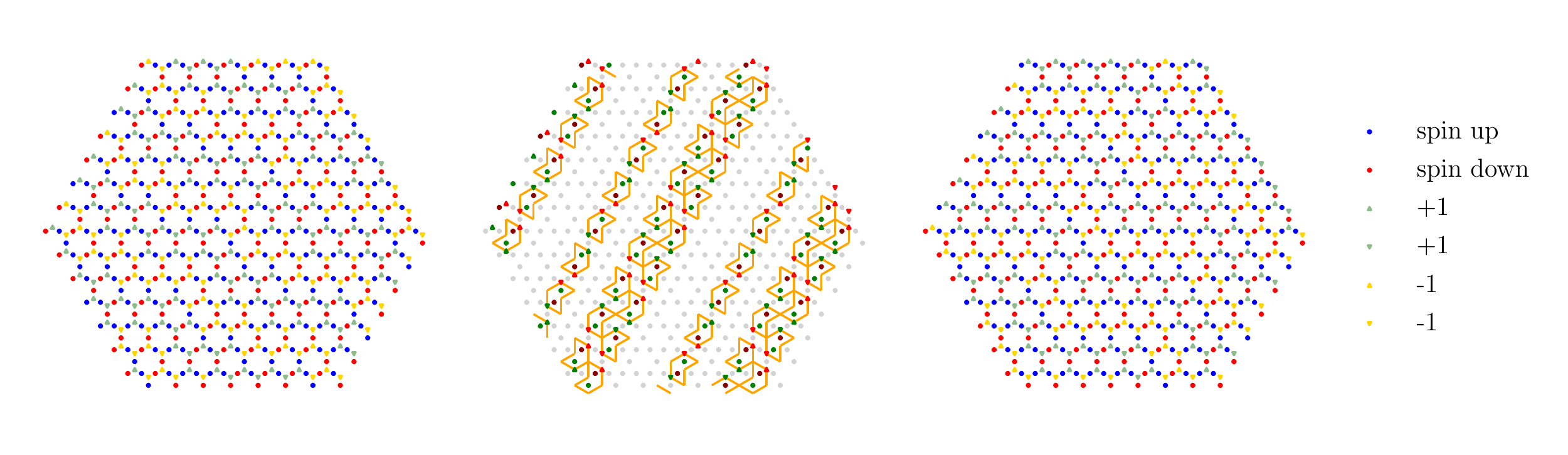}
    \caption{Two examples of ground state configurations for the $J_1-J_2-J_{3||}$ model, including the corresponding charge configuration, for a magnetic field $0 < h < 6J_{3||}$ ($m = 1/9$). The central panel shows the difference between the two configurations, with the following convention: the orange lines delimit the regions where spins are flipped, the green (dark red) dots correspond to spins that are flipped from up to down (down to up), and the red (green) triangles correspond to charges that are changed by +2 (-2).}
    \label{fig:J1J2J3pOneNinthGS}
\end{figure}

As discussed in the main text, the phase diagram of the further neighbour model in a field arises from a selection of a different set of states depending on the value of the magnetic field. In zero field, tensor network results, as well as Monte Carlo results, show that there is a residual entropy even for $0 < J_{3||} < J_2$. Upon introducing the magnetic field, this macroscopic ground state degeneracy is immediately partially lifted. Two examples of such ground states are shown in Fig.~\ref{fig:J1J2J3pOneNinthGS}. The difference between the two configurations illustrates the type of torus-winding updates that can be found (i.e., non-local updates that cross the sample and close in on themselves through the periodic boundary conditions). Each such update can be performed independently. Together with the fact that the tensor network contraction gives zero residual entropy, this provides evidence that the ground state degeneracy is macroscopic but grows exponentially with the linear system size.

\begin{figure}[t]
    \centering
    \includegraphics[width = 0.4\textwidth]{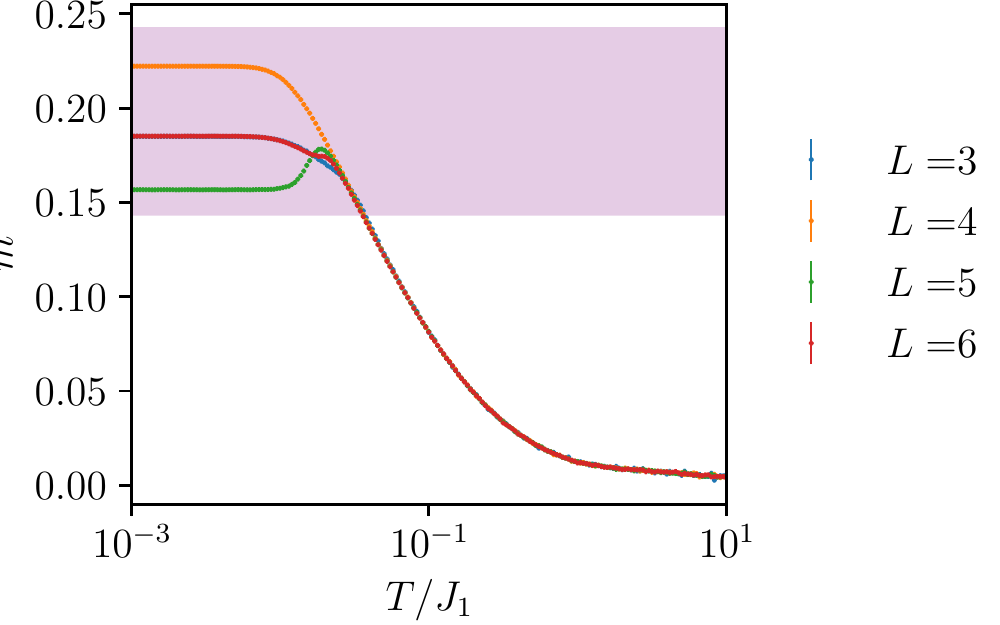}
    \caption{Magnetisation as a function of the temperature for the $J_1-J_2-J_{3||}$ model at $h = 6 J_{3||}$, for various small system sizes. The number of sites is $N = 9 L^2$. The shaded region corresponds to the experimental value. }
    \label{fig:J1J2J3pFirstPTm}
\end{figure}

Upon increasing the magnetic field, there is a first order phase transition in the ground state between the $m = 1/9$ and $m = 1/3$ plateaus. There, the states are a mixture of the $m = 1/9$ ground states and the stripe state of the 1/3 plateau, and the average magnetisation shows a strong dependence in system size (only sizes with a linear size of a multiple of 3 match), Fig. \ref{fig:J1J2J3pFirstPTm}.

In the $m=1/3$ plateau, the charge state is the same as in the corresponding plateau of the nearest neighbour model, but the further neighbour couplings select a long-range ordered stripe state (as long as $J_{3||} < J_2$; for $J_{3||} > J_2$ another long range ordered state is selected).

\begin{figure}[t]
    \centering
    \includegraphics[width = 0.4\textwidth]{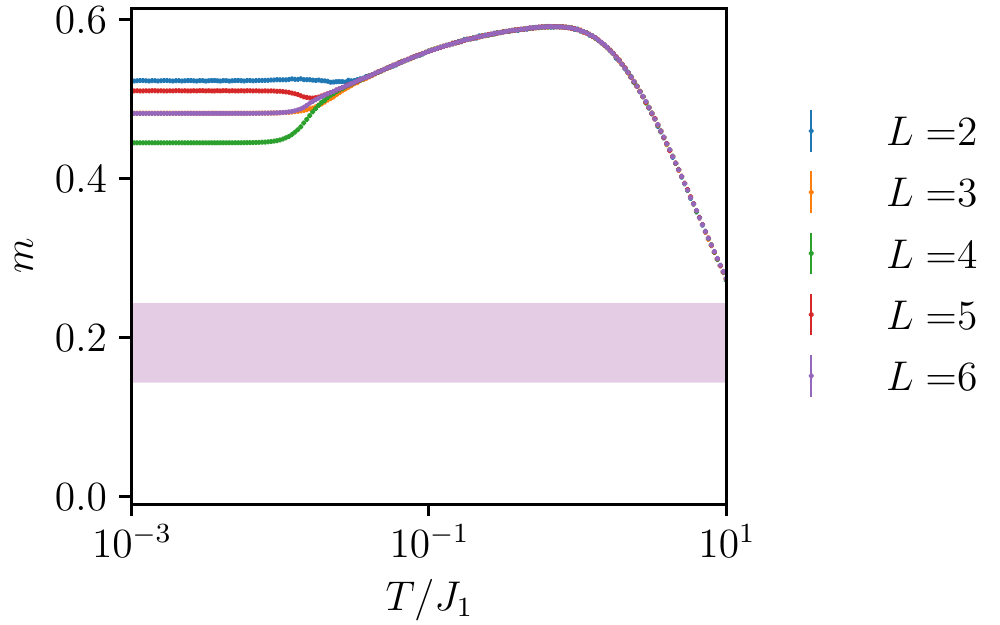}
    \caption{Magnetisation as a function of the temperature for the $J_1-J_2-J_{3||}$ model at $h = 4(J_1+J_2) - 6 J_{3||}$, for various small system sizes. The shaded region corresponds to the experimental value.}
    \label{fig:J1J2J3pSecondPTm}
\end{figure}
At the transition to the $m = 5/9$ plateau, we again find a strong size dependence of the magnetisation (Fig.~\ref{fig:J1J2J3pSecondPTm}). 
\begin{figure}
    \centering
    \includegraphics[width = 0.48\textwidth]{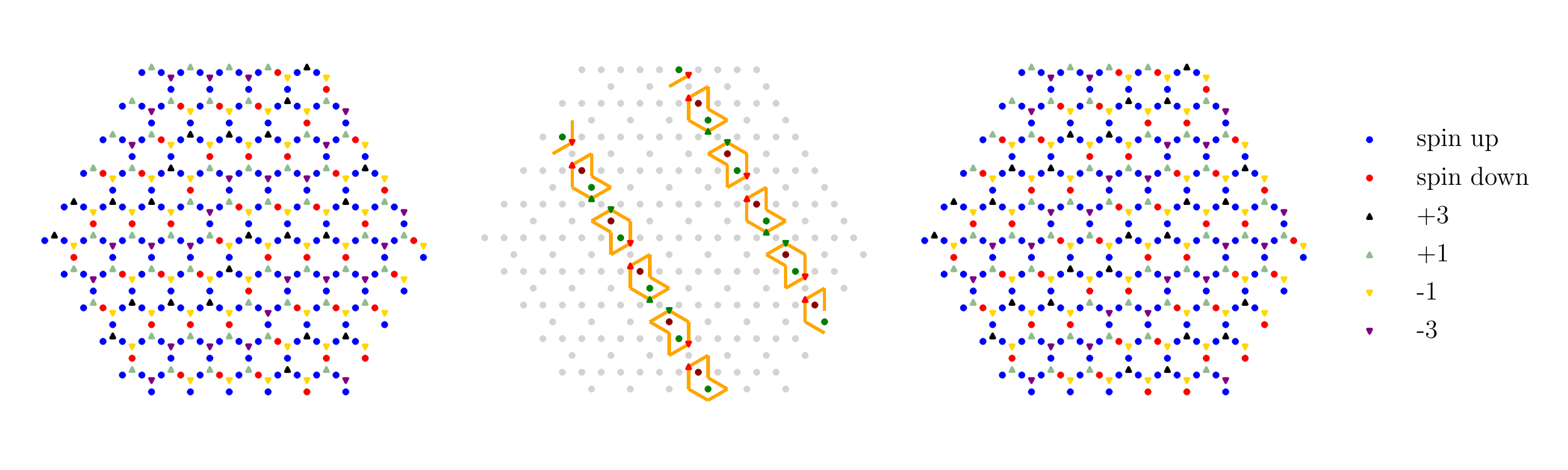}
    \caption{Two examples of ground state configurations for the $J_1-J_2-J_{3||}$ model, including the corresponding charge configuration, for a magnetic field $4(J_1+J_2) - 6 J_{3||}< h < 4(J_1+J_2) - 2 J_{3||}$ ($m = 5/9$). The central panel shows the difference between the two configurations (see the caption Fig.~\ref{fig:J1J2J3pOneNinthGS} for the convention regarding the updates).}
    \label{fig:J1J2J3pFiveNinthGS}
\end{figure}
\begin{figure}
    \centering
    \includegraphics[width = 0.48\textwidth]{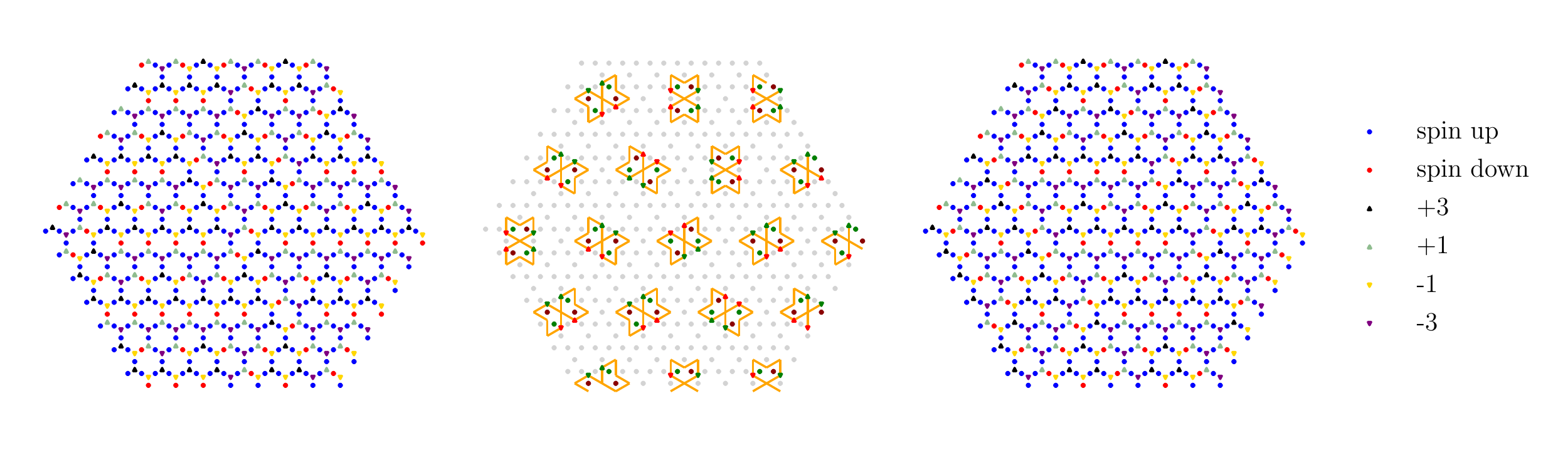}
    \caption{Two examples of ground state configurations for the $J_1-J_2-J_{3||}$ model, including the corresponding charge configuration, for a magnetic field $4(J_1+J_2) - 2 J_{3||}< h < 4(J_1+J_2+ J_{3||})$ ($m = 17/27$). Local updates shown in the central panel bring the system from one configuration to the other (see the caption of Fig.~\ref{fig:J1J2J3pOneNinthGS} for detail).}
    \label{fig:J1J2J3p1727}
\end{figure}
The $m = 5/9$ plateau seems again characterised by a sub-extensive ground state degeneracy, with non-local updates winding the torus (Fig.~\ref{fig:J1J2J3pFiveNinthGS}); however we do not have tensor networks results in this phase. Finally, in the $m = 17/27$ plateau, we find that there are local updates, shown in Fig.~\ref{fig:J1J2J3p1727}, yielding a lower bound for the residual entropy in that phase $S \geq \frac{1}{27} \ln(2)$.\\

The present discussion is a preliminary study of this ground state phase diagram, and much is left to be clarified. Nevertheless, it shows that on the kagome lattice, the residual entropy gets lifted only very progressively - in contrast to the triangular lattice Ising antiferromagnet, for instance. In particular, we want to underline that unlike Refs.~\onlinecite{Mizoguchi2017,Tokushuku2020}, the case where $J_{3||} < J_2$ is \textit{not} at a phase boundary between two ground state phases of the $J_1-J_2-J_{3||}$ model. It is all the more surprising that a finite residual entropy is preserved at some values of the magnetic field.

\section{Data availability}
The experimental data, which served as a basis for this project, as well as the micromagnetic, Monte Carlo and tensor network data produced for this paper, are available on Zenodo~\cite{Zenodo}. The Monte Carlo code for this paper is available upon reasonable request to J.C.

J.C. is the corresponding author for the Monte Carlo and tensor network data, K.H. for the experimental data, and both J.C. and K.H. for the micromagnetic simulations data and analysis.

\bibliography{main}

\end{document}